\documentclass[twocolumn,prd,nofootinbib,superscriptaddress]{revtex4}
\usepackage{color}
\usepackage{amsmath}
\usepackage{mathrsfs}
\usepackage{graphicx}
\usepackage{booktabs}
\usepackage{bm}

\usepackage{adjustbox} %

\newcommand\SkipNeuralNets[1]{}
\newcommand\skipme[1]{}
\newcommand\mysub[1]{\subsubsection{#1}}
\newcommand\E[1]{\left\langle #1\right\rangle}

\newcommand\qmstateproduct[2]{\left\langle#1|#2\right\rangle}

\newcommand\Y[1]{{{}_{#1}Y}}

\newcommand\lnLmarg{ \ln{\cal L}_{\rm marg}{}}

\newcommand\optional[1]{}
\newcommand\unit[1]{{\rm #1}}
\newcommand\editremark[1]{{\color{red}#1}}
\newcommand\response[1]{{#1}}
\newcommand\mc{{{\cal M}_c}}

\def\RIT{Center for Computational Relativity and Gravitation, Rochester Institute of Technology, Rochester, New York
  14623, USA}
\def\UWM{University of Wisconsin-Milwaukee, Milwaukee, WI 53201, USA}
\def\Tokyo{Institute for Cosmic Ray Research, The University of Tokyo,
5-1-5 Kashiwanoha, Kashiwa, Chiba 277-8582, Japan}
\def\Fullerton{California State University Fullerton, Fullerton, CA 92831, USA}
\def\UTAustin{University of Texas, Austin, TX 78712, USA}
\def\GT{Georgia Institute of Technology, Atlanta, GA 30332, USA}
\begin{document}

\title{Expanding RIFT: Improving performance for GW parameter inference}
\author{J. Wofford}
\affiliation{\RIT}
\author{A. B. Yelikar}
\affiliation{\RIT}
\author{Hannah Gallagher}
\affiliation{\RIT}
\author{E. Champion}
\affiliation{\RIT}
\author{D. Wysocki}
\affiliation{\RIT}
\affiliation{\UWM}
\author{V. Delfavero}
\affiliation{\RIT}
\author{J. Lange}
\affiliation{\UTAustin}
\affiliation{\RIT}
\author{C. Rose}
\affiliation{\UWM}
\author{V. Valsan}
\affiliation{\UWM}
\author{S. Morisaki}
\affiliation{\Tokyo}
\affiliation{\UWM}
\author{J. Read}
\affiliation{\Fullerton}
\author{C. Henshaw}
\affiliation{\GT}
\author{R. O'Shaughnessy}
\affiliation{\RIT}
\begin{abstract}
The Rapid Iterative FiTting (RIFT) parameter  inference algorithm  provides a framework for efficient, highly-parallelized
parameter inference for GW sources. 
In this paper, we summarize essential algorithm enhancements and operating point choices for the RIFT iterative
algorithm, including \response{settings} used for analysis of LIGO/Virgo O3 observations.
We also describe other extensions to the RIFT algorithm and software ecosystem.  Some extensions increase RIFT's
flexibility to produce outputs pertinent to GW astrophysics.  Other extensions increase its computational efficiency or
stability.
Using many randomly-selected sources, we assess code robustness with two distinct code configurations, one designed to mimic settings as of LIGO\response{/Virgo} O3 and
another employing several performance enhancements.
We illustrate RIFT's capabilities with  analysis of selected events.
\end{abstract}
\maketitle

\section{Introduction}
Ground-based  gravitational wave (GW) detectors including
 Advanced LIGO  \cite{2015CQGra..32g4001L}  and Virgo \cite{gw-detectors-Virgo-original-preferred,2015CQGra..32b4001A},
\response{now joined by KAGRA \cite{2021PTEP.2021eA101A}}  continue to identify
  coalescing compact binaries
\cite{DiscoveryPaper,LIGO-O1-BBH,LIGO-GW170817-bns,LIGO-O3-NSBH,LIGO-O3-O3b-catalog,LIGO-O3-O3a_final-catalog}.
Many more GW observations are expected as observatories reach design sensitivity \cite{2016LRR....19....1A}, with detection rates expected
to exceed one per day when detectors reach their design sensitivity.   
Their  properties can be characterized via Bayesian inference,  comparing data to the expectations given  different potential sources
\cite{DiscoveryPaper,LIGO-O1-BBH,2017PhRvL.118v1101A,LIGO-GW170814,LIGO-GW170608,LIGO-GW170817-bns,LIGO-O2-Catalog,gwastro-PE-AlternativeArchitectures,gwastro-PENR-RIFT,gw-astro-PE-lalinference-v1}.
At present, a wide variety of phenomenological or interpolated estimates for GW from a merging binary are available
\cite{gwastro-mergers-IMRPhenomP,gwastro-mergers-IMRPhenomPv3,gwastro-SEOBNRv4,2019PhRvR...1c3015V,gwastro-mergers-IMRPhenomXP,2020PhRvD.102d4055O}.
Inferences using these models can be very computationally costly, particularly when using the best available models.
The Rapid Iterative FiTting (RIFT) \cite{gwastro-PENR-RIFT} is one of several parameter inference algorithms
\cite{gw-astro-PE-lalinference-v1,gwastro-pe-bilby-2018} used to produce the initial  interpretation of GW
observations
\cite{LIGO-GW170608,LIGO-GW170817-bns,LIGO-O3-GW190521-discovery,LIGO-O3-GW190521-implications,LIGO-O3-NSBH,LIGO-O2-Catalog,LIGO-O3-O3a-catalog,LIGO-O3-O3b-catalog,LIGO-O3-O3a_final-catalog}.
\response{The most popular} approaches \response{for gravitational wave parameter inference} rely on Markov chains, \response{within either Markov Chain Monte Carlo or nested sampling codes; see \cite{RevModPhys.94.025001} for a recent review. }
\response{By contrast},  RIFT performs Bayesian
inference through Monte Carlo quadrature, combined with an iterative algorithm to successively approximate pertinent likelihoods  \cite{gwastro-PENR-RIFT,gwastro-PENR-RIFT-GPU}.
 RIFT's structure offers novel opportunities to efficiently construct and re-use the outputs needed for Bayesian
 parameter and population inference \cite{gwastro-PENR-RIFT,2020PhRvD.102l4069J}.
RIFT naturally exports a continuous likelihood versus source parameters, valuable for population inference
\cite{gwastro-PopulationReconstruct-EOSStack-Wysocki2019}
and critical when downstream use employs tightly constrained source populations like a concrete nuclear equation of state
\cite{gwastro-PopulationReconstruct-EOSStack-Wysocki2019,gwastro-nsnuc-SteinerJoint-2020}.
RIFT can use these exported likelihoods to produce low-cost, high-accuracy model evidences, allowing for model selection
between different source physics scenarios (e.g., between an aligned or precessing BH binary \cite{LIGO-O3-GW190412},
between a model with and without higher order multipole GW  \cite{LIGO-O3-GW190412}, or
between different models for the nuclear equation of state \cite{LIGO-GW170817-EOSrank}).
RIFT can natively perform  multimodel inference using both interpolated likelihoods and the raw data from which
\response{they are} generated \cite{2020PhRvD.102l4069J}, important given notable modeling systematics.  These multimodel inferences enable  extension and re-use of previous analyses to incorporate additional
modeling as needed.
With companion software \cite{gwastro-RIFT-runmon}, RIFT's workflow can even identify its own  settings,
for example expanding mass prior ranges as needed.
RIFT's structure also offers novel opportunities to perform inference using large-scale distributed computing, while
mitigating the downside of intermittently unreliable computing environments.   
RIFT's computational cost is dominated by an embarrassingly parallel exploration phase, where many source parameters are
independently compared to the data, allowing it to scale to very large computing resources at need. 
Inevitably, these large workflows naturally discover any poorly-configured hardware and software, particularly when
interpreting many candidate sources.  RIFT's iterative structure, however, means that its workflow and settings can be
successively adapted to compensate for infrastructure problems, avoiding poorly-behaved nodes \cite{gwastro-RIFT-runmon}.
RIFT achieves the aforementioned flexibility with low computational cost, based at root on an efficient re-representation of the GW
likelihood \cite{gwastro-PE-AlternativeArchitectures},  combined with GPU-accelerated likelihood evaluation \cite{gwastro-PENR-RIFT-GPU}.  
The RIFT software ecosystem thus provides a robust framework to reduce the overall cost of inferring source parameters, enabling larger-scale analyses
and greater scope to probe waveform systematics. 
\response{Reducing overall evaluation cost and runtime has many potential downstream implications, not least including low-latency parameter inference \cite{PhysRevD.104.104054,PhysRevD.103.104057,PhysRevD.102.104020} needed to facilitate  multimessenger followup observations; see, e.g.,  \cite{2023arXiv230101337Y} and references therein.}

In this paper, we introduce several extensions of the original RIFT implementation, all available through its
open-source code repository \cite{code-rift-newrepo}.  Several of these features were used to interpret
gravitational wave sources during O3, the third observing run of the Advanced LIGO and Advanced Virgo instruments; see,
e.g., \cite{LIGO-O3-O3a-catalog,LIGO-O3-O3b-catalog,LIGO-O3-NSBH,LIGO-O3-GW190521-implications,LIGO-O3-GW190412}.  
This paper is organized as follows.
In Section \ref{sec:review} we briefly review the essential elements of the RIFT algorithm in regular use prior to the O3 analysis.
In Section \ref{sec:o3_version}, we describe essential additions employed in the O3 analysis, and validate our
production setup with standard tests.
In Section \ref{sec:updates}, we describe extensions to RIFT's O3-style approach, to improve its efficiency, flexibility, and
capability for unsupervised operation.   While many of these extensions were first introduced in the RIFT source several
years ago, and have been applied in other work, this paper provides the first detailed description of these updates.
In Section \ref{sec:fiducial_configs}, we enumerate the specific prototype RIFT configurations we recommend for regular
use and which we assess here.
In Section \ref{sec:tests}, we validate several of the key RIFT elements described above with targeted and statistical
tests.
Finally, in Section \ref{sec:reanalysis} we report  on reanalysis of selected real observations, to highlight RIFT's improved
performance and capability.
Several of our demonstrations are performed on real gravitational wave data, available from the Gravitational Wave Open
Science center \cite{2021SoftX..1300658A}.
Our study provides a backward- and forward-looking description of  RIFT code development, as needed for long-term
sustainable reproducibility of its GW inference results.

\section{RIFT review}
\label{sec:review}

A coalescing compact binary in a quasicircular orbit can be completely characterized by its intrinsic
and extrinsic parameters.  By intrinsic parameters we refer to the binary's  masses $m_i$, spins, and any quantities
characterizing matter in the system.
By extrinsic parameters we refer to the seven numbers needed to characterize its spacetime location and orientation.  
We will express masses in solar mass units and
 dimensionless spins in terms of Cartesian components $\chi_{i,x},\chi_{i,y}, \chi_{i,z}$, expressed
relative to a frame with $\hat{\mathbf{z}}=\hat{\mathbf{L}}$ and (for simplicity) at the orbital frequency corresponding to the earliest
time of computational interest (e.g., an orbital frequency of $\simeq 10 \unit{Hz}$).  We will use $\lambda,\theta$ to
refer to intrinsic and extrinsic parameters, respectively.

\begin{figure*}
\includegraphics[width=\columnwidth]{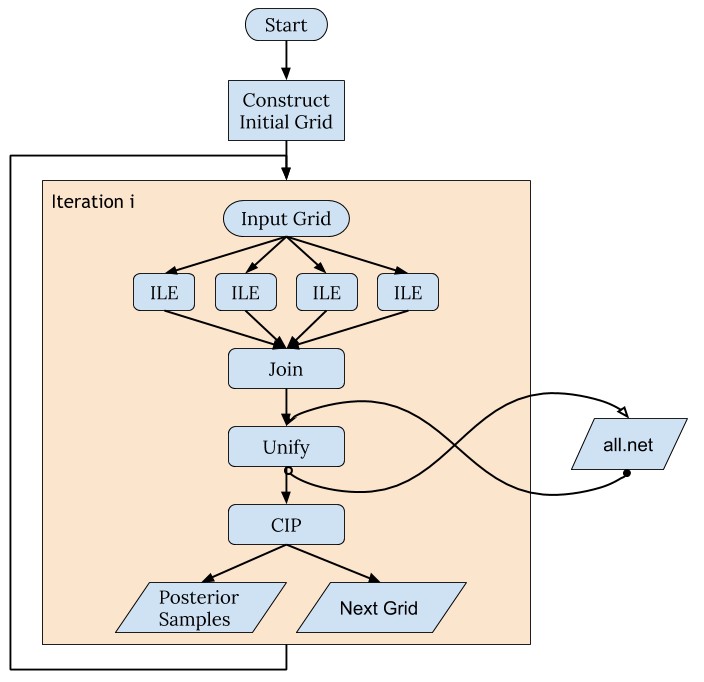}
\includegraphics[width=\columnwidth]{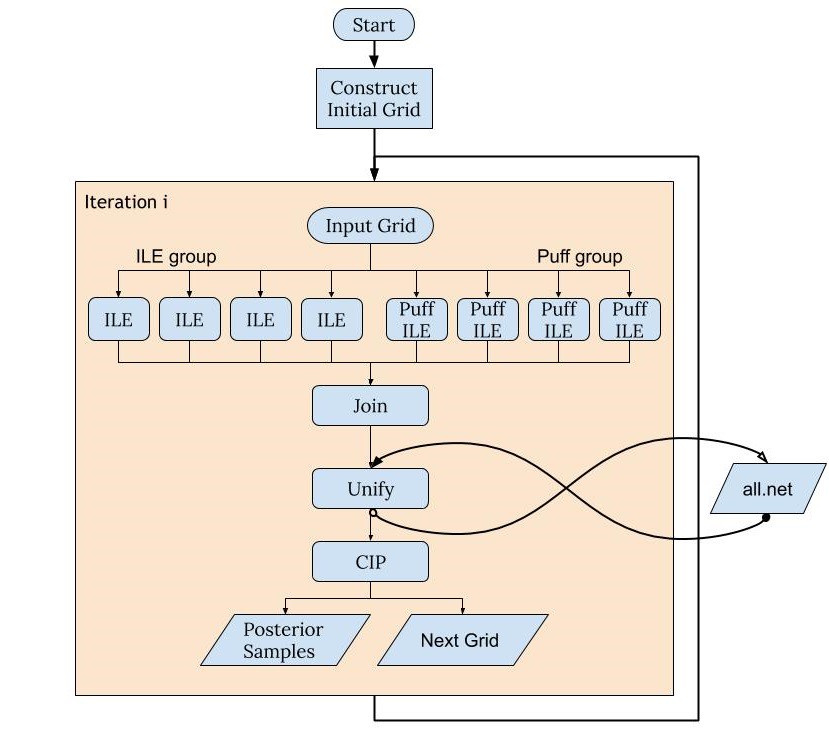}
\caption{\label{fig:workflow}\textbf{RIFT workflow}:  A flowchart of the RIFT iterative workflow.  The individual 
  worker jobs (ILE, \response{an acronym derived from ``integrate likelihood:extrinsic''}) compute the marginalized likelihood; the likelihood fitting and posterior-generation code (CIP,
  \response{an acronym derived from ``construct intrinsic posterior''}) uses
  these marginal likelihoods to estimate the posterior distribution.  The join stage combines the results from ILE workers for a
  specific iteration; the ``unify'' step accumulates results over multiple iterations into a single file.  
  The  iterative  workflow starts with an initial grid, encompassing
  some estimate of initial parameters.    Not shown here explicitly are (a) the \response{jittering or} ``puffball'' step, which
  generates a mirror grid with a subset of parameters randomly perturbed, accordingly to the samples' covariance matrix,
  whose results are used as the inputs to the ``puff group'' shown on the right; (b) the option for multiple CIP instances,
  which work in parallel to more efficiently generate slowly-converging posteriors; 
  and (c) convergence testing, to assess when to terminate the iterative process.
}
\end{figure*}

As illustrated in Figure \ref{fig:workflow}, RIFT principally consists of a two-stage iterative process to interpret gravitational wave observations $d$ via comparison to
predicted gravitational wave signals $h(\bm{\lambda}, \bm\theta)$.   In one stage, implemented by many workers in parallel
(denoted by ILE in the diagram), RIFT computes a marginal likelihood 
\begin{equation}
 {\cal L}\response{({\bm \lambda})}\equiv\int  {\cal L}_{\rm full}(\bm{\lambda} ,\bm\theta )p(\bm\theta )d\bm\theta
\end{equation}
from the likelihood ${\cal L}_{\rm full}(\bm{\lambda} ,\theta ) $ of the gravitational wave signal in the multi-detector network,
accounting for detector response; see  \cite{gwastro-PE-AlternativeArchitectures,gwastro-PENR-RIFT} for a more detailed specification.  
In the second stage, denoted by CIP in the diagram, RIFT performs two tasks.  First, it generates an approximation to ${\cal L}(\lambda)$ based on its
accumulated archived knowledge of marginal likelihood evaluations 
$(\lambda_\alpha,{\cal L}_\alpha)$.  Second, using this approximation, it deduces the (detector-frame) posterior distribution
\begin{equation}
\label{eq:post}
p_{\rm post}=\frac{{\cal L}(\bm{\lambda} )p(\bm{\lambda})}{\int d\bm{\lambda} {\cal L}(\bm{\lambda} ) p(\bm{\lambda} )}.
\end{equation}
where prior $p(\bm{\lambda})$ is the prior on intrinsic parameters like mass and spin.

\subsection{Evaluating the marginalized likelihood}
As described in previous work \cite{gwastro-PE-AlternativeArchitectures}, RIFT's likelihood uses physical insight to carry out its evaluation particularly
efficiently for binaries with similar intrinsic parameters but different extrinsic parameters.  
At a high level, RIFT relies on a decomposition of arbitrary gravitational wave signals $h(t)$ into physical basis signals $h_{lm}(t)$, associated
with a (spin-weighted) spherical harmonic decomposition of radiation in all possible emission directions.  This
decomposition allows RIFT to compute  cross-correlations between this basis and each detector's data; the likelihood for
arbitrary source orientations,  sky positions, and distances follows by a weighted average of these cross-correlation timeseries.
Recently, Wysocki and collaborators described a very efficient GPU-accelerated implementation of the likelihood,
enabling significant speed improvements \cite{gwastro-PENR-RIFT-GPU}.

Given the likelihood ${\cal L}_{\rm full}(\bm\lambda,\bm\theta)$, RIFT evaluates the marginal likelihood via an adaptive Monte Carlo
integrator:
\begin{align}
\label{eq:lnL:MonteCarlo}
{\cal L}(\bm\lambda) \simeq \frac{1}{N} \sum_k {\cal L}_{\rm full}(\bm\lambda,\theta_k) p(\bm\theta_k)/p_s(\bm\theta_k)
\end{align}
Inherited from its progenitor \cite{gwastro-PE-AlternativeArchitectures}, RIFT  performed this Monte Carlo integrator using an (adaptive) sampling prior $p_s$ which has
 product form, consistent with standard Cartesian adaptive integrators \cite{Lepage:1980dq,2021JCoPh.43910386L,book-mm-NumericalRecipies}.  
After a large block of evaluations, each one-dimensional marginal sampling prior can be updated to more
closely conform to the support of the integrand, based on a (smoothed) fixed-size one-dimensional histogram for each
adaptive dimension.   
While very powerful, this adaptive integrator limits RIFT for  two common applications.  First, its proposed sampling prior is
extremely inefficient when the integrand exhibits strong correlations between many dimensions.  Second, 
its python-based implementation
generates random numbers with its CPU, which must be transferred back and forth to and from the GPU when evaluating the likelihood.
In this work, we will examine two alternatives which alleviate each limitation in turn.

\subsection{Likelihood interpolation and posterior distributions}
To estimate ${\cal L}$ from discrete samples $\lambda_\alpha,{\cal L}_\alpha$, RIFT used Gaussian process regression.
Following the RIFT paper, 
for brevity and to be consistent with conventional notation, in this section we denote
  $\bm{\lambda}_\alpha$ by $x$ and $\lnLmarg_\alpha$ by $y$.
In this approach, we estimate the expected value of  $y(x)$ from data $x_*$ and values $y_*$ via
\begin{eqnarray}
\E{y(x)} = \sum_{\alpha,\alpha'} k(x,x_{*,\alpha}) (K^{-1})_{\alpha,\alpha'} y_{*,\alpha'}
\end{eqnarray}
where $\alpha $ is an integer running over the number of training samples in $(x_*,y_*)$ and where the matrix  $K =
k(x_\alpha,x_\alpha')$ $y_*$.     We employ a kernel function $k(x,x')$ which allows  for uncertainty in each
estimated training point's value $y_{*,\alpha}$ due to Monte Carlo integration, as well as a conventional squared
exponential kernel to allow for changes in the functions versus parameters:
\begin{eqnarray}
\label{eq:gp:k}
k(x,x') =  \sigma_o^2 e^{-(x-x')Q(x-x')/2} + \sigma_n^2 \delta_{x,x'}
\end{eqnarray}
The hyper-parameters of this kernel ($\sigma_o,\sigma_n$ and the positive-definite symmetric matrix $Q$) are chosen to
minimize the likelihood of our training data $x_k,y_k$ with covariance matrix $K$:
\begin{align}
\ln \ell(y) =  -\frac{1}{2} (y/\sigma)^T K^{-1} (y/\sigma)  - \ln \sqrt{ \frac{|K|}{(2\pi)^N \sigma^2}}
\end{align}
where $\sigma_k$ are the individual estimated uncertainties in each $y_k$ and $\sigma^2\equiv \prod_k \sigma_k^2$.
We perform all Gaussian process interpolation with widely-available open-source software \cite{scikit-learn}.  
The computational cost of full-scale Gaussian process optimization and evaluation increases rapidly with the dimension $D$ of the
matrix $K$, as $D^3$ and $D^2$ respectively. 

Given the likelihood, fair samples from the posterior distribution are generated by the following two-step process,
described in the RIFT paper.
First, using the likelihood estimate $\hat{\cal L}_{\rm marg}$ and the same adaptive Monte Carlo integrator described above, 
we perform the Monte Carlo integral $\int d{\bm \lambda} \hat{\cal L}_{\rm marg} p({\bm \lambda})$, producing sample
locations ${\bm \lambda}_k$ and associated weights $w_k= \hat{\cal L}_{\rm marg} p(\lambda)/p_s(\lambda)$.  
Second, we make a fair draw from these weighted samples.

\subsection{Exploring the parameter space}
\label{sec:sub:explore}
For expedient convergence, RIFT has two additional methods to explore the parameter space: dithering and incremental
dimensionality.  

After the posterior is produced and a candidate grid generated, RIFT can optionally produce a second candidate grid
derived from and supplementing the first.  In this second grid, points are generated by performing dithering on (or
randomization of) arbitrary
combinations of parameters, then rejecting unphysical combinations.  For example, the candidate points may have small
(correlated) offsets in chirp mass, $\eta$, and $\chi_{\rm eff}$  added, with offset covariance matrix set by the
covariance matrix of the input candidate grid.    
Particularly after several iterations, this dithering  can remedy a significantly-offset initial grid which misses the true
likelihood maximum.    This dithering also insures good sampling outside the boundaries of the target point.
The original RIFT paper \cite{gwastro-PENR-RIFT} only implemented correlated dithering based on sample covariance.  Later in this paper, we
describe incremental improvements to the dithering process which further improve performance.

RIFT can also employ different parameterizations at each stage.  In particular, as explained in the RIFT paper, RIFT can employ likelihood models with increasing
numbers of parameters, starting with the dominant degrees of freedom (e.g., $\mc$, $\eta$, and $\chi_{\rm eff}$ for
massive BHs) and adding in subdominant degrees of freedom in subsequent iterations.   
This approach helps address a tradeoff between cost and complexity.   For the first few iterations, RIFT needs to
identify the peak likelihood, as characterized by the dominant parameters.  Using all model parameters can be highly
counterproductive, as fits with all degrees of freedom require overwhelming numbers of evaluations  $\lambda_k, {\cal
  L}_k$ in order to avoid overfitting/under-resolving.  (With too few evaluations and several irrelevant parameters included, the Gaussian process behaves pathologically.)    By reducing the number of poorly-constrained parameters early
on, we can employ far fewer points early on.  Because of the computational cost of GP regression in high dimensions and
with many  points, this was essential for
handling complex sources like precessing BH binaries with the original GP likelihood estimate. 
The appropriate dimensional hierarchy depends on the physics involved (e.g., configurations with high
mass; BHNS with strong precession; NS-NS binaries with tides; et cetera) but is well-motivated from simple Fisher matrix
arguments.
Specifically, the component masses and a measure of aligned binary spin (e.g., $\chi_{\rm eff}$) approximately
characterize the dominant degrees of freedom for nearly-nonprecessing binaries, particularly when organized as the chirp
mass $\mc$ and symmetric mass ratio $\eta$.  As most observed binaries exhibit
nearly no precession, these variables form a natural set to adopt for the first iterations.  As transverse and other
spin degrees of freedom have a subdominant impact on the marginal likelihood, we can add these incrementally, after
obtaining a converged estimate for the behavior for nonprecessing degrees of freedom.
Prior to O3, these choices were made by humans, and the iteration plan assembled by hand and adjusted at need.

\subsection{Convergence testing}
The RIFT paper \cite{gwastro-PENR-RIFT} introduced a procedure to assess convergence: for each marginal 1d distribution, compute the
KL divergence between successive iterations, $D_{KL}(p|q) = \int dx p \ln p/q$ where $p,q$ are 1-dimensional probability
densities.   A fiducial convergence threshold was $O(10^{-2})$ for each variable.  At the time, these KL divergences were evaluated using  KDE-based
estimates of each 1d marginal distribution.   
Subsequently, Delfavero \cite{gwastro-mergers-RIFT-DelfaveroThesis} introduced and assessed a simpler and more stable 1d convergence diagnostic:
the net $L^1$ difference between each one-dimensional CDF: 
\begin{eqnarray}
\label{eq:test_metric:L1}
D_{L^1} = \max_x|\hat{P}(x) - \hat{Q}(x)|
\end{eqnarray} where $P,Q$ are empirical
CDFs associated with the two sample sizes.   The $L^1$ norm has been well-studied in the context of KS tests.  Delfavero
proposed a convergence threshold of $-\frac{1}{N}\ln \alpha$, where $\alpha$ was the desired confidence level of the
test and $N$ is the common sample size.

Finally, to better capture correlations in our convergence tests, we have also implemented a simplified
multi-dimensional convergence test, which compares the empirical means $\mu_1,\mu_2$ and covariance matrices $\Sigma_1,\Sigma_2$ associated with two
sets of fair samples under the assumption that both characterize a Gaussian distribution:
\begin{align}
\label{eq:test_metric:KL}
D_{KL} &= \frac{1}{2} \left[ (\mu_2-\mu_1)\Sigma_2^{-1}(\mu_2-\mu_1)    \right  . \nonumber \\
 & \left.  + (\text{Tr}(\Sigma_2^{-1}\Sigma_1)-d)   + \ln \frac{|\Sigma_2|}{|\Sigma_1|} \right] 
\end{align}  
where $d$ is the dimension of the problem.    Like the 1d KL divergence test, we adopt a  fiducial convergence threshold
of $10^{-2}$.  The user can select any subset of variables (and any coordinates) with which to evaluate this joint test,
though we recommend using $\mc,\eta,\chi_{\rm eff}$ at a minimum.
Unless otherwise noted, we adopt and report on runs using the latter convergence diagnostic below.

\subsection{Limitations}

To recap, RIFT organizes Bayesian inference as an iterative two-stage process.  In one stage, it rapidly evaluates a marginal likelihood ${\cal L}$ for compact
binary source parameters  $\lambda$, via a Monte Carlo integral.  In another stage, it uses its accumulated knowledge of
previous likelihood evaluations $(\lambda_k,{\cal L}_k)$  to estimate
${\cal L}$ as a function of arbitrary $\lambda$; from this estimate, it draws samples  for the posterior for $\lambda$, again via a Monte
Carlo integral.  The output of the second stage is passed back to the first, until the results converge.

While in principle effective, in practice this strategy relied principally on high-dimensional fits and dithering to explore the model
space.  The Gaussian process fits employed previously, however, were excessively parsimonious outside the previously
trained domain, prohibiting exploration.  Dithering was an essential but occasionally fragile element of our procedure
to explore the parameter space.   For this reason, in Section \ref{sec:o3_version} below we introduced several
additional techniques to  automate  exploration of the binary parameter space, particularly by improved dithering and by systematically
hierarchically adding degrees of freedom
with increasingly subdominant effects on typical likelihoods.

Additionally, as originally implemented,  key elements of the RIFT parameter inference strategy had notable sources of inefficiency.
  For example, the adaptive Monte Carlo integrator inherited from Pankow et al \cite{gwastro-PE-AlternativeArchitectures} is both
relatively slow and algorithmically inflexible, not well-suited to sample distributions with strong correlations which
its adaptive algorithm's built-in assumptions can't efficiently replicate.   More painfully, the standard interpolation implementation adopted
(Gaussian process regression) scaled very inefficiently with the number of input likelihood evaluations, placing severe
limits on the scale of problems that could be usefully addressed.
Described at greater length below, these defects are being addressed by the methods first described in this work.

\section{RIFT during O3}
\label{sec:o3_version}
In the O3 era, production-scale RIFT calculations employed several additional operating-point choices and features which
have not previously been described in the literature.

\subsection{Waveform support}

RIFT inference is performed using the spin-weighted spherical harmonic waveforms  $h_{lm}(t)$ or $h_{lm}(f)$ \cite{gwastro-PE-AlternativeArchitectures,gwastro-PE-AlternativeArchitecturesROM,gwastro-PENR-RIFT,gwastro-PENR-RIFT-GPU}, usually computed from binary parameters
through the \texttt{lalsimulation} library.  During the O3 analysis era and publications
\cite{LIGO-O3-O3a-catalog,LIGO-O3-O3b-catalog}, commonly-used estimates for the gravitational waves emitted from quasicircular  binary merger included 
IMRPhenomD \cite{2016PhRvD..93d4006H,2016PhRvD..93d4007K}, IMRPhenomPv2 \cite{gwastro-mergers-IMRPhenomP}, IMRPhenomXPHM \cite{gwastro-mergers-IMRPhenomXP}, and
SEOBNRv4PHM \cite{2018PhRvD..98h4028C,2020PhRvD.102d4055O}.  
While the illustrations and tests presented in this work draw upon these established source models, we point out  that
RIFT's likelihood-based approach enables transparent visualization, calculation,  and mitigation of the impact of
waveform systematics
\cite{2020PhRvD.102l4069J,gwastro-PENR-RIFT}.

\subsection{Dithering and exploration }
\label{sec:puff}

 RIFT originally generated candidate future samples using an estimate $\hat{{\cal L}}_{\rm marg} $ of the marginal likelihood.  While this method could
 very efficiently explore the parameter space, it would often  only explore within the neighborhood already
explored, even with high likelihood on the edge of the previously-explored set.  
In O3, we  therefore added a simple dithering algorithm, to supplement candidate points with a companion set, where each
companion point was drawn from the original sample but had added  random uncertainty in selected parameters.
\response{We chose random uncertainties centered on but with larger than the covariance than the available training data
  $x_\alpha$,  with the goal of enveloping the posterior and its marginal-significance tails, to stabilize our estimate
  of the log likelihood in a region not well served by  RIFT's normal approach for selecting training data (draws from the posterior).
  [In a sense, we use an ``overdispersed''
  investigation of training data $x_\alpha$ to avoid an ``underdispersed'' final posterior.]  }
We further generalized our
dithering algorithm in three ways: by rejecting dithered samples based on proximity; by
allowing the user to eliminate cross terms in the covariance matrix used for dithering; and by allowing the user to
request random candidates in any subset of parameters, instead of simple dithering.  

\begin{figure}
\includegraphics[width=\columnwidth]{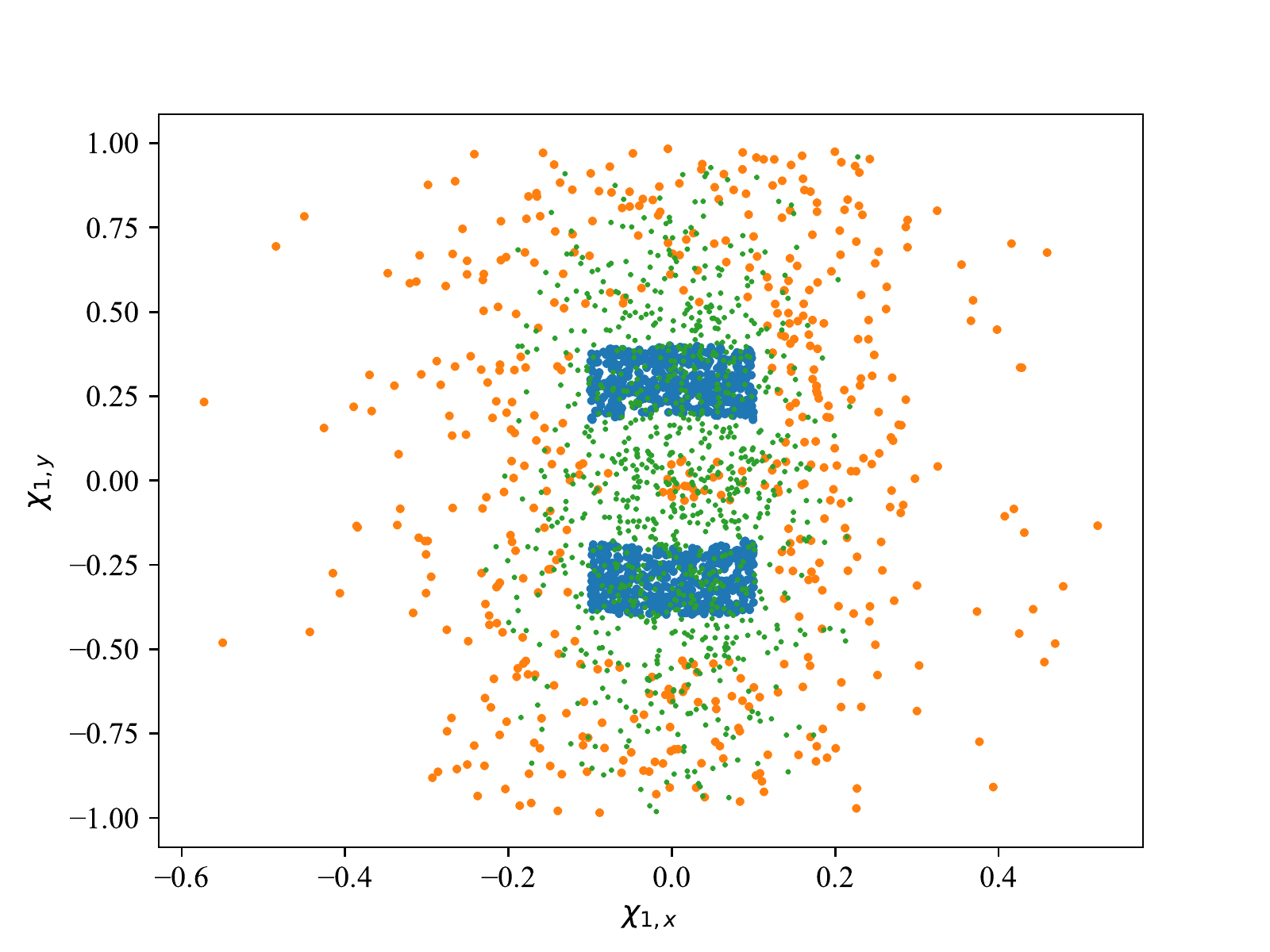}
\caption{\label{fig:Puff}Demonstration of dithering algorithms: The blue points show synthetic input  data which is uniform in
  $\chi_{1x},\chi_{1,y}$ in $|\chi_1|\in[0.2,0.4]$ and $\chi_{1,x}\in[-0.1,0.1]$.  The green points show the
  orginal dithering procedure with $F=1$ and $\epsilon=0$.  The orange points show the result with
  $F=3,\epsilon=0.01$.  The modified algorithm more efficiently explores the outskirts of the otherwise-sampled area.
}
\end{figure}

Our first  dithering algorithm 
 draws $\delta x_\alpha$ for a specified set of variables $\alpha$, based on the empirical covariance matrix
 $\Sigma_{\alpha\beta}$ for those variables.  The offsets are  randomly drawn from a multivariate normal distribution
 with  covariance matrix $F^2 \Sigma$ with $F$ a user-specified
 factor (chosen as $F=3$ by default).
Without rejection, this method  frequently produces candidate points in extremely close
proximity to previously densely-explored regions.         We therefore allowed the user to specify a threshold $\epsilon$,
such that 
dithered samples with distance $\delta x_\alpha \delta x_\beta \Sigma^{-1}_{\alpha\beta} / d$ smaller than $\epsilon$ to any previously-evaluated point
would be removed, where $d$ is the dimension of parameters being dithered.  
As a result, the dithered samples at late times better
explore the outskirts of the posterior.
Figure \ref{fig:Puff} shows an example of the two methods, applied to a toy problem.  The modified method does not
generate samples inside regions otherwise being explored by the input grid. 
The current implementation adopts the same $\epsilon$ for all iterations, and  performs \response{dithering} and rejection based on
the most recent samples rather than all past history.

RIFT's original dithering algorithm would also extend only along the principal axes of existing correlations in its
targeted input variables.  While GW observations can produce strong correlations near the peak likelihood, farther away
from the peak the likelihood surface can exhibit other correlations.  As a result, RIFT's original dithering algorithm
would not enable efficient identification of subdominant correlations and extended, correlated tails in the posterior
distribution.   To address this deficiency, we provide the capability to employ two core modifications, though neither
is active by default.   In the first, the user can modify the covariance matrix used to dither the input samples by
requiring $\Sigma_{\alpha,\beta}=\Sigma_{\beta,\alpha}=0$ for any list of user-specified pairs $(\alpha,\beta)$.
In the second, the user can request that any specific coordinate $\alpha$ is drawn at random, uniformly over a
user-specified range. 
The latter method is extremely useful for marginally-accessible degrees of freedom (e.g., subdominant tidal parameters),
which are prone to overfitting.

\subsection{Physics-inspired iterative architecture}
\label{sec:arch}

RIFT's final results are produced from a set  of likelihood evaluations $\{\lambda_\alpha,\lnLmarg_\alpha\}$.
Nominally, RIFT uses the same prior and likelihood model to produce final results and accumulate these likelihood evaluations, in its iterative
process.   However, RIFT can also accumulate these 
likelihood evaluations with any prior, and using iterations with fit estimates which omit known-subdominant degrees of
freedom.   These choices, denoted as architectures in the text below, can significantly reduce the latency or even
overall computational cost, as noted in Section \ref{sec:sub:explore}.    We used them in O3  because
higher-dimensional likelihood models require more input data and often higher computational (gaussian-process) cost; whenever plausible,
lower-dimensional likelihood models were desirable for exploratory iterations.

\noindent \emph{Architectures for binary black holes}: For all compact binaries without matter, RIFT's O3-era unsupervised
approach was very conservative.   Specifically,  the O3-era  RIFT starts with 3 iterations
fitting using $\mc,\eta,\chi_{\rm eff}$, with a volumetric spin prior; 2 iterations fitting with $\mc,\eta,\chi_{\rm
  eff}$ and $\chi_{\rm -} = (m_1 \chi_{1,z}-m_2 \chi_{2,z})/M$ and a volumetric spin prior; 2 iterations fitting with $\mc,\eta, \chi_{\rm
  eff},\chi_{i,x},\chi_{i,y}$ for $i=1,2$ with a volumetric spin prior; and (if adopting conventional priors) 3 iterations using the same parameters, but
with a spin prior that is uniform in spin magnitude.  The validation study for this approach is described in Section \ref{sec:sub:validate}
with Figure \ref{fig:PrecessingPP}.

This strategy was designed to  characterize the massive BH binaries with $\mc > 20 M_\odot$ that were
relatively common in O1O2, and was particularly targeted to identify signatures of strong precession. 
 With relatively few cycles in contemporary ground-based instruments, BH binaries
with $\mc > 20 M_\odot$ have posterior distributions which only weakly constrain intrinsic parameters except for
$\mc,\eta,\chi_{\rm eff}$.  Particularly at very high mass, the transverse spins in particular usually have minimal impact on the posterior
distribution.  As a result, when investigating massive and possibly precessing BH binaries, we can adopt an architecture
which increases in complexity, where the first few iterations use a fit with only $\mc,\eta,\chi_{\rm eff}$; the next
few iterations add an antisymmetric aligned spin $\chi_{-}$; and the last few iterations use all spin degrees of freedom.

For most massive BH binaries, this unsupervised architecture was massive overkill.
However, this configuration is also robust and efficient when the true signal parameters are not covered by
the initial candidate grid.  Frequently, real GW searches identify parameters well-seperated from the final posterior
distribution.   Similarly, due to strong model systematics, the true parameters may be recovered with substantial bias
with an alternative model.  

\vspace{10pt}
\noindent \emph{Architectures for matter}:
While this paper will describe all pertinent updates to RIFT,  we have chosen to emphasize binary black holes and largely eschew matter effects, for clarity deferring 
new demonstrations of our current and extended matter-related capabilities to future work. 
When performing an unsupervised analysis with matter, the O3-era RIFT algorithm adopted  the same  architecture choices as massive binary black holes, in
particular assuming the likelihood at leading order only depends on $\mc,\eta, \chi_{\rm eff}$ and tides.  As with binary black holes, we adopted a lower-dimensional model early on, assuming
 the marginal likelihood depends on the dimensionless tidal deformabilities $\Lambda_i$ only through $\tilde{\Lambda}$
 for most iterations; see \cite{LIGO-GW170817-bns,LIGO-GW170817-EOSrank} for discussion of these parameters.  To account for degeneracies, we perform
 correlated \response{dithering} in $\mc,\eta,\chi_{\rm eff}, \tilde{\Lambda}$.  
To further ensure the low-$\Lambda$ region is well-explored, in O3 we  adopted a non-uniform prior on $\Lambda_i$ which favors
small $\Lambda$.   RIFT results using this approach have been previously presented, including a novel population study 
\cite{gwastro-PopulationReconstruct-EOSStack-Wysocki2019}.

As with the BH-BH case, this architecture is motivated by the physics of binary inspiral.  
Working to leading order, we characterize the gravitational effects of compact
objects with matter by a dimensionless tidal deformability parameter $\Lambda_i$.  Following convention when presenting results not
conditioned on other observations or theory, we
adopt a uniform prior on these $\Lambda_i$, extending from $0$ to $5000$ independent of compact object mass.  This prior
is not well-suited to exploring the tidal parameter space, because real compact objects are subject to an equation of state
and thus $\Lambda(m)$ relation which depends strongly on mass, is typically much less than $5000$,   and goes to zero at
high mass.  Additionally, tidal effects are highly subdominant and enter at leading order  through a single
mass-weighted combination $\tilde{\Lambda}$.   Exploring the tidal parameter space using the default prior with   RIFT
is exceptionally inefficient, because the prior strongly disfavors the low-$\tilde{\Lambda}$ configurations associated
with the (weak) peak in the marginal likelihood, particularly for very massive NS with extremely small $\Lambda(m)$.
Conversely, the approach described above was demonstrably sufficient to enable a multi-event population analysis to
recover a proposed NS equation of state and mass/spin distribution from synthetic GW observations \cite{gwastro-PopulationReconstruct-EOSStack-Wysocki2019}.

\subsection{Well-motivated initial grids}

When given a good starting grid $\{\lambda_k\}$, RIFT converges well.  For unsupervised analysis of binary black holes
in O3, we used a hypercube in
$\mc,\eta,\chi_{\rm eff}$ chosen based on the search-reported candidate parameters.  
The chirp mass region was chosen over a logarithmic region  of width $\Delta \ln \mc =\pm (1.5)^20.3(v/0.2)^7/\rho$
centered on the reported chirp mass $\mc_*$, where 
$v$ is the smaller of $(\pi \mc_* f_{\rm min})^{1/3}$ or $0.2$, $\rho$ is the search-reported signal to noise, and $f_{\rm
  min}$ is the minimum frequency used for parameter inference.   The $\eta$ extent covers from $\eta_{\rm min}$ to
$1/4$.  If the trigger symmetric mass ratio $\eta_*>0.1$, then  $\eta_{\rm min}$ was (by default) the larger of $0.1$ and
that value $\eta_1$ such that $m_2=1M_\odot$ given $\mc_*,\eta_1$.  If the trigger symmetric mass ratio is more extreme
($\eta_*<0.1$), then $\eta_{\rm min}=0.25 \eta_*$. 
Finally, the $\chi_{\rm eff}$ interval was chosen to be $\chi_{\rm eff,*}\pm 0.3/\rho$.
This wide region in mass, mass ratio,  and aligned  spin helped compensate for the often-large biases between  search
trigger parameters and the true posterior.

For unsupervised investigations involving matter, 
motivated by
plausible nuclear equations of state, we adopt an initial grid which uniformly covers a region in the neighborhood of a
fiducial analytic estimate $\Lambda_{fid}(m)$, assumed to be $20$ for $m>2.2 M_\odot$ and $3000
((2.2-m/M_\odot)/1.2))^2$ otherwise.  Specifically, we uniformly sample 
 $\Lambda \in
[\lambda_{min,fid},\lambda_{max,fid}]$  where $\Lambda_{min,fid}$ is the smaller of $50$ and $0.2 \Lambda_{fid}$ and
$\Lambda_{max,fid}$ is the smaller of $0.2 \Lambda_{fid}$ and 1500.  
With this starting grid, we can recover tidal parameters for realistic NS over a wide range of masses.  
For example, these settings were adopted in our detailed systematics study about jointly fitting the nuclear equation of
state and BNS population \cite{gwastro-PopulationReconstruct-EOSStack-Wysocki2019}.

\optional{
\subsection{More robust distributed operation}
\label{sec:pipeline}

RIFT is a natively parallel algorithm, naturally operating  asynchronously within each stage.  To achieve full benefits
of parallelization, however, we need to access large-scale distributed computing resources, requiring  the creation of special
  workflows (and software environments) which are well-adapted to these resources' available computing power, inter-node
  bandwidth, and common access to persistent storage.    Appendix \ref{ap:pipelining} describes some of the workflow issues.
}

\subsection{Validation of O3 configuration}

Among other tests, we validated RIFT O3-era code configurations against generic, randomly chosen merging binaries using a
standard probability-probability (PP) plot test \cite{mm-stats-PP,gwastro-skyloc-Sidery2013}.
Using RIFT on each source $k$, with true parameters $\mathbf{\lambda}_k$, we estimate
the fraction of the posterior distributions which is below the true source value $\lambda_{k,\alpha}$   [$\hat{P}_{k,\alpha}(<\lambda_{k,\alpha})$] for each intrinsic parameter $\alpha$.  After reindexing the sources so $\hat{P}_{k,\alpha}(\lambda_{k,\alpha})$ increases with $k$ for some fixed $\alpha$,  a
plot of $k/N$ versus $\hat{P}_k(\lambda_{k,\alpha})$ for both mass parameters can be compared with the expected result
($P(<p)=p$) and binomial uncertainty interval.

\begin{figure}
\includegraphics[width=\columnwidth]{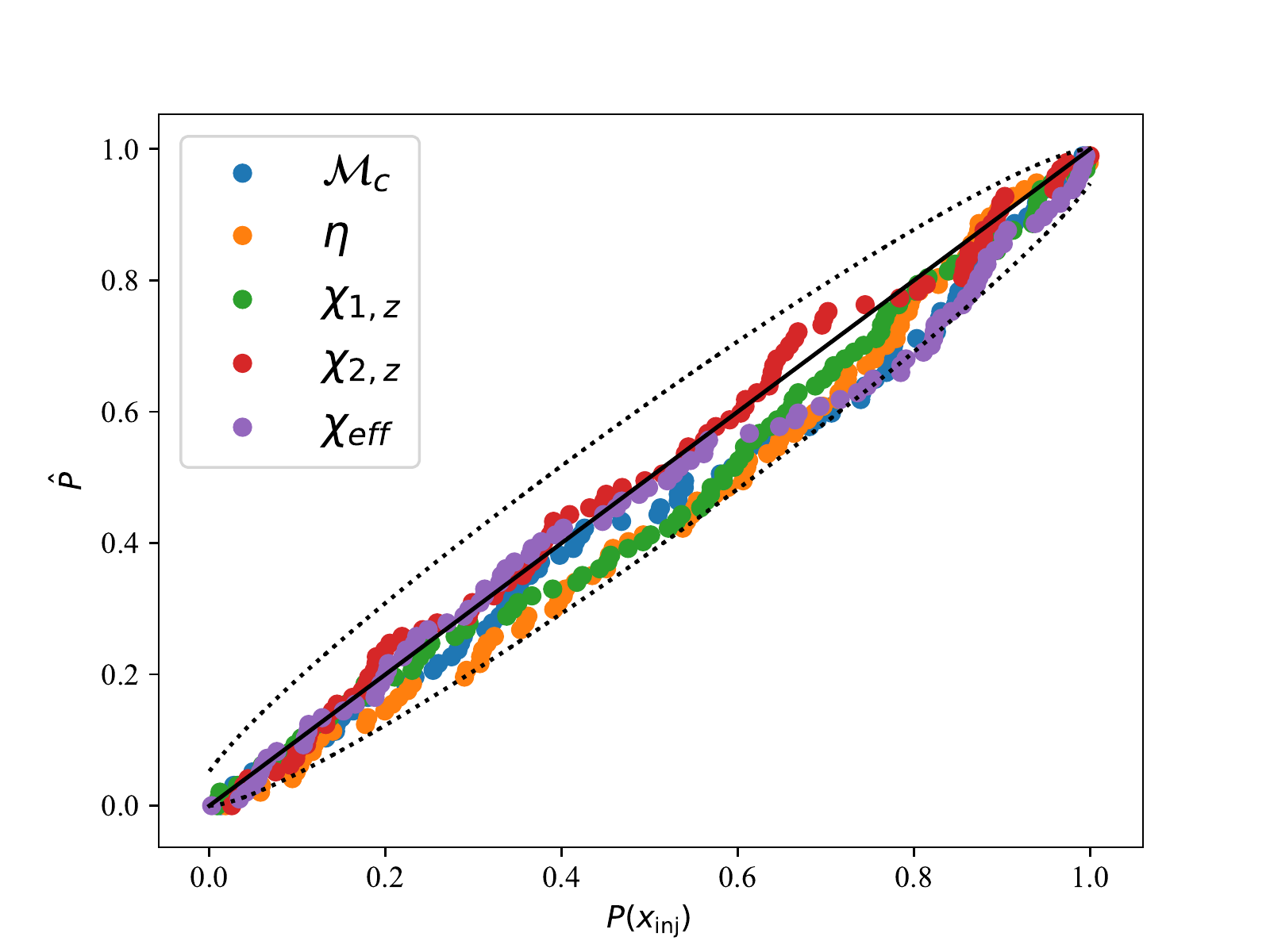}
\caption{\label{fig:NonprecessingO3PP}\textbf{Nonprecessing PP plot, early-O3 code}: Probability-probability plot to validate recovery of
  precessing synthetic signals, following the RIFT methods paper \cite{gwastro-PENR-RIFT}. The synthetic sources and parameter inferences are constructed with IMRPhenomD in
  gaussian noise with presumed known PSDs,
 for 3-detector networks, starting the signal at 20 Hz and using a 4096 Hz sampling rate.  Detector-frame masses are drawn uniformly  in the
  region bounded by ${\cal M}/M_\odot \in [30,60]$ and $\eta \in [0.2,1/4]$, and sources are drawn volumetrically between $1.5$ and $4
  \unit{Gpc}$.   Both BH dimensionless spins are drawn uniformly in $\chi_{i,z}$. 
  For computational efficiency, all sources in this  specific test have a fixed and presumed known sky location.
}
\end{figure}

Other previously-published studies have already reported on comparable  PP plot tests, in the context of waveform
systematics \cite{gwastro-RIFT-systematics-AnjaliAasim-2020}, using the O3-era code.  
Conversely, Section \ref{sec:tests} describes more comprehensive tests and PP plots applied to the current edition of
the code.
However, for completeness, Figure  \ref{fig:NonprecessingO3PP} shows the result of one such contemporary O3-era test for intrinsic degrees
of freedom.

\subsection{Inefficiencies and Limitations of the O3 configuration}
\label{sec:sub:o3_problems}

RIFT's development up to O3 was tightly constrained, needing to be completed and assessed well before any O3 analysis,
resulting in  occasionally fragile and suboptimally efficient but still extremely
portable and reproducible tool.   RIFT's O3 configurations have many completely arbitrary limitations, introduced both
by the difficulties inherent in our software environment and our timeline.
The foremost difficulty in operating RIFT remains its  organization: multiple independent command-line scripts,
communicating information via files, orchestrated into a pipeline via \texttt{condor} \cite{condor-practice,BOCKELMAN2020101213,1742-6596-664-6-062003}.  For O3, we did not have
sufficient time to implement more than the most naive control logic: a fixed number of iterations to investigate
the intrinsic variables, followed by an
(optional) step to extract extrinsic variables, with some fixed $O(1)$ set of extrinsic samples associated to each  intrinsic point.
Convergence diagnostics were only used post-facto by the end user, to characterize run quality.  
With convergence tested only by humans in postprocessing, we needed to run every analysis for an extended period,  usually employing many
more iterations than necessary, to ensure  almost
all problems would be well-converged without human intervention.  The most challenging and unanticipated problems,
however, would  require human intervention.

The deployment timeline also introduced additional unavoidable development requirements, occurring often  simultaneously
with ongoing efforts to refine our workflow during O3a.
For the first half of O3 (O3a), RIFT had to be refactored into a python package (\texttt{pypi} and \texttt{conda} in
particular), so it could be integrated into the LVK's standard software infrastructure.  To support this refactoring
specifically and code portability in general, we also had to create a continuous-integration test suite.  
Prior to the second half of O3 (O3b), RIFT had to be ported  to python 3. 

Despite its fragility and overkill, RIFT was extremely successful in O3.  In O3a, RIFT was extensively used to analyze
the GWTC-2  events using models with higher-order modes \cite{LIGO-O3-O3a-catalog}.   In O3b (GWTC-3), RIFT was also used to
analyze events with a costly model including higher-order modes \cite{LIGO-O3-O3b-catalog}, and was also operated principally by external
groups through large-scale automated software (\texttt{asimov}) \cite{code-asimov}.

\begin{figure}
\includegraphics[width=\columnwidth]{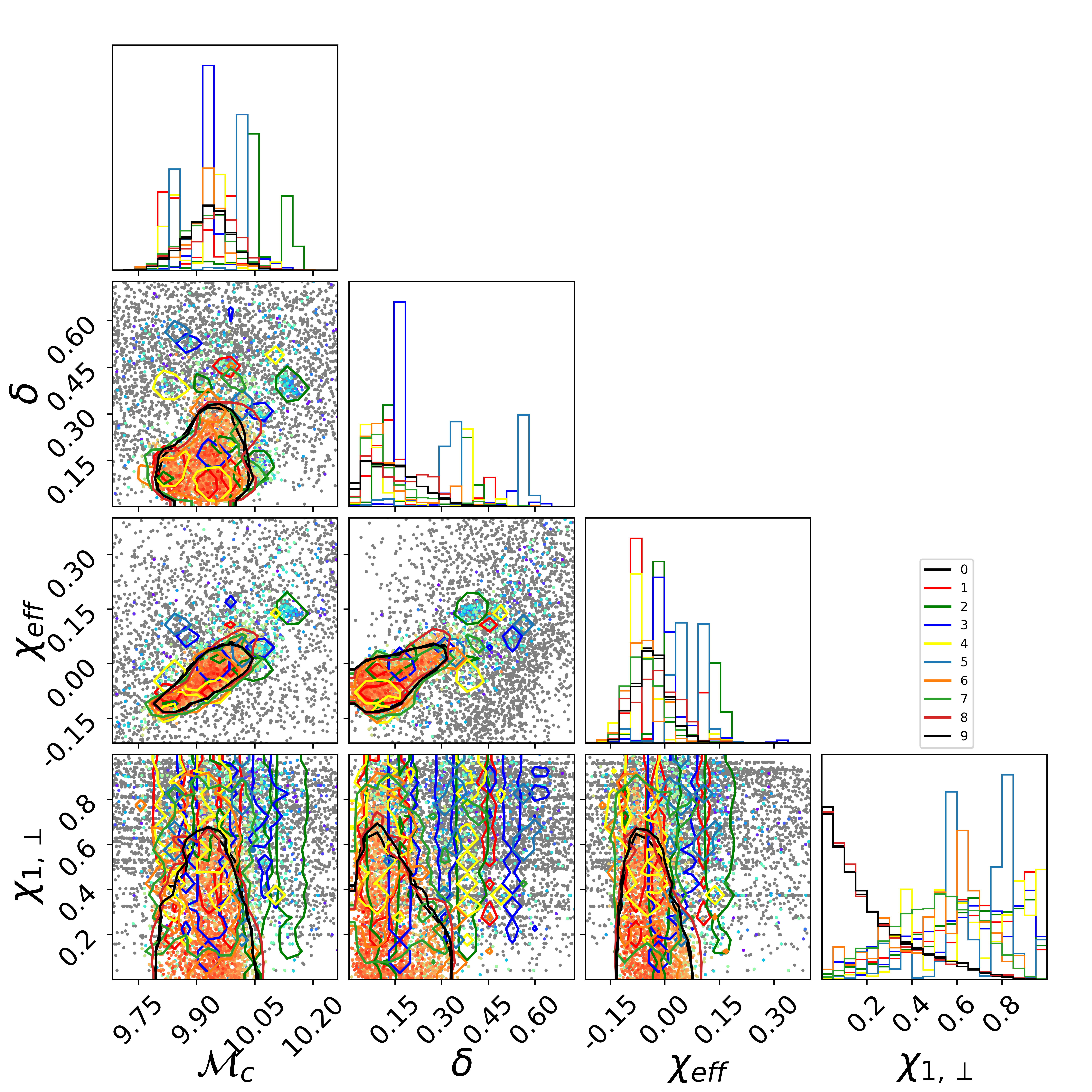}
\caption{\label{fig:unpleasant_o3_run}
\textbf{Illustration of O3  architecture limitations for  unsupervised inference of low-mass binaries}: A corner plot
  with 90\% credible interval quantiles for different RIFT iterations, indicated by different line colors.  The color
  scale shows the likelihood range, over a dynamic range $\Delta \ln {\cal L}\le 15$; gray points indicate likelihood
  evaluations below this range.   This figure shows an unsupervised analysis of GW190707,
  evaluated using the old O3 convergence architecture (with SEOBNRv4PHM).  \response{The panels show marginal
    distributions in chirp mass $\mc$, asymmetric mass ratio $\delta = (m_1 - m_2)/M$, inspiral effective spin
    $\chi_{\rm eff}$, and the magnitude of the transverse dimensionless spin of the primary $\chi_{1,\perp} = \sqrt{\chi_{1,x}^2+\chi_{1,y}^2}$.}
  Section \ref{sec:sub:o3_problems} enumerates the many algorithmic limitations of the O3 analysis highlighted by this
  kind of  analysis.
}
\end{figure}

Above and beyond the severe limitations introduced by using a fixed number of iterations, without convergence test
integration within our control logic,  our O3 experience suggested
several additional elements of RIFT needed improvement.  \response{Figure} \ref{fig:unpleasant_o3_run} provides an example of an
unpleasant but typical O3 RIFT inference of a low-mass binary, illustrating many of the problems described below.
First and foremost, our Gaussian-process fitting implementation using fiducial coordinates became almost unusably slow when
trained with many inputs, requiring as much as a day to generate for several challenging problems of astrophysical
interest.  In fact, in O3, we implemented several workarounds to prevent our fit from ever using too
many training points.  However, particularly for low-mass binaries with strong inter-parameter correlations in its posterior, our fits were also prone to misidentify suitable length scales [i.e., the diagonal
elements of $Q$ in Eq. (\ref{eq:gp:k})], leading to patchy and irregular posteriors when the fits were not informed by
overwhelmingly large data volumes; see Figure \ref{fig:unpleasant_o3_run} for examples.   Since our technique required many iterations of fitting, often one to several weeks
could be required to interpret the most interesting precessing binaries.

Second, our adaptive  Monte Carlo integration algorithm did not effectively exploit extremely strong and well-understood
correlations in the posterior distribution of chirping binaries.  For BHNS binaries in particular, the natural error
ellipsoids are extremely long and narrow; see, e.g., \cite{gwastro-mergers-HeeSuk-FisherMatrixWithAmplitudeCorrections,gwastro-mergers-HeeSuk-CompareToPE-Precessing}.   For these extreme binaries, 
our intrinsic posterior Monte Carlo integration (performed in CIP) typically completed with  a ratio  $n_{\rm eff}/n$, 
 which roughly measures the
number of independent sample points per proposed Monte Carlo trial, often smaller than $10^{-8}$.  Even typical
low-mass binaries had low values of $n_{\rm eff}/n$.   Combined with the
relatively long evaluation time of Gaussian processes, all low-mass binaries were uncomfortably difficult to investigate.

Third, our choice for how to explore precessing DOF was not well-adapted to investigate the low-mass binaries which
nature provides, whose overall spins (and transverse spins) seem small.   Instead,  motivated by
discovery potential, in O3 we used a volumetric prior (in a hypercube) for the component spins $\boldsymbol{\chi}_i$ to initially explore both  aligned and transverse
degrees of freedom.  
This prior was used in most of the initial iterations, when we adopted a fitting ansatz based only on the  \emph{aligned} degrees of freedom,
under the assumption that precession effects were small, to populate presumed-subdominant precessing degrees of freedom.
We used this prior for several iterations, including early iterations where the likelihood fit did
not include and could not adapt to transverse degrees of freedom.  However, as became  apparent later in  O3, none of
the low mass events had significant support for nonzero transverse spin.  By contrast, our prior frequently generated points with
large transverse spins, which fit poorly (because these configurations would have many easily-observable precession
cycles at low mass).  At best, our choice of volumetric prior wasted time that could have been spent exploring the
transverse degrees of freedom more efficiently.  At worst, the volumetric prior points
actively impeded convergence early on.   In practice, we sometimes needed the final fully-precessing iterations even to
get a plausible posterior at all for low-mass binaries such as NSBH \cite{LIGO-O3-NSBH}.  

Figure
\ref{fig:unpleasant_o3_run} provides an example showing how our inference only stabilized after adopting a
uniform-spin-magnitude prior in the final iterations.
In this analysis, the first several iterations incorrectly adopted a likelihood
completely independent of transverse spin, despite frequently sampling large transverse spins which corresponded to
substantial precession.  This O3-era combination of extreme-spin prior and no-transverse likelihood model was extremely
difficult to fit and sample, particularly in the old coordinate system which lacked awareness of the strong $\mc,\eta,\chi_{\rm eff}$ correlations
expected from  leading-order post-Newtonian inspiral.    The  last few iterations (here labelled 6,7,8) however employ
transverse spins in their fits and, combined with a more suitable spin prior, finally recover a smooth 
posterior.
For this and similar posteriors, where  only the last few iterations are well-behaved, manual additional
investigation was required, extending the existing run to assess if RIFT had indeed converged.

Finally, given the many inefficiencies already limiting our performance and limited development time, we left many
elements of RIFT in highly unoptimized forms.  For example, the coordinate conversions within CIP used to implement generic
chart transformations between  fitting and sampling coordinates used a generic but slow data structure instead of fast
vectorized once-and-for-all transformations.
CIP workers were operated such that, if any one failed, all were rerun.  This poor choice required considerably more processing
and longer latency when very large numbers of workers were needed to handle BHNS binaries, for example.
Some of our integration algorithms were  insufficiently overflow-protected, causing errors when even modest-amplitude
signals' likelihoods were evaluated directly (as opposed to only as a logarithm).
Finally, users had  few guarantees about the effective sample size of their output.  For a handful of low-significance
events in particular, the small and unpredictable sample
size was intermittently a challenge in O3b, during which  several postprocessing resampling stages were applied to
RIFT's output via  \texttt{asimov} to change the distance prior and
add  calibration marginalization.

\section{Updates}
\label{sec:updates}
In this section, we describe several extensions to the way RIFT was used during O3, improving its likelihood approximation; integration; and
workflow.    Where appropriate, we also provide simple (PP) tests to validate specific modules.  After describing these
many possible additions, 
in the next Section \ref{sec:sub:validate}  we describe how we downselect between  these options:  by measuring RIFT perfomrance  when
interpreting two fiducial sources: a synthetic binary black hole and GW190412.
Having downselected between the many available configurations, that section also provides targeted validation studies
using synthetic binary black hole and binary neutron star sources.

\subsection{Added coordinate systems (and priors)}
\label{sec:coordinates}
All algorithms used within RIFT --  interpolation, posterior generation, grid placement, dithering, convergence tests,
et cetera  -- perform better in coordinate
systems which are well-adapted to the  likelihoods of real gravitational wave sources.
For example, all our current and new unstructured interpolation algorithms inherit some implicit or explicit
dependence on the coordinate system used to formulate them.

\begin{figure}
\includegraphics[width=\columnwidth]{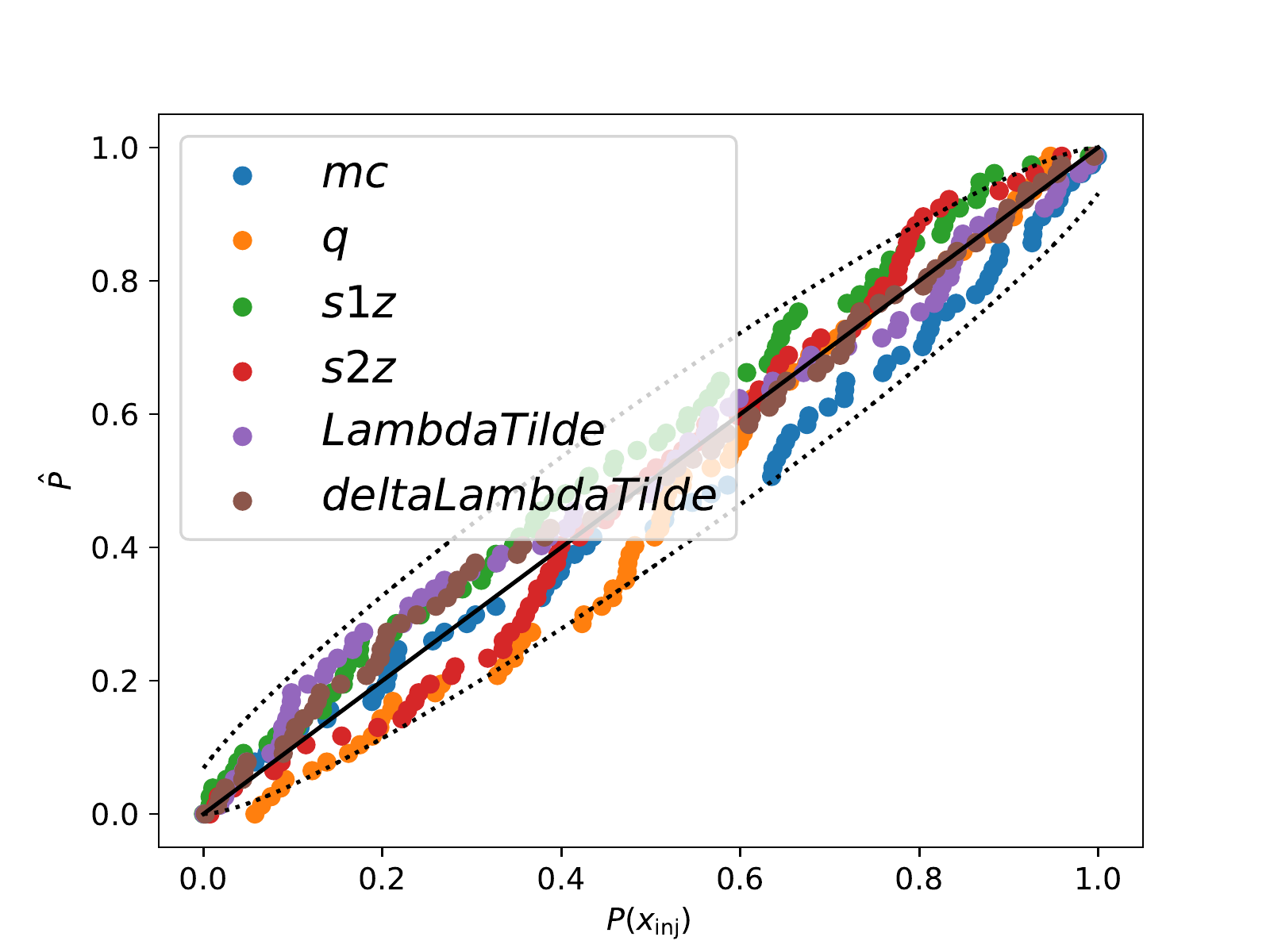}
\caption{\label{fig:MucoordPP}\textbf{RIP PP plot}: Probability-probability plot to validate the rotated inspiral phase coordinates. The synthetic sources and parameter inferences are constructed with NRHybSur3dq8Tidal ($\ell=5$) in
  gaussian noise with presumed known PSDs,
 for 3-detector networks, starting the signal at 30 Hz and using a 4096 Hz sampling rate.  Detector-frame masses are drawn uniformly  in the
  region bounded by ${\cal M}/M_\odot \in [1.2,1.4]$ and $\eta \in [0.2,1/4]$, and sources are drawn volumetrically between $90$ and $240  \unit{Mpc}$.   Both NS dimensionless spins are drawn uniformly and volumetrically from [0,0.05]. 
  For computational efficiency, all sources in this test have a fixed and presumed known sky location. 
}
\end{figure}

\noindent \emph{Rotated Inspiral-Phase (RIP) coordinates} Particularly for low-mass systems, the  neigborhood
of near-peak likelihood is  best characterized (and quite ellipsoidal) in well-chosen, instrument-dependent coordinates.  While optimal
local coordinates can always be derived by  Fisher matrix methods, in general the orientation of this optimal ellipsoid
relative to an underlying generic coordinate chart is highly source-dependent.   However,  using a fiducial contemporary ground-based network to eliminate (minimal)
ambiguity about the appropriate detector network, recently Lee and collaborators \cite{gwastro-mergers-SoichiroCoordinates-2022}
introduced a global coordinate system remapping $\mc,q,\chi_{1,z},\chi_{2,z}$ into a \emph{global} coordinate system
which is well-suited to model the likelihood for \emph{arbitrary} sources.

The RIP coordinate system is motivated by the leading-order post-Newtonian expressions for the gravitational wave strain
emitted by a nonprecessing binary in the $\ell,|m|=(2,2)$ mode.  Using standard techniques \cite{lrr-Blanchet-PN,gw-astro-mergers-approximations-SpinningPNHigherHarmonics}, the frequency-domain gravitational wave
phase  $\tilde{h} = |h(f)|\exp(-i \Psi(f))$), evaluated at some reference
frequency can be expressed in terms of several post-Newtonian parameters
\begin{subequations}
\label{eq:rip:psi}
\begin{align}
\psi_1 &= \frac{3}{128} (\pi  \mc f_{\rm ref})^{-5/3} \\
\psi_2 &= \frac{55}{384}(\eta + \frac{743}{924}) \eta^{-2/5} (\pi \mc f_{\rm ref})^{-1} \\
\psi_3 &= \frac{3}{32} (\beta -4\pi)\eta^{-3/5} (\pi \mc f_{\rm ref})^{-2/3}
\end{align}
\end{subequations}
where for convenience we adopt $G=c=1$ units in our expressions, where the PN parameter $\beta$ is defined by
\cite{1995PhRvD..52..848P}
\begin{align}
\beta = \frac{1}{12}\sum_{i=1}^2 \left [ 113 (m_i/M)^2 + 75 \eta \right] \chi_{i,z}
\end{align}
and where $\chi_{i,z} \equiv \hat{L} \cdot {\mathbf S}_i/m_i^2$ is the projection of the dimensionless spin along the
instantaneous angular momentum axis. 
The \response{rotated inspiral-phase coordinates $X_i$  } follow from a coordinate
transformation $\boldsymbol{X}=U \boldsymbol{\psi}$ where $U$ an instrument-dependent $3\times 3$ matrix derived from
the Fisher matrix expressed in terms of these coordinates.  To be concrete, we follow Lee et al and adopt a single fiducial choice for $U$:
\begin{align}
U=\begin{bmatrix}
0.97437198&  0.20868103&  0.08397302 \\
     -0.22132704&  0.82273827&  0.52356096 \\
     0.04016942& -0.52872863&  0.84783987 \\
\end{bmatrix}
\end{align}
Unless otherwise noted, we  employ a nominal $f_{\rm ref}=200\unit{Hz}$ to define this transformation for all masses.
Despite the relatively high reference frequency, this  transformation remains well-behaved even for very massive black
hole binaries, implying the coordinates can be employed throughout the observed (detector-frame) space of compact binary parameters.

Following  Lee et al (their Section III.C), we  define the RIP coordinate system $\mu_1,\mu_2,\delta,\chi_{2,z}$.  
Within the framework described so far, this coordinate system can be employed within the fitting and posterior
generation stage (CIP) in two ways.  On the one hand, we can use RIP as a coordinate system well-adapted to \emph{fitting} the
likelihood.  In this approach, after re-expressing our training points $\boldsymbol{\lambda}$ in the RIP coordinate system,
our unstructured interpolation code produces an approximation $\hat{\cal L}(\boldsymbol{\lambda})$ in terms of those
coordinates.  Aside from this modest change, CIP can be used the same way, for example using any coordinate system and
Monte Carlo integration technique to perform posterior generation.
Figure \ref{fig:MucoordPP} shows an end-to-end validation study of RIFT when RIP coordinates are  employed within CIP, using a suite of many
synthetic aligned-spin sources drawn with  random intrinsic and extrinsic parameters.

On the other hand, we can also use RIP as a coordinate system well-adapted to exploring and \emph{sampling} the
likelihood.  To do so we must define an effective sampling prior $p_{s,eff}(\boldsymbol{\lambda})$ for the RIP coordinates
$\boldsymbol{\lambda}$.  Though nominally simple, the  complicated nonseperable boundaries associated with this sampling
prior have so far  complicated our ability to employ this seemingly simple prior within our existing frameworks based on
purely seperable physical priors.  We
defer use of these accelerated coordinates to future work on very-low-latency analysis.

\noindent \emph{Rotated detector-network-frame sky coordinates} To more efficiently sample the sky, we provide users the
option to use a coordinate system for the sky where the nominal north pole corresponds to  a vector connecting two of
the interferometers.  As discussed in many previous implementations of this transformation (see, e.g., \cite{2020MNRAS.499.3295R,gw-astro-PE-lalinference-v1}), in this coordinate system
the posterior distribution will be aligned with lines of constant nominal declination, enabling more efficient adaptive sampling.

\noindent \emph{Pseudo-cylindrical coordinates for spheres}: Posterior generation of precessing spins can be
computationally costly in the most straightforward spin coordinate system; spherical polar coordinates for each spin.  
For context, for  nonprecessing binaries we can adopt carteisan aligned spins $\chi_{i,z}$, and use an adaptive
integration method which captures correlations between $\chi_{1,z},\chi_{2,z}$.  By contrast, in spherical polar
coordinates, these simple and strong correlations are distributed among many more parameters.   Similarly, for most
massive BH binaries observed so far, the \emph{transverse} spin components $\chi_{i,\perp}$ are extremely weakly
constrained.  However, in spherical polar coordinates, the transverse and aligned spin components are strongly mixed.

To improve the prospects for our adaptive integrators to better reflect the correlations among spin parameters, we
introduce a  coordinate transformation mapping a sphere to a cylinder: $(R,z)= (\bar{R}\sqrt{1-\bar{z}^2},\bar{z})$
for $\bar{R},\bar{z}\in[0,1]$.  In these coordinates,  the overall spherical volume element $d\phi \wedge R dR \wedge
dz$ can be recovered with the seperable sampling priors  $p_V(\bar{R}) = 2\bar{R}$ and $p_V(\bar{z}) =
3(1-\bar{z}^2)/4$.  These coordinates enable efficient sampling of the unit sphere with a volumetric prior
$p_V(\bar{R})p_V(z)/2\pi$ using ccoordinates well-adapted to
the typical constraints afforded by GW observations. 

While sufficient for volumetric sampling, however, the discussion in Appendix \ref{ap:mc} suggests more singular
sampling priors will enable better sampling of the fiducial uniform-spin-magnitude spin prior (i.e., $ d^3\chi/3|\chi|^2$).  In this common scenario,
we adopt a more singular pseudo-radial sampling prior $p_s(\bar{R}) = \bar{R}^{-3/4}/4$, or equivalently a 
uniformly-sampled radial coordinate $\bar{u}=\bar{R}^{1/4}$.  Numerical experiments similar
to those in Appendix \ref{ap:mc} demonstrate improved scaling  relative to
naively reweighting volumetric samples.  To be concrete, these modified pseudo-cylindrical coordinates represent a dimensionless
spin vector $\boldsymbol{\chi}$ as
\begin{align}
\label{eq:pseudo_cylinder}
\boldsymbol{\chi} &= \hat{\mathbf{z}} \chi_z + \bar{\chi}_{u}^4\sqrt{1-\chi_z^2}[ \cos \phi \hat{\mathbf{x}}+ \sin \phi \hat{\mathbf{y}}]
\end{align}
In terms of these coordinates  $\bar{\chi}_u,,\phi, \chi_z$, a volumetric prior follows from $p_V$ by change of
coordinate and jacobian:  $p_{UV}=15 \bar{\chi}_u^7 (1-\chi_z^2)/8\pi$.  The corresponding uniform spin magnitude prior is
$1/3|\boldsymbol{\chi}|^2$ times this function.

\noindent \emph{Generalized precession coordinate}: The originally proposed precession parameter $\chi_p \equiv \max
\left(\chi_1 \sin{\theta_1}, \tilde{\Omega} \chi_2 \sin{\theta_2}\right)$ characterizes the largest dynamical spin in
the binary, but fails to account for the effects of dual-misalignment. This deficiency, which manifests in systems with
equal mass ratio and large transverse spins, is resolved by the parameter $\langle \chi_p \rangle$, which averages over
all spin angles on the precession timescale \cite{Gerosa_chipavg_2021}. The initial implementation of $\langle \chi_p
\rangle$ is detailed in \cite{precession_RIFT}, where events from O3b were analyzed as a post-processing step using
existing samples to compute posteriors for both $\chi_p$ and $\langle \chi_p \rangle$. Additionally this parameter is a
constant of motion at 2PN order on the spin-precession timescale (and nearly conserved on the radiation-reaction timescale), making it a good candidate for a fitting coordinate when computing posteriors for
analyses that assume precession. By computing an approximate ${\cal L}(\lambda)$ from the archived marginal likelihood
calculations in the $\langle \chi_p \rangle$ coordinate, we then assign a uniform prior in the domain $0 \leq \langle
\chi_p \rangle \leq 2$ to compute the posterior distribution. Note that although the $1 < \langle \chi_p \rangle \leq 2$
domain is exclusive to binaries with two misaligned spins, there are spin morpholigies in the $0 \leq \langle \chi_p
\rangle < 1$ domain for which $\langle \chi_p \rangle$ differs strongly from $\chi_p$. Allowing the prior to cover this
space leaves the analysis agnostic to the fully precessing behavior. This functionality has now been implemented in RIFT
as part of the CIP subroutine, and the efficacy of this parameter is currently being tested via injection study, the
results of which will be discussed in a forthcoming publication; see also \cite{2022arXiv220700030D}.

\subsection{More robust and efficient likelihood approximations}
\label{sec:likelihood}

In this section, we summarize several different techniques to approximate the marginal likelihood.

\mysub{Random forests}
First used in  RIFT for interpolation in  \cite{2020arXiv200101747W},
random forests interpolate generic functions by  constructing a family of many random decision trees, with piecewise
constant approximations of the form $y(x)= \sum_k w_k \mathbf{I}_k(x)$ where $\mathbf{I}_k$ is unity inside the selected
volume and zero elsewhere \cite{2001MachL..45....5B,book-Murphy-MachineLearning,mm-stat-ExtraTrees}.  Customarily, each choice in the decision tree decides between one (randomly selected)
coordinate in the variable $x$; as a result, each decision tree selects a sequence of rectangular cartesian regions.
Random forests construct an ensemble of trees, each with randomly chosen decision points.  We employ the ExtraTrees
algorithm  \cite{mm-stat-ExtraTrees}, as implemented in scikit-learn  \cite{scikit-learn}.
In the limit of extremely deep and random trees, this algorithm converges to a piecewise linear and continuous
approximation  \cite{mm-stat-ExtraTrees}.

Because random forests' basis functions $\mathbf{I}_k$ are step functions aligned with the coordinate axes of $x$,
random forests can be  sensitive to the choice of coordinates, particularly when the posterior exhibits strong
correlations between multiple parameters.   Our investigations suggest RF fits robustly perform well in all coordinates for sources with broad,
uncorrelated posteriors (e.g., massive binary black holes).  By contrast, for high-mass-ratio sources in particular RF
fits should only be used with specialized coordinate systems like the RIP coordinates above.
Even more so than gaussian processes, random forests do not extrapolate well outside of their domain, and as a result
posteriors which extend to sharp prior boundaries can introduce undersampling or even pathologicar behavior.  As a concrete example, RF-based
posterior generation for binary neutron star observations with nonprecessing binaries with  uniform priors on
$\chi_{i,z}\in[-1,1]$ can behave extremely poorly; uniform spin magnitude sampling for nonprecessing BNS with RF fits should always employ a
tightly restricted spin prior.   As a second example, RF-based posteriors for the transverse spin
require extensive sampling near $\chi_{i,\perp}\simeq 0$ to explore this region well, hence the pseudo-cylindrical
coordinates of Eq. (\ref{eq:pseudo_cylinder}). 
As a third example, RF-based posterior generation near the
equal-mass line can be prone to under-predicting the region near $q\simeq 1$, though  suitable mass ratio sampling
coordinates could mitigate this effect.

\mysub{Sparse gaussian processes}
RIFT initially adopted   conventional Gaussian Process (GP) regression to estimate the marginal likelihood versus intrinsic parameters,
    with a full rank (squared exponential) kernel $k(x,x')$ as provided by \textsc{scikit-learn}.
Straightforward GP regression techniques are costly since they involve matrix inverses, with nominal cost scaling as
    $n^3$ for a full-rank matrix  \cite{book-Rasmussen-GP}.
This scaling severely limited our ability to increase model dimension or to use more training data.
    Sparse kernels or approximations have been widely explored in the GP literature 
    \cite{2016arXiv160604820B,2009arXiv0912.3268A,2013arXiv1309.6835H,2019arXiv191007123J}.
To perform GP regression more efficiently, we have implimented a 
    piecewise polynomial covariance function with compact support \cite{book-Rasmussen-GP}.
These basis functions are guarenteed to be positive definite, and the
    covariance between points becomes zero as their distance increases, 
    and are given as $K_{ppD,q}(r)$.
\begin{align}
K_{ppD,0}(r) &= (1-r)^{j}_{+} \\
K_{ppD,1}(r) &= (1-r)^{j + 1}_{+} ((j + 1)r + 1) \\
K_{ppD,2}(r) &= \frac{(1-r)^{j + 2}_{+} ((j^2 + 4j +3)r^2 + (3j +6)r + 3)}{3} \\
K_{ppD,3}(r) &= (1-r)^{j + 3}_{+} 
  \left ((j^3 + 9j^2 + 23j + 15)r^3  \nonumber
   \right. \\ & \left. + (6j^2 + 36j + 45)r^2  \nonumber
  \right. \\ & \left. + (15j + 45)r +15) 
  \right )\div 15
\end{align}
Where $j = \lfloor\frac{D}{2}\rfloor + q + 1$, $D$ is the dimensionality of your data set.
    $q$ is chosen such that the sample function is $2q$ times differentiable.
We have chosen $q = 1$, and added a whitenoise kernel as well.
We have seen that the sample time for this function scales only with $n$ for high $n$.

\mysub{Quadratic and gaussian estimates for placement}
During initial exploration the posterior for tightly constrained events,  particularly for precessing binaries, relatively few points  $\bm{\lambda}_\alpha$
will have high  likelihood ${\cal L}_\alpha$.   With limited training data in these iterations, our most flexible and efficient
interpolation methods in practice can spuriously identify overly-complicated likelihood estimates, with complex
isocontours and mutiple extrema.   For several future applications, we introduce two  simple 
likelihood approximations, both 
using some pre-determined threshold ${\cal L}_{cut}$ to  identify the subset of training data
$\bm\lambda_\alpha$ with ${\cal L}_\alpha > {\cal L}_{cut}$.  
In the  mean-covariance approximation, we compute the sample mean $\bar{\bm{\lambda}}$ and
sample covariance $\bar{\bm\Sigma}$, then  adopt the ansatz
\begin{align}
\ln \hat{\cal L}_{cov} = \ln {\cal L}_{max} -  \frac{1}{2} (\lambda - \bar{\lambda})_p (\lambda - \bar{\lambda})_q \bar{\Sigma}^{-1}_{pq}
\end{align}
In the quadratic approximation, by contrast, we perform a least-squares quadratic form fit to $\ln {\cal L}$ versus
$\bm\lambda$, then use the expression
\begin{align}
\label{eq:quad}
\ln \hat{\cal L}_{cov} = \ln {\cal L}_{maxfit} -  \frac{1}{2} (\lambda - {\lambda}_*)_p (\lambda - \lambda_*)_q \Gamma_{pq}
\end{align}
where $\ln {\cal L}_{\rm maxfit},\lambda_*,\gamma$ are all identified by the quadratic fit.
We provide these simple approximations for testing, for potential use in ultra-low-latency analysis,  and to better extract simple approximate results (e.g., Gaussian
approximations) from detailed analyses.   These two approximations are not included in the operational recommendations
for long-latency offline inference presented later in this work.

\subsection{Improved  integration}
\label{sec:integrate}

RIFT uses Monte Carlo integration in both stages of its iterative process, for posterior generation (CIP) and likelihood
marginalization (ILE).  Beacuse of the dynamic range, sometimes sharp
features, and strong correlations present in the likelihood integrand, RIFT uses custom implementations of adaptive Monte Carlo
integration.    In this section, we primarily  describe alternative Monte Carlo integration implementations which meet some
of our design goals.   Appendix \ref{ap:mc} describes how RIFT and other codes characterize sampling size: RIFT
customarily uses  $n_{\rm eff}$ while $n_{\rm ESS}$ is used by many other inference codes. Figure \ref{fig:TestGMM:Toy} illustrates how the two new Monte Carlo integration methods compare
to our previous approach, for the purposes of estimating posterior distributions via weighted samples.  
The code used to generate this figure (and thus test the integrators at a variety of target resolutions) is disseminated
with the source and run as part of our continuous integration suite.  
Also disseminated with the RIFT source is the code used to generate Figure \ref{fig:TestInt:Rosenbrock}, the inferred
sample distribution implied by the Rosenbrock likelihood  \cite{rosenbrock,2020MNRAS.497.5256F}.

RIFT's low-level likelihood evaluation can use direct quadrature over some extrinsic degrees of freedom.  In this
work, we specifically describe how RIFT can now use a fast numerical quadrature over distance.

\subsubsection{Adaptive sampling with gaussian mixture models}
Because many of our integrands have strongly correlated dimensions, seperable sampling priors are often very
inefficient.  To identify correlations, we provide an alternative adaptive sampler, such that $p_s$ is built from
Gaussian mixture models.   

In the simplest and default form, we continue to assume a seperable sampling prior.  For adaptive dimensions, we  adopt one-dimensional sampling
distributions 
\begin{align}
p_{s,k}(\theta_k)  = \sum_\alpha w_\alpha p_n(\theta_k|\mu_\alpha,\sigma_\alpha)
\end{align}
where $p_n$ is a standard normal distribution with  mean $\mu_\alpha$ and standard deviation $\sigma_\alpha$; where
$w_\alpha$ are weights associated with each gaussian component; and where for simplicity we fix the number of components
a priori.  Each adaptive iteration, we use the expectation-maximization algorithm to re-assess the weights and
covariance \cite{mm-stat-EM,mm-stat-EM-theory,mm-stat-EM-book}, organizing the calculation to enable fast iterative updates; see the Appendix for
details.   
For example, each iteration the integral result $I$ and variance $\sigma^2_I$ are updated with a running
average using the previous values $I,\sigma_I^2$ and the values over the current subsample $I_{new},\sigma_{Inew}^2$, according to 
\begin{align}
    I' = {{n_s I + I_{new}} \over {n_s + 1}}\\
    \sigma^2_{I'} = {{n_s \sigma^2_{I} + \sigma^2} \over {n_s + 1}}.
\end{align}
where $n_s$ is the number of previous iterations.
To gracefully handle finite boundaries, we use truncated normal distributions $p_{n,t}$ in place of normal distributions
$p_n$ in our mixture model.

More commonly, we employ correlated sampling in subsets of dimensions: $p_s$ is no longer seperable.
The same algorithm applies.   To handle finite boundaries, we use fast rejection sampling to identify valid
configurations; see the Appendix.
Our implementation allows the user to specify at runtime which (if any) dimensions will use correlated sampling.

\begin{figure}
\includegraphics[width=\columnwidth]{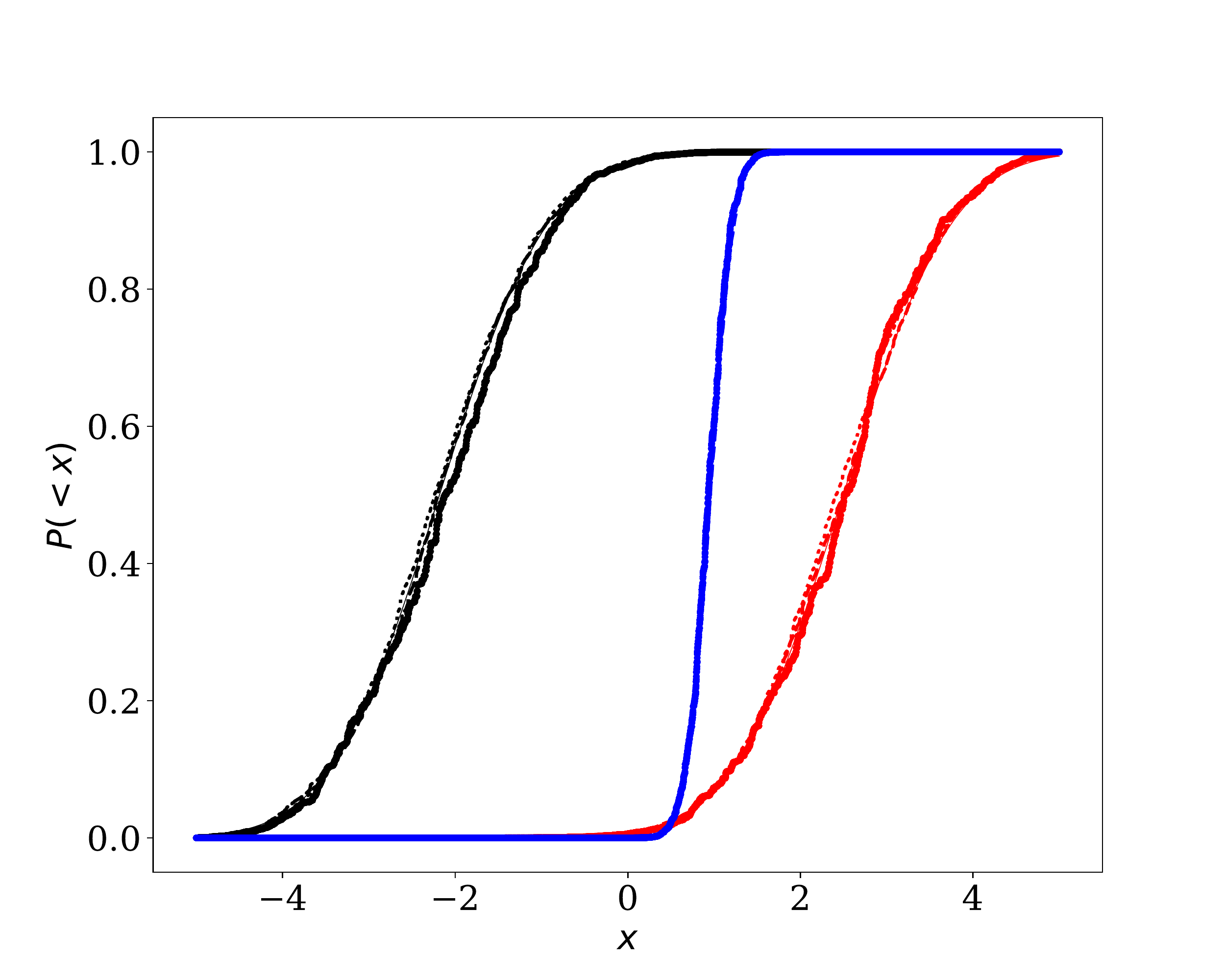}
\caption{\label{fig:TestGMM:Toy}\textbf{Integrating random multidimensional gaussian}: Cumulative distribution functions inferred for a
random uncorrelated  three-dimensional gaussian likelihood, different adaptive weighted Monte Carlo integration techniques.   The horizontal axis indicates the dimension $x_i$ and the vertical
axis our estimate of $P(<x_i)$.  Colors indicate the dimension being rendered (e.g., $x_1,x_2$ or $x_3$), line styles indicate the method used to generate the
  CDF.  Solid lines show the true CDF; dashed lines show a CDF obtained with the original MCMC integrator;  thin dotted
  lines show the corresponding estimate from the \response{gaussian mixture model (GMM)} integrator; and heavy dots show the results with the adaptive
  cartesian integrator.  All three estimates are derived from the same number of 
  function evaluations ($2\times 10^4$) and produce comparable $n_{\rm eff}\simeq O(50)$; both are chosen to be small
  enough (and hence our resolution poor enough) so that the
  reader can differentiate between the different curves in the figure. 
}
\end{figure}

\begin{figure}
\includegraphics[width=\columnwidth]{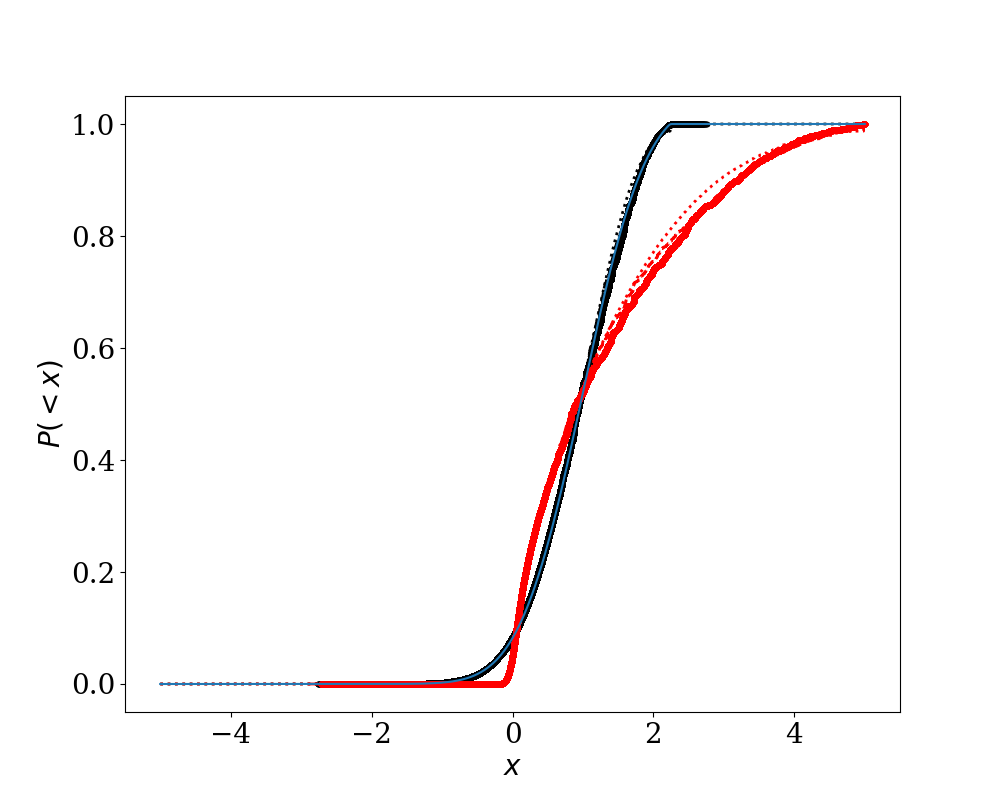}
\caption{\label{fig:TestInt:Rosenbrock}\textbf{Integrating the rosenbrock likelihood}: One-dimensional marginal
  distributions for the two-dimensional rosenbrock likelihood inferred using our three standard integrators with
 a fixed number $10^6$ random adaptive likelihood evaluations using $10^4$ samples per adaptive batch.  The solid
  line shows an independent solution derived from one-dimensional quadrature \cite{2020MNRAS.497.5256F}.   All three
  integrators predict an integration error and recover the quadrature-based evidence ($\simeq -5.804$)  to within $\simeq
  10^{-2}$ for this fixed sample size.  For these settings we find $n_{\rm ESS} \simeq 5800, 1400, 5800$ ($n_{\rm
    eff}\simeq 3000,130,300$) for the
  default,  \response{adaptive cartesian (AC)}, and \response{gaussian mixture model (GMM)} integrations respectively.
  The one-dimensional Jensen-Shannon (JS) divergence between inferred marginal distributions for
  each parameter is comparable to or smaller than a fiducial threshold \cite{2021MNRAS.507.2037A} of $10/\text{max}[n_{\rm ESS,A},n_{\rm ESS,B}]$ for the integrators A,B respectively
  (e.g., for the default and AC integrators it is $<10^{-3}$, well below this threshold; for any other integrator and the GMM integrator, it is
  less than $5\times 10^{-2}$, or modestly above the target).  Note that since RIFT convergence is assessed using $n_{\rm eff}$,
  the GMM integration would be run many times longer in a normal RIFT run.
}
\end{figure}

 To illustrate how  this new sampler compares to the original implementation in controlled circumstances, we employ both to produce
independent samples from an underlying correlated three-dimensional gaussian likelihood function.   
Figure \ref{fig:TestGMM:Toy} shows the true and estimated one-dimensional cumulative distribution functions, after a
fixed number of likelihood evaluations.  As expected, the new GMM-based integrator  recovers the true
distribution more accurately at fixed cost.  More extensive tests with a wider range of sample sizes and reference distributions
corroborates this anecdotal example.

\subsubsection{GPU-accelerated Monte Carlo integration}
The ILE likelihood is dramatically more efficient when implemented on GPUs.   The previous adaptive integrator, however, performed
all random number generation with a CPU, then transferred large numbers of random samples
to the likelihood  evaluator on the GPU.  The overhead associated with the Monte Carlo integrator can limit
ILE's performance. We therefore re-implemented a simplified version of the previous Monte Carlo integration algorithm,
using \textsc{cupy}/\textsc{numpy} to allow the same source code to drive both CPU-only and GPU-enhanced mode.
The end-user can request this integration algorithm in both CIP (in CPU-only mode) and ILE.
As in the initial implementation, we assume a seperable sampling prior $p_s(\bm{\theta})=\prod_k p_{s,k}(\theta_k)$.
For dimensions that do not benefit from adaptive refinement, we use fixed priors.
For adaptive dimensions, $p_{s,k}(\theta)$ is revised based on the recent past history of $N_{adapt} =n_{adapt}*n_{chunk}$
samples.  The adapted sampling distribution after refinement is a histogram
\begin{align}
p_{s,k}(x) = \frac{1}{\Delta \Theta_k N_{adapt}} \sum_\alpha n_\alpha {\cal S}(x|x_\alpha,\Delta X/N_{adapt})
\end{align}
where $n_\alpha$ is the number of samples in the past history with $\theta_k$ between $\theta_\alpha$ and $\theta_\alpha
+\Delta \Theta/N$, so $\sum_\alpha n_\alpha=N_{adapt}$;  and where $S(x|x_*,\Delta x)$ is a unit step function equal to $1$ between $x_*$ and $x_*+\Delta x$ and
zero elsewhere.  To minimize fine-tuning and the need for costly conditional statements, following the original
implementation we we employ a fixed number $n_{bins}=100$ bins in adaptive dimensions.

\begin{figure}
\includegraphics[width=\columnwidth]{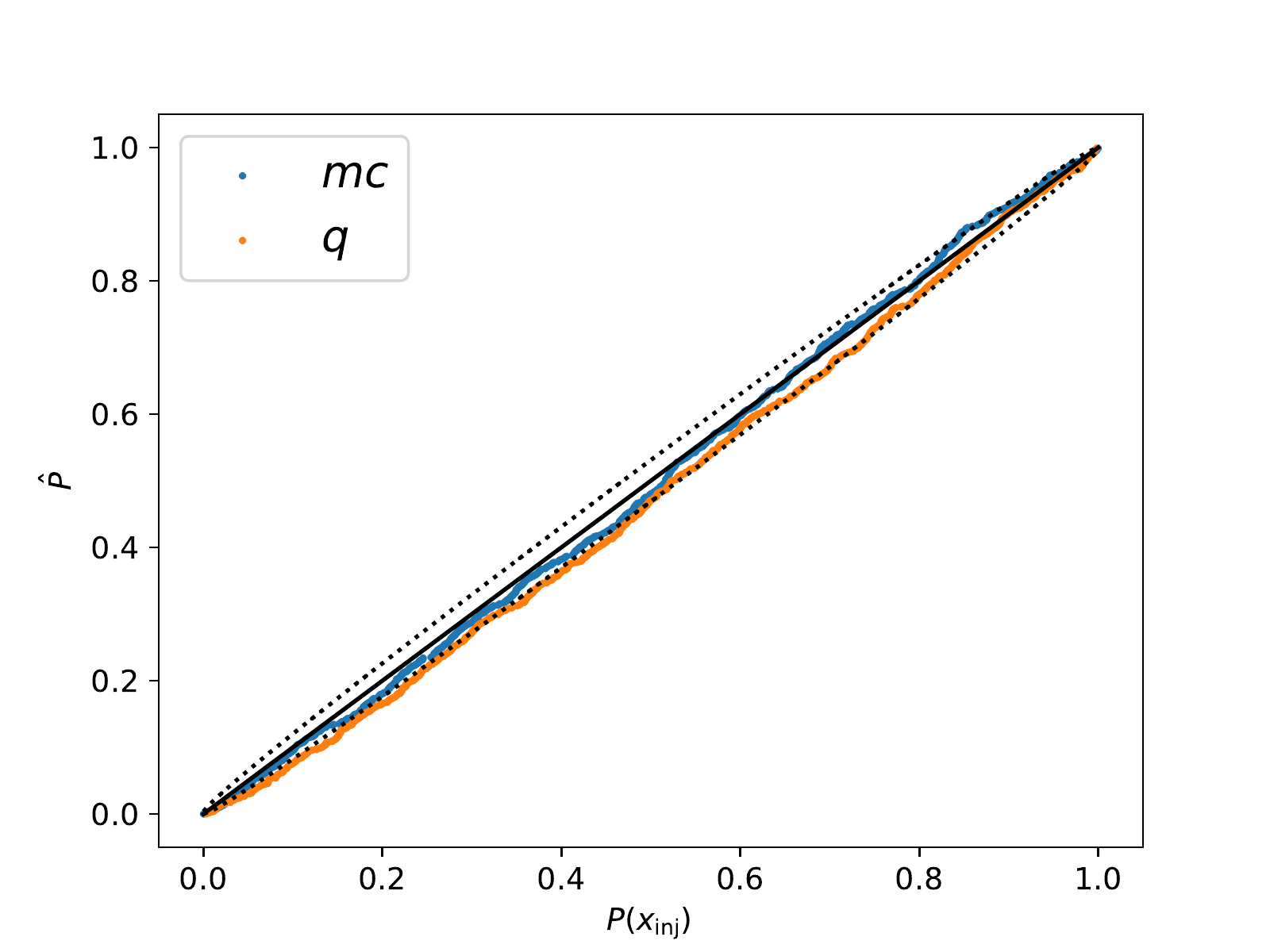}
\caption{\label{fig:AC-gpu_pp}{\textbf{End-to-end test with GPU-accelerated Monte Carlo integration}}
Similar to Figure \ref{fig:NonprecessingO3PP}, a probability-probability plot constructed with synthetic zero-spin
injections.  In this figure, the underlying calculations used marginalized likelihood calculations evaluated using the GPU-accelerated Monte Carlo integration code.
}
\end{figure}

When combined with RIFT's GPU-native likelihood function, all elements of the Monte Carlo integration can be performed on
the GPU board,  with minimal data transfer as needed to orchestrate the integration.  As a result, this fully-GPU marginal
likelihood evaluates very quickly, even for models involving many higher-order modes.  All of the essential coordinate
transformations described previously which accelerate ILE are compatible with this implementation, including distance
marginalization and rotated sky coordinates. 
Figure \ref{fig:AC-gpu_pp} shows an end-to-end validation study of RIFT when  adaptive cartesian integration  is employed within ILE, using a suite of many
synthetic zero-spin sources drawn with  random intrinsic and extrinsic parameters.

\subsubsection{Distance marginalization}
\begin{figure}
\includegraphics[width=\columnwidth]{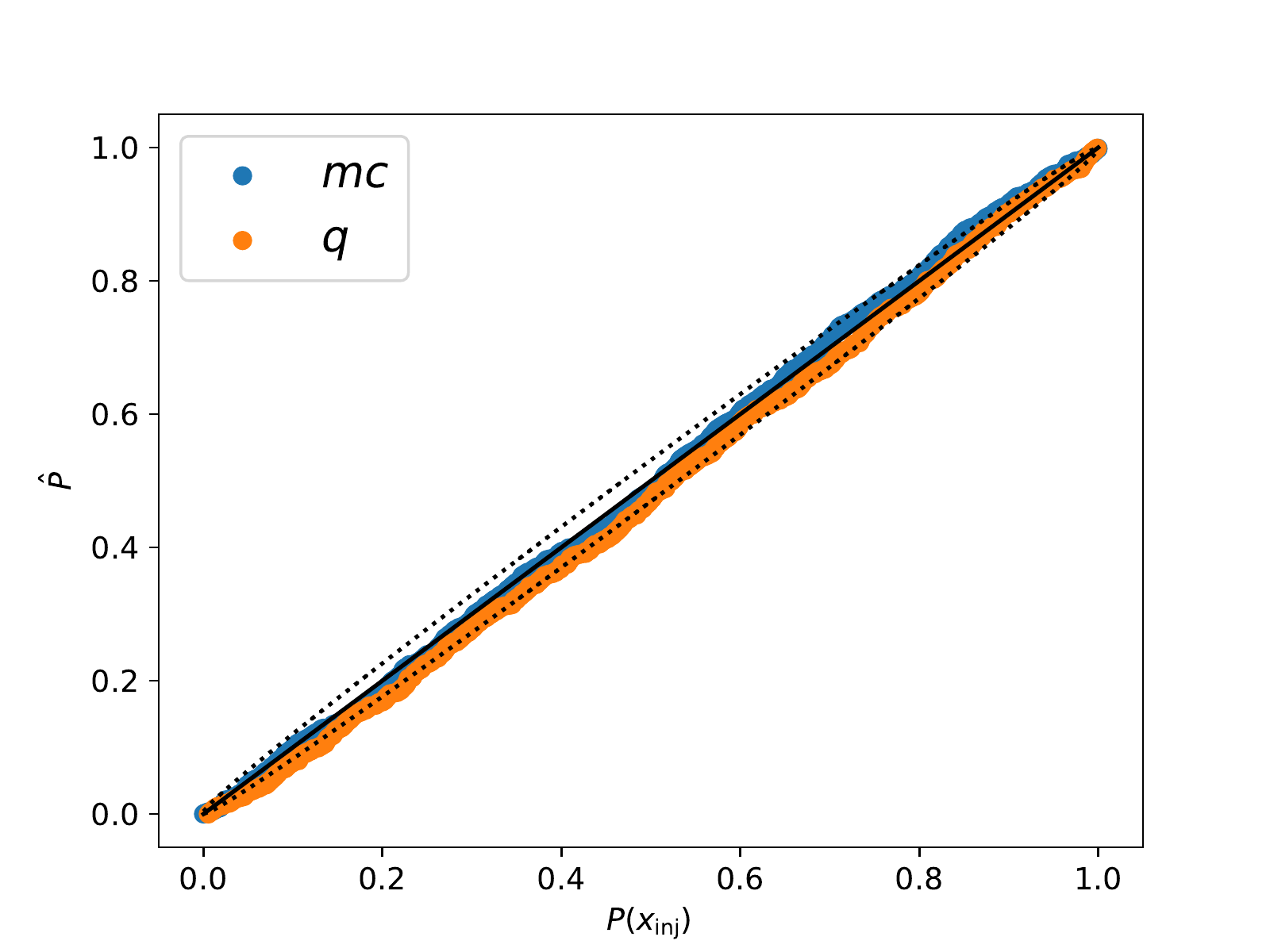}
\caption{\label{fig:dmarg_pp}{\textbf{End-to-end test with distance marginalization}}
Similar to Figure \ref{fig:NonprecessingO3PP}, a probability-probability plot constructed with synthetic zero-spin
injections.  In this figure, the underlying calculations used marginalized likelihood calculations evaluated using the distance marginalization code.
}
\end{figure}

Following previous work \cite{2020PASA...37...36T}, Morisaki developed a concrete technique to directly marginalize over
distance\cite{gwastro-mergers-Soichiro-dmarg}. 
An implementation of this technique by Morisaki and Wysocki is now available within ILE.
Figure \ref{fig:dmarg_pp} shows a large-scale end-to-end test of this code, to demonstrate it preserves the statistical
purity of our recovered intrinsic parameter distributions.
Directly marginalizing in distance reduces the computational overhead of the Monte Carlo integration step, allowing
notably faster performance at fixed target accuracy.
Figure \ref{fig:dmarg_pp} shows an end-to-end validation study of RIFT when  distance marginalization  is employed within ILE, using a suite of many
synthetic zero-spin sources drawn with  random intrinsic and extrinsic parameters.

\subsubsection{Gaussian resampling of gaussian likelihoods}
When suitable, a  gaussian likelihood approximation [Eq. (\ref{eq:quad})] allows us to refactor our Monte Carlo integration technique:
rather than draw samples $x_k$ from a \emph{sampling prior} $p_s$ and computing the expectation of ${\cal L}(x) p(x)/p_s(x)$, we
instead draw samples from the \emph{normal likelihood} [modulo boundary truncation effects] and compute the expectation
of ${\cal L}_{max} p(x)/p_g(0)$ where $p_g$ is the appropriate truncated normal distrbution evaluated at its peak.
The posteriors deduced with a Gaussian likelihood can be surprisingly close to the full answer, even allowing for large
model  dimensions \cite{gwastro-mergers-GaussianLikelihoods-Delfavero2021}.
This reweighting-based technique can also be performed extremely quickly, with the corresponding calculations generally
limited by infrastructure  (e.g., starting up an interpreter and loading libraries; file input and output).
A subsequent companion study will outline the reliability and performance of various ultra-low-latency strategies,
including the reliability and efficiency of this approach.

\subsection{Updated convergence architecture}
\label{sec:sub:new_arch}
\begin{figure*}
\includegraphics[width=\columnwidth]{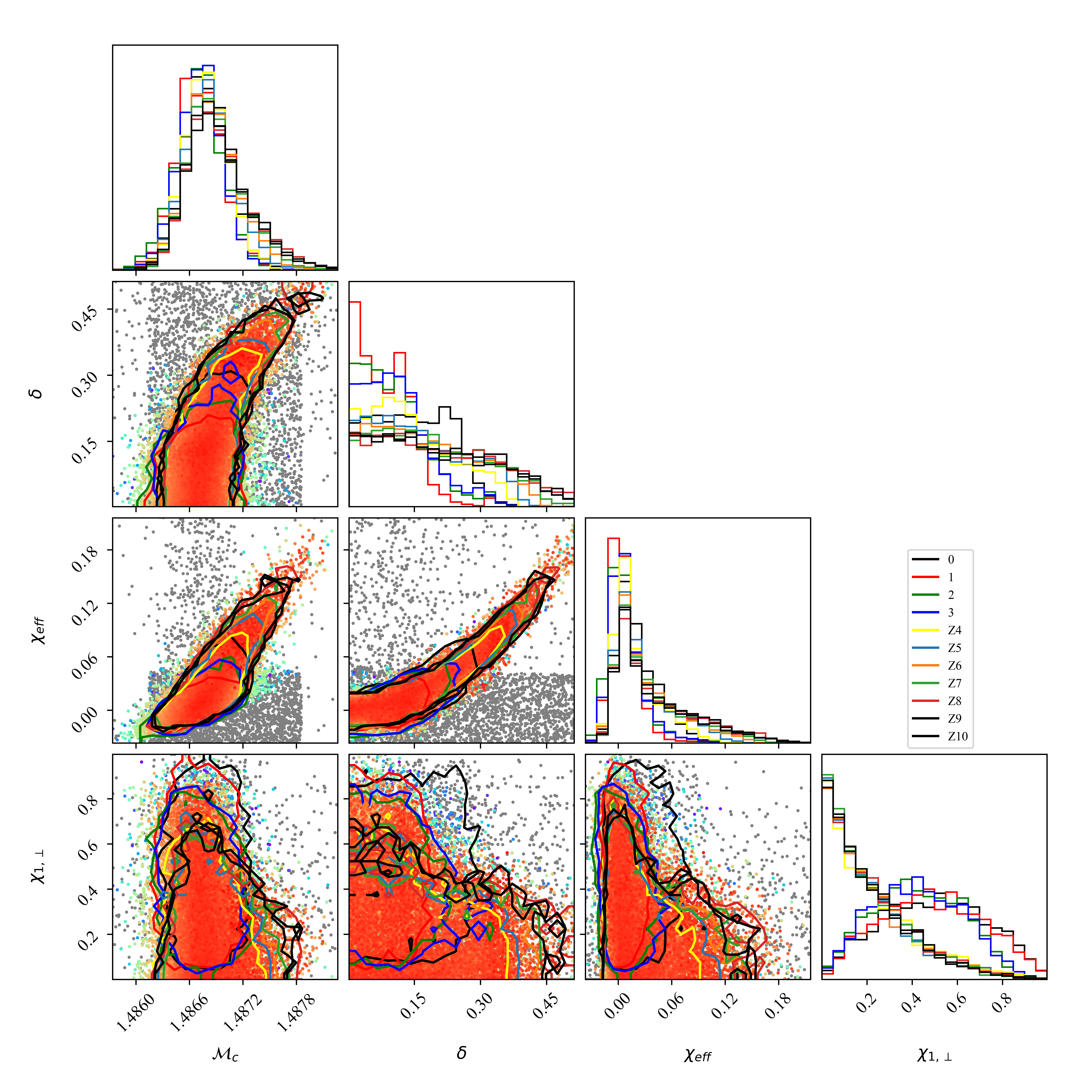}
\includegraphics[width=\columnwidth]{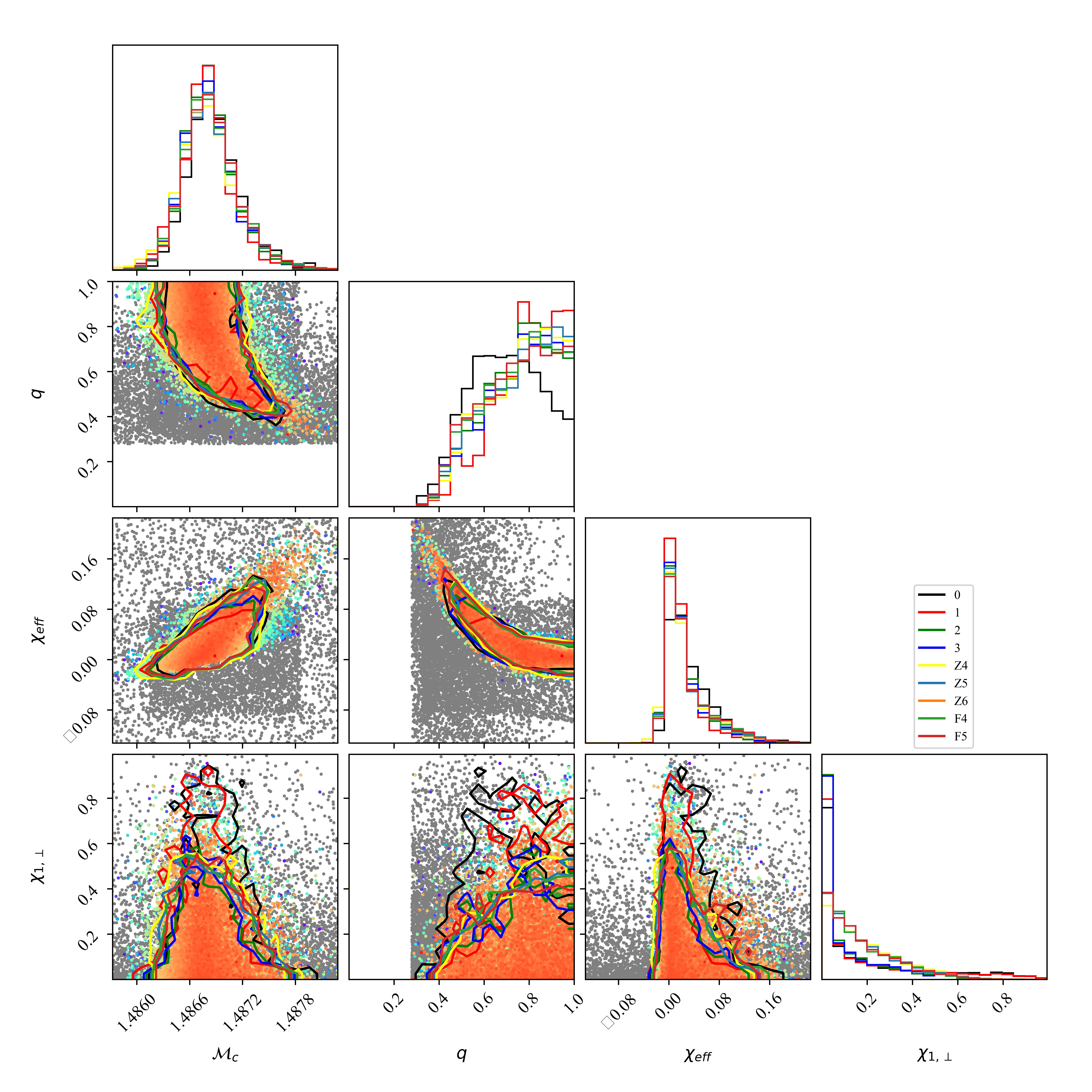}
\caption{\label{fig:demo_new_arch}\textbf{GW190425, inferred with different choices for initial priors}: Both panels show corner plots
  with 90\% credible interval quantiles inference for GW190425 using IMRPhenomPv2 using binary black hole priors (i.e.,
  $|\chi_i|$ uniform in magnitude between 0 and 1) for different RIFT iterations, indicated by different line colors; the color
  scale shows the likelihood range, over a dynamic range $\Delta \ln {\cal L}\le 15$.    The two panels adopt the same
  architecture, differing only in   the coordinates used for spin sampling [Eq. (\ref{eq:pseudo_cylinder})] and in the
  prior adopted for transverse spins in the first four iterations. 
}
\end{figure*}

\begin{figure}
\includegraphics[width=\columnwidth]{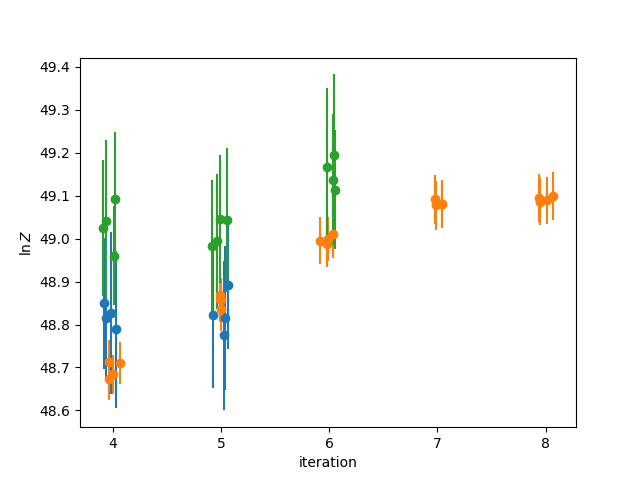}
\caption{\label{fig:Z_190425}\textbf{Intermediate evidence estimate versus iteration}: For the RIFT analyses of 190425
  shown in Figure \ref{fig:demo_new_arch},  this figure shows our \emph{internal} evidence estimates
  $Z=\int {\cal L}_{\rm marg} p(\lambda)d\lambda$ produced by each CIP worker versus iteration number, while iterating
  towards convergence.  Points show the  Monte Carlo estimate, and error bars are estimated 90\% \emph{statistical} confidence intervals
  (i.e., $1.64$ times the Monte Carlo integration error estimate).  We have slightly jittered the $x$ coordinates in
  this figure to better differentiate overlapping error bars.
  Orange and blue indicate the analyses performed in the left and right panels of Figure \ref{fig:demo_new_arch} respectively.
}
\end{figure}

As described in Section \ref{sec:arch}, RIFT adopts different settings in different iterations, to leverage our
experience with  hierarchically  exploring  compact binary parameter space.  Particularly for unsupervised operation, RIFT's
initial grids  often only explore a three-dimensional subset of  nonprecessing binary parameters.   We therefore adopt a
sequence of settings for each iteration's use of the CIP code, which both performs fits  and generates the posterior via weighted Monte Carlo
integration.   In O3, these settings were chosen to gradually increase the sampling and fitting  dimensionality, with
the hope of identifying and characterizing many strongly precessing BH-BH binaries.    However, as described
in Section \ref{sec:sub:o3_problems}, the overly conservative and inflexible choices adopted for unsupervised operation in O3 were extremely
inefficient for low-mass or highly-asymmetric binaries.  

In this work, we introduce a new architecture which (conbined with the previously-reported coordinates and integrators)
efficiently and reliably recovers the properties of low-mass and asymmetric binaries.
Specifically, we first perform two iterations using $\mc,\delta,\chi_{\rm eff}$ as fitting parameters, sampling uniformly in
mass; uniformly in $\chi_{i,z}$;  and with a modified prior for $\chi_{i,\perp}$.    These iterations capture the
dominant aligned-spin degrees of  freedom for most BH binaries, while populating the transverse spins.
We  next perform two iterations using $\mc,\delta,\chi_{\rm eff},\chi_p$ as fitting parameters, with the same priors as
before.  By adding transverse spin dependence, we capture the (dominant)  impact of transverse spin, particularly
important for low masses or asymmetric binaries when transverse spins are so frequently constrained to be nearly zero.   
Finally, we iterate to convergence, using $\mc,\delta,\chi_{\rm eff},\chi_{-},\chi_{i,\perp}$ as fitting parameters, and
uniform-in-spin-magnitude sampling parameters.  Extrinsic parameters are extracted from the final converged iteration.
For comparison, we will also describe results derived using an otherwise similar architecture, but adopting
the pseudo-cylindrical coordinates for spin provided by Eq. (\ref{eq:pseudo_cylinder}).   In these alternative analyses, the
first four iterations adopt a uniform prior on $\bar{\chi}_{i,u}$, which densely samples the region with
$\boldsymbol{\chi}_{i,\perp} \simeq 0$.

This new approach is enabled by adding a notable missing feature for RIFT: iteration until convergence.  RIFT users can
now request a specific CIP configuration be used repeatedly, in a recursively-generated sub-workflow, until the
posterior converges according to the user-specified convergence diagnostic.  

Figure \ref{fig:demo_new_arch}  illustrates unsupervised operation for GW190425, interpreted with IMRPhenomPv2.  The left panel shows our
default new architecture, where the initial prior over transverse spins is  well-adapted to discovering and characterizing large transverse spins, while the
right panel employs a more concentrated initial spin prior (i.e., the uniform-in-$\bar{\chi}_{i,u}$ prior).  Both demonstrations perform dramatically better than 
the low-mass analysis shown in Figure \ref{fig:unpleasant_o3_run}, with  a steady increase in understanding as our exploration and likelihood model
adapts as necessary to model the posterior given the adopted priors.  However, these two analyses' small differences
highlight the importance of adopting initial priors well-suited to the event and final objective. 

In the left panel of Figure \ref{fig:demo_new_arch}, the RIP and spin coordinate systems enable
the first two iterations to rapidlly identify pertinent aligned degrees of freedom.  The next two iterations then
correctly contrain the transverse spin, while refining an estimate  for masses and $\chi_{i,z}$ that is appropriate for
our initial prior.  However, when we adopt the final spin prior and iterate to convergence, the code (correctly)
increasingly identifies an extended region with higher spin and mass ratio, smoothly connected to the main posterior but
now identified as pertinent given the new prior.  Iterations cease when the code  converges.

In the right panel of Figure \ref{fig:demo_new_arch}, we repeat our analysis using the alternative configuration above,
differing only in the spin coordinates used throughout the analysis and in the transverse spin prior adopted for the first few iterations.
As exemplified by the analysis from the left panel, nature so far has provided binary black holes consistent with zero
transverse spin.  For low mass binaries, the transverse spins are well constrained to be near zero.  As a result, the
analysis shown on the right converges much more quickly to our final result.
The contrasting performance of the two analyses shown in Figure \ref{fig:demo_new_arch} highlights the dangers of simply
reweighting an existing result to a new prior; see also Appendix \ref{ap:mc} for further discussion.  Figure
\ref{fig:Z_190425} provides another way to  quantify the impact of our initial prior choices on convergence, using the
multiple Monte Carlo estimates of the evidence  $Z=\int {\cal L}_{\rm marg} d\lambda$ and their error reported by each
CIP worker.\footnote{ This internal-use evidence during these intermediate iterations has substantially larger
  statistical errors than final evidence, which is evaluated using much longer iterations during the final iteration.}
In the orange points, showing the analysis using our default transverse prior for the first initial iterations, we see
the evidence estimate systematically evolves upward as the posterior approaches our final converged result.  As
expected given Figure \ref{fig:demo_new_arch}, the
statistical errors estimated from Monte Carlo integration substantially understate the systematic error in the evidence.
By contrast, the green traces immediately identify the final overall evidence, again as expected given Figure \ref{fig:demo_new_arch}.

\subsection{Automated information transfer between analyses}
\label{sec:fetch}

Due to its iterative nature and reliance on archived likelihood evaluations $\{\lambda_k,{\cal L}_k\}$, RIFT has unique
capabilities to use information from previous or even concurrent analyses with different models and configurations \cite{gwastro-PENR-RIFT}.
Though these capabilities are particularly powerful when adopting the same waveform model and data analysis settings
(i.e., the likelihoods themselves can be re-used),  they can also be very powerful tools even between waveforms.
As a concrete example, RIFT analyses performed using multiple waveform models can efficiently marginalize over waveform uncertainty
\cite{gwastro-RIFT-systematics-AnjaliAasim-2020}.
As another example, rapid analyses with simpler waveform physics (e.g., no precession) or faster waveform models can
feed directly into an ongoing RIFT analysis, by supplying additional target points for likelihood evaluation.

The RIFT workflow has always had natural stages where external information can be conveniently inserted (e.g., adding
likelihood evaluations, 
or proposing new points for next-step evaluation).  Where previously we had ad hoc procedures to manually edit or supply
the necessary files, with the latest generation of RIFT we introduce the \texttt{fetch} process, designed to  retrieve
candidate points (or likelihoods) from any external run.   [In fact, we even retrieve information from the recursive
iterate-to-convergence stage via this same framework.]

We foresee three natural use cases for the \texttt{fetch} framework.  
First, this framework enables a particularly efficient run hierarchy for modest-latency analysis over the first few
minutes, hours, and days.  Fast analyses (with RIFT and other codes) using simplified physics (e.g., without precession)
seed longer-timescale analysis with more physics.  Within and between stages,  RIFT supplies an approximate posterior
distribution.
Second, building on this approach for offline followup, this framework enables efficient analysis with multiple
approximations, where these approximations' analyses may have different timescales owing to their computational cost.
The two analyses can inform each other, if simultaneous, or the fast analysis can inform the slower one if computational
costs are significantly different (e.g., due to the incorporation of many higher-order modes).
Finally, by fetching from previous work, RIFT can most efficiently complete final production-quality analyses, building
on previous experience.

\subsection{Adaptive mesh refinement}
In conventional RIFT, the fitting- and posterior-generation stage is the most serial and time-consuming, particularly
for low-mass sources.  Rose et al  \cite{2022arXiv220105263R} introduced an adaptive mesh refinement (AMR) for
gravitational wave parameter inference, a strategy which very
efficiently finds and explores the (intrinsic, marginal) likelihood over modest dimensionality  (i.e. $d\le 4$, corresponding to the
nonprecessing intrinsic degrees of freedom).     Below, we describe one way that RIFT can use AMR without employing additional
external information supplied by searches or precomputed overlap tables.

Our default AMR approach is initiated with a coordinate hypercube in one of a few blessed groups of parameters, such as
$\mc,\delta,\chi_{i,z}$.  The  AMR engine then successively retrieves information about likelihoods on grid nodes;
assesses grid cells which require refinement; and identifies new node centers for subsequent evaluation.
At each specific grid level, cells are identified as needing refinement based on a threshold $p$.  A specific cell $n$ out of $N$ cells is selected if (after sorting all cell likelihoods ${\cal L}_k$
and forming the cumulative sum  ${\cal S}(n) =\sum_{k=1}^n {\cal L}_k$) the sum satisfies $S(n)/S(N)>1-p$.  Roughly speaking, this
threshold associates each cell (at any refinement level) with equal AMR  probability mass, and performs refinement of the
most significant fraction $p$ of the nominal AMR probability mass.   Each successive grid level fully refines all areas
requested for refinement.  No prior coordinate-dependent or refinement-level
information is used to guide the refinement choices. %

To assess convergence of our refinement, we have two natural diagnostics: the integrated likelihood and the distribution
of likelihoods.  For the first, at each level we can  estimate the unweighted evidence  $Z = \int {\cal L} dx$ with successive Riemann
integral estimates
$Z_\ell = \sum_k {\cal L}_{l,k} \prod_\alpha \Delta x_{\alpha}/\ell$ where ${\cal L}_{\ell,k}$ are the likelihood values in level $\ell$ with a
top-level grid spacing of $\Delta x_\alpha$ for each dimension $\alpha$.
For the second, we can use the distribution of $2\ln {\cal L}_{\ell,k}$ at each $\ell$.  When AMR is nearly
converged, the inter-evaluation seperations will be small, and the distribution should  be roughly consistent with a $\chi^2$
distribution with roughly $d$ degrees of freedom, depending on the number of well-constrained parameters being
simultaneously explored.

Lacking the need to interpolate the likelihood or sample a posterior, the AMR engine operates within seconds.
Operationally, the AMR engine behaves like  a drop-in replacement for CIP: the code can effectively iterate to
convergence using just AMR.  Thus, the AMR engine provides an extremely rapid way to explore the likelihood.  
As desired, we can also run conventional CIP in parallel, during  postprocessing, or even as part of a parallel non-AMR analysis
with more degrees of freedom to identify a fully-interpolated posterior distribution.
This latter approach in particular offers an extremely powerful technique to bootstrap inference for the most
challenging low-mass, high-mass ratio sources.

\subsection{Single-event EOS inference with pretabulated equation of state}
\label{sec:eos}

RIFT already has at least two frameworks to constrain the nuclear equation of state (EOS).  On the one hand, given any
tabulated EOS, RIFT can efficiently compute an evidence for that EOS, based on integrating the interpolated  marginal likelihood
${\cal L}(\lambda)$ while accounting for the unique relationship between NS mass $m$ and tidal deformability
$\Lambda$ that this EOS allows \cite{LIGO-GW170817-EOS}.  On the other hand, RIFT can also similarly constrain a parameterized
equation of state, constructing a posterior for its hyperparameters  \cite{gwastro-PENR-RIFT}.
However, single-event inferences have two significant limitations.  First, near-future
measurements must simultaneously constrain the EOS and NS mass and spin distribution, to avoid
introducing biases into the recovered EOS.   Second and more pertinent here, any single-event inference ignores
substantial prior knowledge about the nuclear EOS obtained from previous analyses.  The extension described below
provides a simple remedy to this situation suitable for near-future investigations. 

Several studies have adopted nonparametric approaches to EOS inference, relying on concrete tables of many EOS
realizations \cite{2020NatAs...4..625C,2019PhRvD..99h4049L,2021PhRvD.104f3003L,2022arXiv220411877G}.  These EOS libraries can be weighted to better fit any observation (e.g., gravitational wave, NICER, or
pulsar mass constraint) and as needed resampled to impose desired priors (e.g., uniform in maximum mass, $R_{1.4}$, et
cetera). 
The most precise but computationally intensive RIFT strategy involves brute force: compute the EOS evidence for each
tabulated  EOS.
A simpler albeit more approximate strategy involves an order statistic $S_\alpha$ defined for every tabulated EOS
$\alpha$.  We have adopted $S=\lambda(\mc*2^{1/5})$ as our ordering statistic: the tidal deformability of  each neutron
star in a symmetric binary, such that the chirp mass is consistent with the observed (detector-frame) chirp mass.
Because in practice the mass ratio of NS binaries can't be differentiated from unity, this quantity is a good estimate
for the dominant impact  ($\tilde{\Lambda}$) that the EOS has on the inspiralling binary, evaluated at masses
appropriate for the binary.    Each iteration,  CIP can construct a posterior in $X=(m_i,\boldsymbol{\chi}_i,S)$ and thus
proposed synthetic binaries $\lambda_k$, where the binary tidal deformabilities $\Lambda_i$ associated with each $X_k$ are estimated using the
EOS with the closest order statistic $S_\alpha$ to $S_k$ (i.e., $\Lambda_i(S_{\alpha})$).  
This approach allows us to quickly employ any EOS tabulation conditioned on any previous measurements as part of our usual
iterative inference technique.

\section{Selecting Fiducial RIFT configurations}
\label{sec:fiducial_configs}
RIFT's modular organization offers immense operational flexibility.  Before providing detailed validation studies for
selected configurations, in
this section we briefly describe several code configurations and report on their performance, to illuminate our choices
behind the specific configurations.

\subsection{Selecting between integration and fitting algorithms: A matrix of configurations}
RIFT has several modules for integration and interpolation.    
To simplify the process of
discriminating between and validating all of the principal code configurations, we for simplicity focus the most
well-behaved scenario:  massive binary black holes, similar to those frequently identified by  \response{binary black hole}  searches
in advanced LIGO and Virgo data during O3.  This choice for fiducial profiling tests, anecdotal examples, and PP plots allows us to assess
these configurations in the best possible light, and is appropriate for most observed sources.

Specifically, we summarized three integration techniques for CIP (default, GMM, and adaptive cartesian or AC) and three  fitting
methods (\response{gaussian process (gp), random forest (rf), and sparse gaussian process}).   Additionally, some of these CIP techniques can be employed with multiple refinements
(e.g., different parameter correlations allowed for GMM; different coordinate systems; et cetera).  
Both new integration techniques can potentially also be used in and accelerate ILE.   Being GPU-accelerated, the AC
method is particularly well-suited for ILE, since its other costly likelihood-evaluation operations are already
performed on-GPU.
In subsequent sections, we will exclusively employ AC integration within ILE.
With a focus only on seleting between different  CIP configurations, in this section we fix our ILE settings, employing
the previous default Monte Carlo integrator, and only report the impact on overall CIP runtime.

\begin{figure}
\includegraphics[width=\columnwidth]{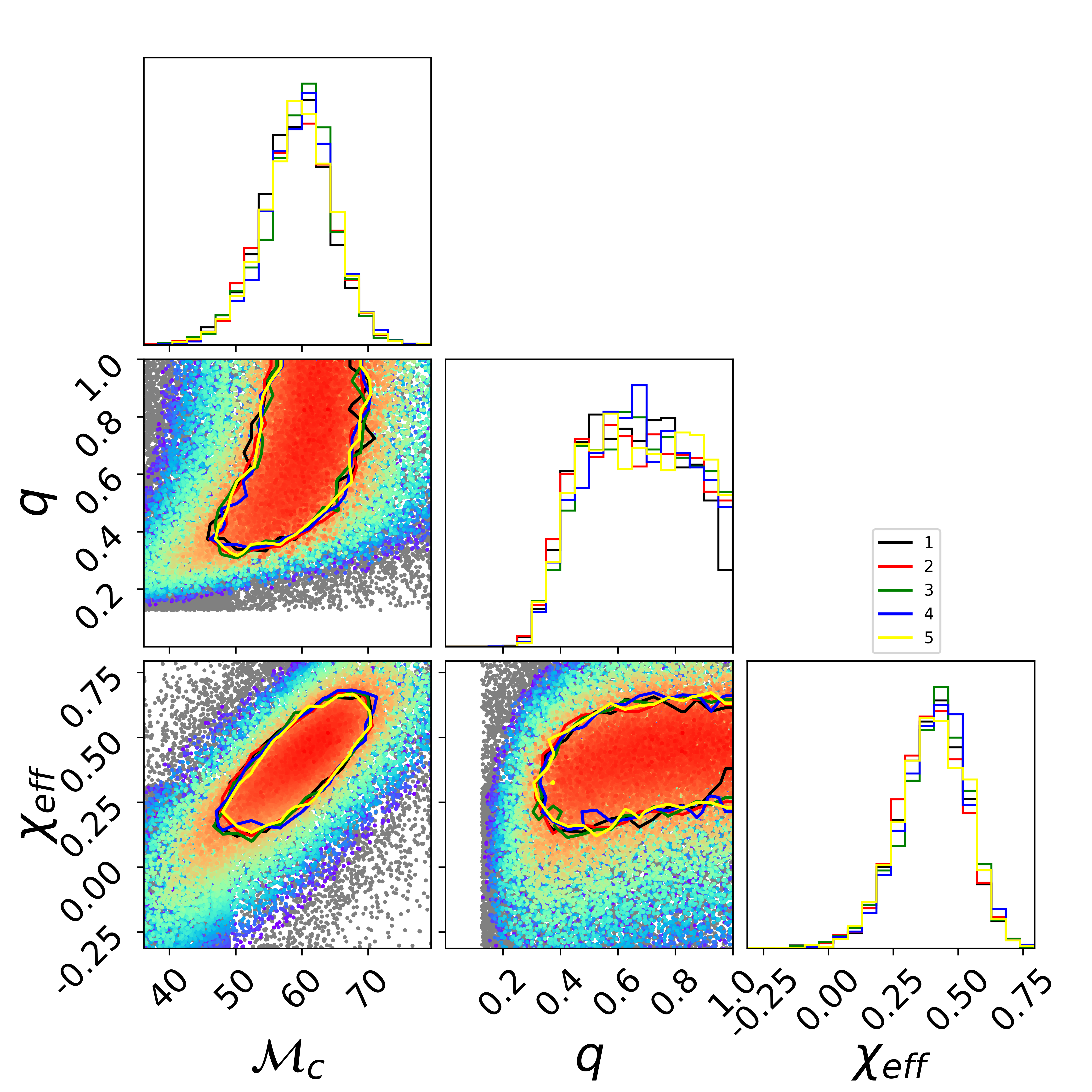}
\caption{\label{fig:show:profile_190620}\textbf{Example of results from fiducial GW190620 analysis:} Posteriors
  derived from all five benchmark analyses of GW190620 \response{enumerated in Table \ref{tab:profile_190620}}  using
  IMRPhenomD and uniform priors on $\chi_{i,z}$.  All agree.
\response{  The run labels 1,2,3,4,5  matches the row order used in Table \ref{tab:profile_190620}.}
}
\end{figure}

Table \ref{tab:profile_190620} illustrates changing overall code resource use from an analysis of GW190620 with a
straw-man configuration: a nonprecessing IMRPhenomD model with uniform priors on $\chi_{i,z}\in[-1,1]$.  All
configurations adopt the same architecture: RIP coordinates for spin; an initial grid of 1500 points;  two iterations
omitting the subdominant spin, followed by iteration to convergence with the subdominant spin included.
As demonstrated by Figure \ref{fig:show:profile_190620}, all analyses converge to a comparable-quality result.
Considering all possible pairs of these 5 analyses, the mean  one-dimensional JS divergence for $\mc_z,q,\chi_{\rm eff}$
are
$(2,1.8,1.9)\times 10^{-3}$, respectively, dominated by comparisons with the O3 configuration (GP/default)  and consistent with the target threshold and sample size produced by this
experiment: 4500 samples
produced from each analysis, based on $n_{\rm eff}$.  
Despite intentionally adopting the most favorable circumstances for the default configuration, with very few
points and model complexity for the gaussian process interpolator and fiducial integrator, this intentionally simplified
example  shows that even in this simplest of cases, 
using our new interpolation and integration methods produce overall better performance.

\begin{table}
\begin{tabular}{|c|c|c|c|c|c|}
\hline
Fit & Integrator & Total (h)  & CIP (h)  & ILE (h) & $T_W$(100,3) \\
\hline
GP & Default & 14.25 & 3.8 &  10.5 &  1.4\\
GP & GMM & 11.8 & 3.5 &  8.25  & 1.2\\
RF & Default & 7.5 & 0.17 & 7.5  & 0.13\\
RF & GMM & 7.3 & 0.17 & 7.1  & 0.13 \\
RF & AC & 5.5 & 0.17 & 5.3 & 0.11 \\
\hline
\end{tabular}

\caption{\label{tab:profile_190620}\textbf{Example of total analysis costs: GW190620}: Runtime costs (in hours)
  for a complete, converged analysis of GW190620 using IMRPhenomD and   five different RIFT
  configurations: two different interpolators and three different integrators.   The first two columns indicate the CIP
  interpolation and integration configuration; the third is the total CIP resource usage (here, over all 3 simultaneous CIP
  instances); and the fourth is the total ILE resource usage, which is smaller for more efficient posterior exploration
  and modeling.  The last column [Eq. (\ref{eq:resources}), evaluated with $N_I=100$ and $N_C=3$] estimates the average
wait time if 100 ILE instances and 3 CIP instances are always available for RIFT inference.
}
\end{table}

Figure \ref{fig:Reproduce} %
illustrates  code performance on inference of a single zero-spin binary black hole using a nonprecessing IMRPhenomD
model.   

\subsection{Fiducial production-quality configuration}
\label{sec:recommend}
Motivated by the above, we recommend the following settings for our production analysis.
For marginal likelihoods (ILE), we use the AC integrator with distance marginalization, using a target $n_{\rm
  eff}\simeq 10$.  Only the skymap is adapted; other degrees of freedom are sampled by brute force.   When assessing
batches of points by a single ILE worker, we freeze the skymap after the first iteration.
For posterior generation (CIP), we use the GMM sampler with an RF fit, using the previously-described convergence
architectures including iterating to convergence, with at least 3 CIP workers contributing to the overall posterior in
each iteration.  We use correlated sampling among $\mc,\delta$ and the cartesian spin
components early on, to accelerate sampling.

\section{Tests}
\label{sec:tests}
\label{sec:sub:validate}

\subsection{Anecdotal end-to-end unsupervised operation }
To insure that these alternative algorithmic components \response{do not} change RIFT's inferences, and to obtain profiling information to
characterize their performance, we performed a large suite of analyses on real events throughout and after the
development process.
Figure \ref{fig:demo_new_arch} shows a concrete example: an analysis of GW190425 with IMRPhenomPv2 with $\ell \le 2$
modes.
Our default test suite included  GW151226, GW170829, GW190412, GW190425, GW190814, GW190620, and GW200115. 
Almost all worked without human supervision throughout the development process; in our final code configuration,
GW190814 converges quickly as well. 
We systematically tested  ILE
with GPU acceleration and distance marginalization; CIP with a random-forest fit and correlated GMM sampling, with 3
workers; and a workflow with a convergent subdag and customary precessing iterative structure. 
Selected examples from these validation studies appear elsewhere in this work.

\subsection{Illustrative example}
\begin{figure}
\includegraphics[width=\columnwidth]{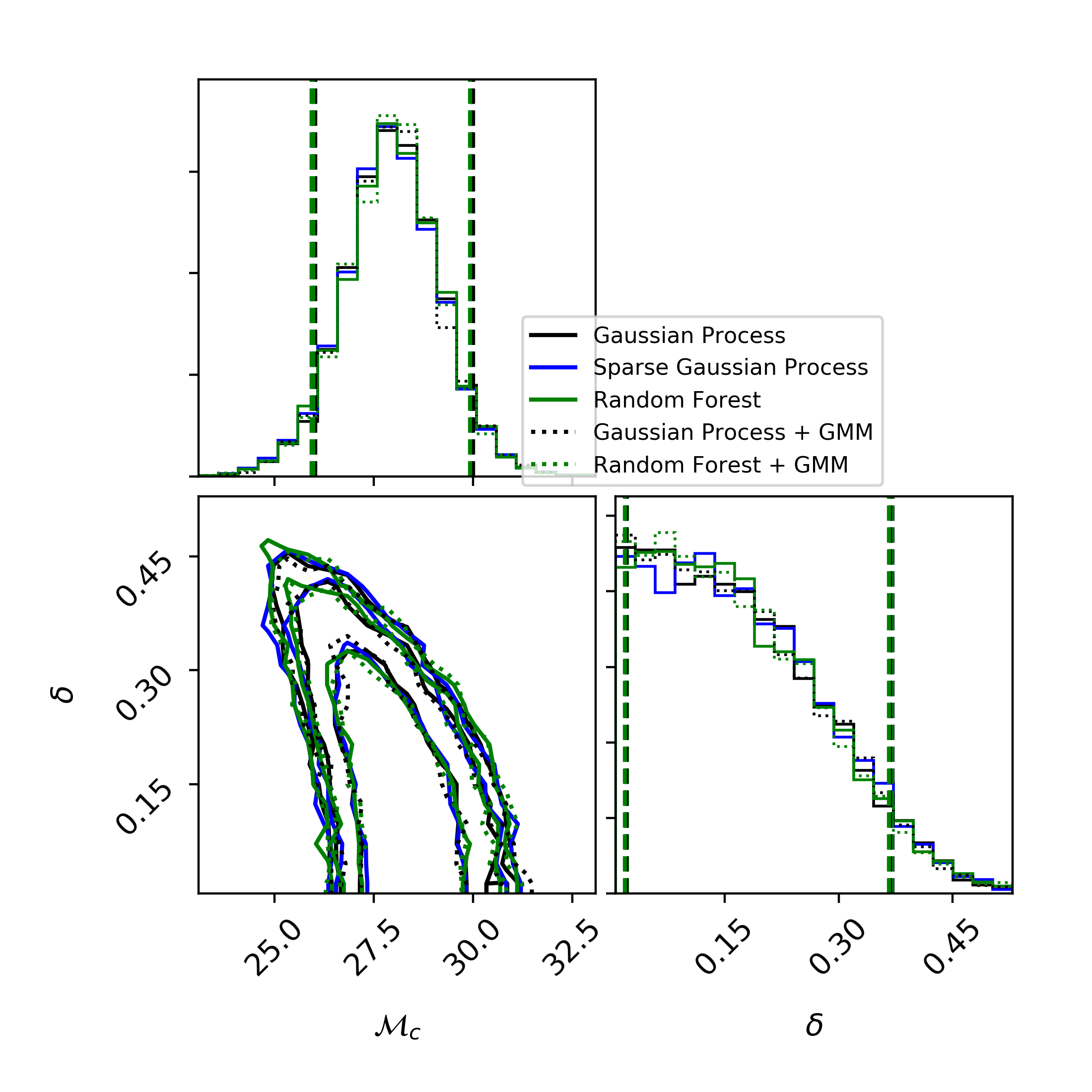}
\caption{\label{fig:Reproduce}\textbf{Recovering a fiducial posterior distributions: \response{binary black hole}}: %
Results of
  an analysis of a \response{binary black hole} model with aligned spins.  Posterior distribution for $\mc$ and
  $\delta=(m_1 -m_2)/M$ obtained using RIFT with different interpolation methods (colors) and integration methods (line
  styles).  All source parameters and analysis settings agree precisely with the original RIFT-GPU paper \cite{gwastro-PENR-RIFT-GPU}.
The black curves indicate conventional GP interpolation with a squared exponential kernel, as in the original RIFT
paper; the blue, green, and red lines indicate sparse-GP interpolation (SGP), random-forest interpolation (RF).  Colors
and line styles indicate the integration and fitting methods used.
}
\end{figure}

To provide systematic, controlled, quantitative tests of our algorithmic changes, we employed two fiducial sources:
GW190620,  as discussed with Figure \ref{tab:profile_190620} and Table \ref{fig:show:profile_190620} above, and a fiducial synthetic
zero-spin source, shown in 
Figures  \ref{fig:Reproduce}.  We employ six variants of our algorithms,
changing the fitting method (line colors) and the MC integrator (line styles).  
Likelihood interpolation methods shown are the original aproach (black), our sparse GP code (blue), and the random
forest code (green).  
Monte Carlo integration options for CIP are the original MC method (solid) and the GMM adaptive integrator (dashed).  
Unlike the previous example, which used a contemporary adaptive architecture, the synthetic source used an O3-style configuration: a fixed number of iterations, using a fixed (and complete) coordinate
system to characterize the likelihood. 

This anecdotal example consists of a zero-spin BH, shown as the first panel in  Figure \ref{fig:Reproduce}.  In
this example, RIFT uses the same setup as the ILE-GPU paper \cite{gwastro-PENR-RIFT-GPU}: we
perform 7 iterations, starting with a 100-point uniform grid in $\mc,\delta$; each iteration has \editremark{5000} evaluation
points.   Each of these tests use a \response{jittering} factor of 3 and force-away parameter of 0.05. We
 show posterior distributions obtained with RIFT using each combination of settings.  All agree.

\subsection{Component performance on many  randomly-selected sources  }

\begin{figure}
\includegraphics[width=\columnwidth]{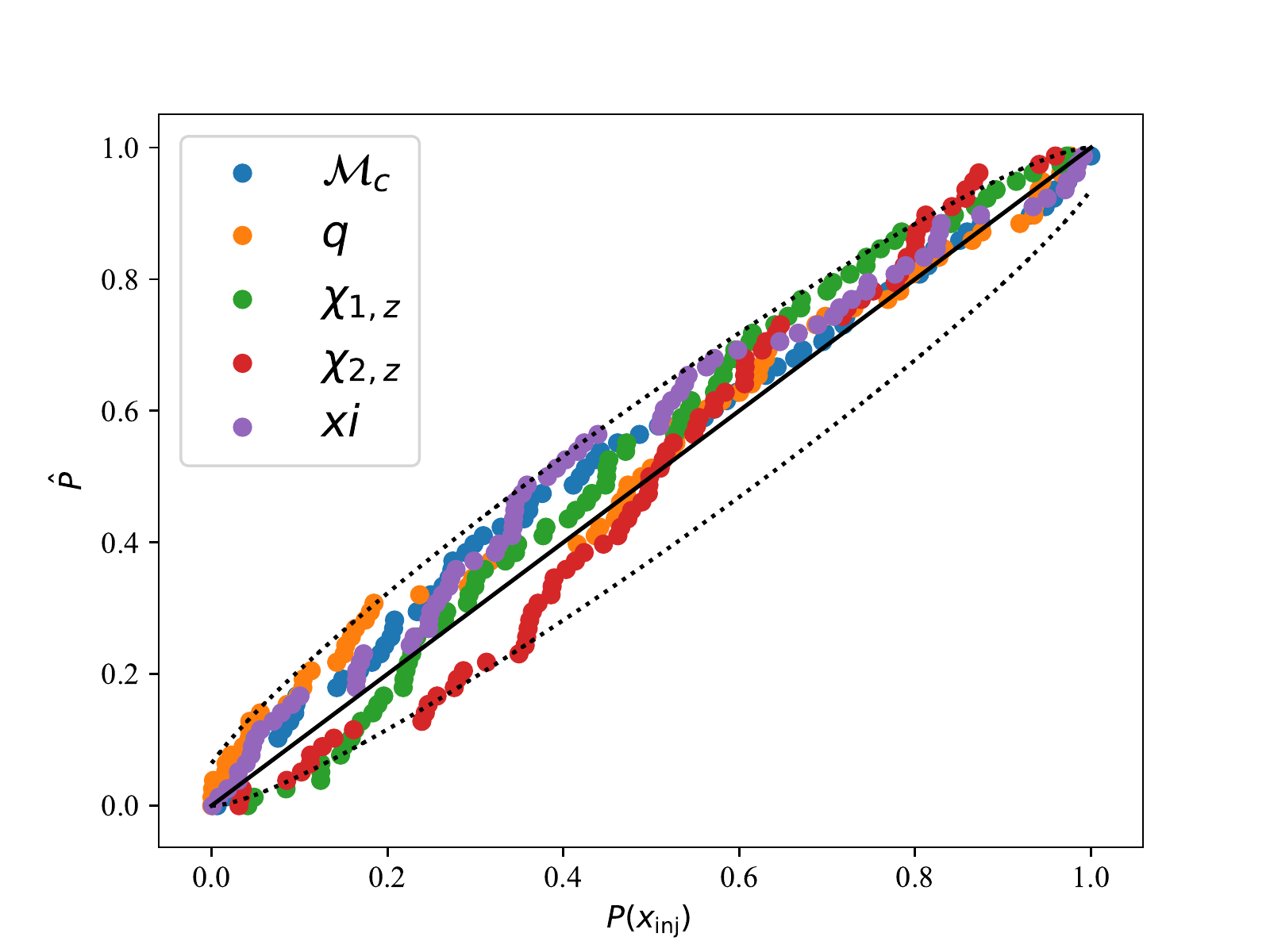}
\caption{\label{fig:PrecessingPP}\textbf{Precessing PP plot}: Probability-probability plot to validate recovery of
  precessing synthetic signals, following the RIFT methods paper \cite{gwastro-PENR-RIFT}. The synthetic sources and parameter inferences are constructed with IMRPhenomPv2 in
  gaussian noise with presumed known PSDs,
 for 3-detector networks, starting the signal at 20 Hz and using a 4096 Hz sampling rate.  Detector-frame masses are drawn uniformly  in the
  region bounded by ${\cal M}/M_\odot \in [30,60]$ and $\eta \in [0.2,1/4]$, and sources are drawn volumetrically between $1.5$ and $4
  \unit{Gpc}$.   Both BH dimensionless spins are drawn uniformly and volumetrically from the unit sphere. 
  For computational efficiency, all sources in this test have a fixed and presumed known sky location. 
}
\end{figure}

We have also validated several of the new code configurations with probability-probability (PP) plot tests, using models
of varying complexity.    Several of these PP plot tests have already appeared earlier, in sections describing and
validating individual module components: 
Figure~\ref{fig:MucoordPP}, a PP test for RIP coordinates and nonprecessing PE for NS with tides (and higher-order
modes); 
Figure~\ref{fig:AC-gpu_pp}, a PP test for the AC integrator;  and Figure~\ref{fig:dmarg_pp}, a PP test for distance
marginalization. 
Figure \ref{fig:PrecessingPP} shows yet  another test of multiple new code components -- here, random forest fits and
GMM integration in CIP.
For this figure,  we constructed 200 random synthetic sources with precessing black
hole spins, then estimated their parameters using the IMRPhenomPv2 waveform model. 

To more sharply validate that our different code configurations for CIP produce identical results on a large sample of
synthetic sources, we  compared two code configurations on 100 random synthetic injections with
zero spin in distinct realizations of random gaussian noise: the fiducial code configuration used in O3, and a version
using a random forest fit and GMM sampler in CIP.  
We extended each analysis until our KL-divergence-based diagnostic on $\mc$ fell below $10^{-2}$.
We find that each pair of analyses of the same data produces the same results, as measured by our KL-divergence-based
diagnostic
 Figure \ref{fig:ReproducePP} shows another measure of agreement between the two algorithms: the difference between the
quantiles $\hat{P}(x_j)$, versus $\hat{P}$.   Small random differences $\simeq 1/\sqrt{10^4}$ between the two inferred
$\hat{P}$  are expected because both probabilities are derived from $10^4$ posterior samples.    By construction, this test shows both codes produce indistinguishable PP plots. 
In short, all the extensions described produce indistinguishable results,
differing only in their efficiency.

\optional{
\begin{figure}
\includegraphics[width=\columnwidth]{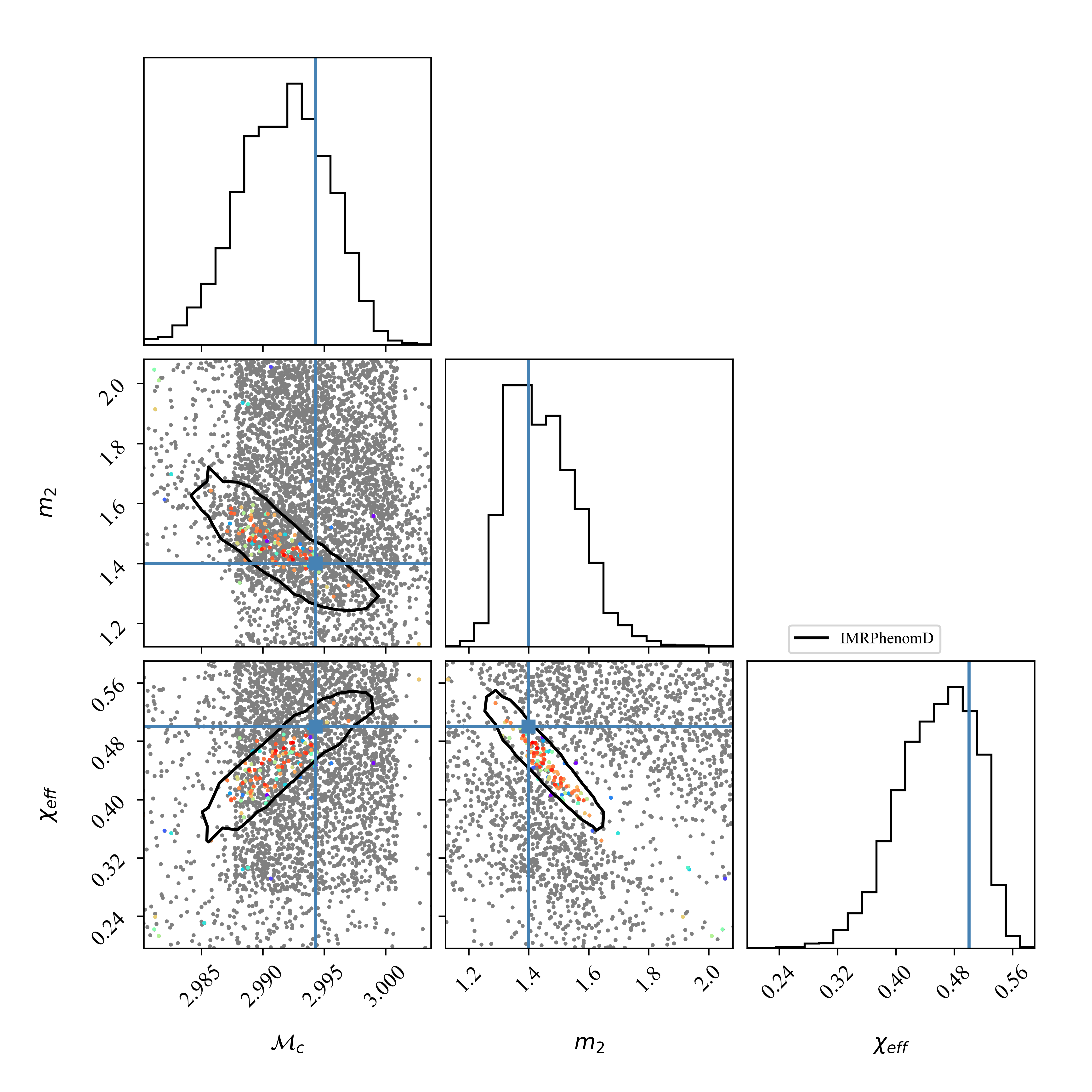}
\caption{\label{fig:BHNS_fiducial}{\textbf{Demonstration for an NSBH}: Corner plot of a BH-NS source $10M_\odot+1.4
    M_\odot$ with non-zero aligned spins, using the RF+GMM methods.  As discussed in the text, this analysis would be
    cost-prohibitive with several other code configurations.}}
\end{figure}
}

\begin{figure}
\includegraphics[width=\columnwidth]{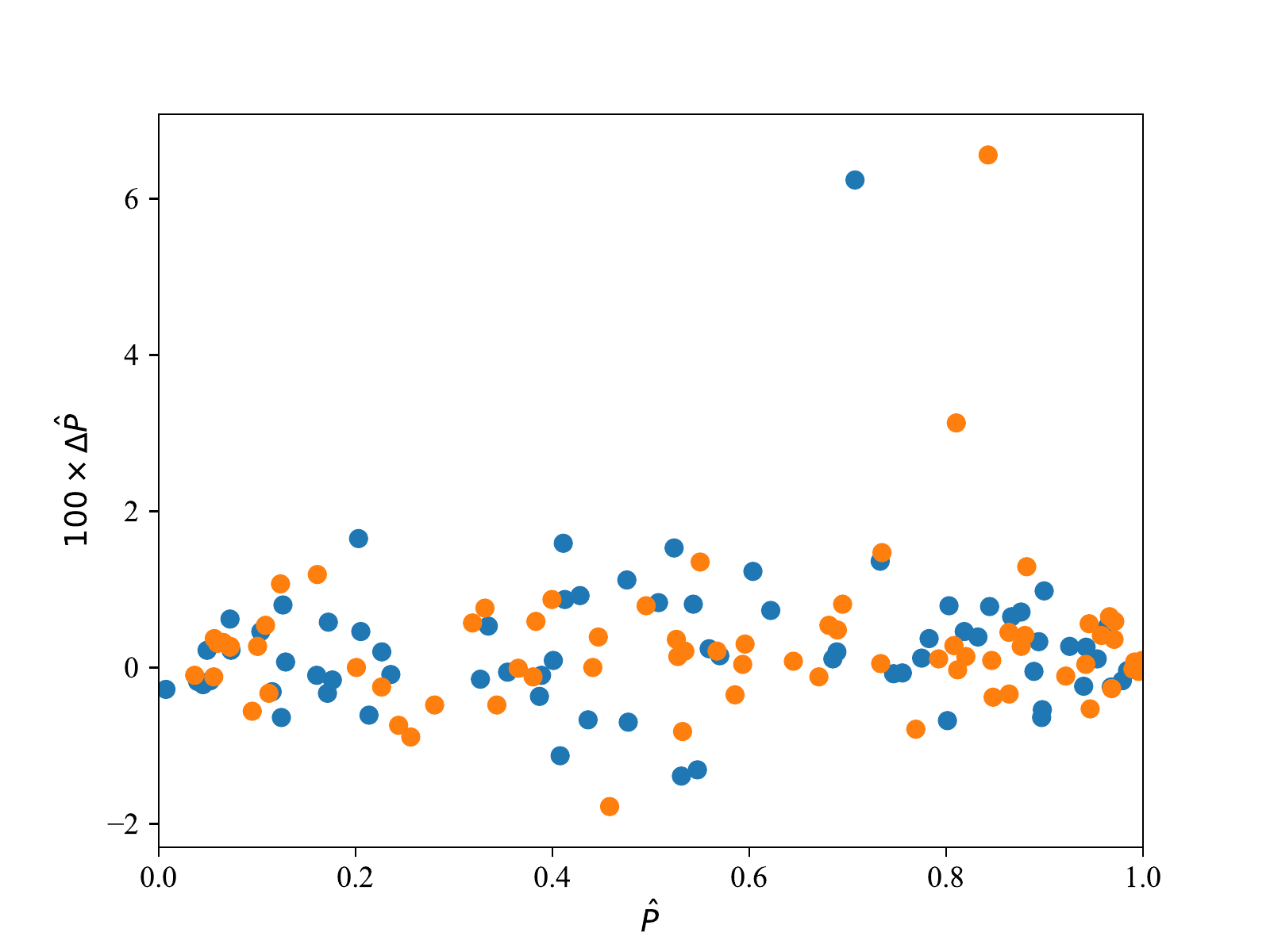}
\caption{\label{fig:ReproducePP}\textbf{Recovering many random posterior distributions}: For the two code configurations
  a plot of inferred  $\Delta \hat{P}\equiv \hat{P}_1(x)-\hat{P}_2(x)$ versus $\hat{P}_1(x)$ for $j$ indexing the synthetic
  event, where $\hat{P}$ corresponds to the  empirical estimate of the posterior CDF deduced from posterior samples
  (here using $10^4$ samples), $1,2$ refer to the two code configurations, and where $x$ refers to chirp mass (blue) or
  mass ratio (orange).    The two codes assign nearly the same quantile to the injected value for all injections. 
}
\end{figure}

\section{Analysis of recent events}
\label{sec:reanalysis}

RIFT has been extensively used to analyze GW observations in O1 \cite{NRPaper}, O2  \cite{LIGO-O2-Catalog}, and O3
\cite{LIGO-O3-GW190412,LIGO-O3-GW190521-implications,LIGO-O3-O3a-catalog,LIGO-O3-O3a_final-catalog,LIGO-O3-O3b-catalog}.
In this section, we briefly reanalyze some recent notable observations with RIFT, to highlight the performance
advantages of the configurations and extensions described in this paper.    Additionally, we also examine  selected events which other groups have
prioritized for their own reanalysis.  
While we adopt largely consistent data conditioning settings and priors as used in previously published work, we do not
attempt to rigorously reproduce any previously published work, for simplicity adopting the algorithms  described
above without any added non-RIFT extensions.  (For example and by contrast, the headline results presented in recent LVK analysis of O3 adopt a
different fiducial distance prior and attempt to
marginalize over calibration uncertainties \cite{LIGO-O3-O3a_final-catalog,LIGO-O3-O3b-catalog}.  In this work we adopt
the customary $d_L^2$ distance prior and do not include marginalization over data processing uncertainties.)
Rather, the illustrative results presented below in part reflect the reasonable differences expected between groups
adopting different analysis choices.

\subsection{GW190814, GW190412}
\begin{figure}
\includegraphics[width=\columnwidth]{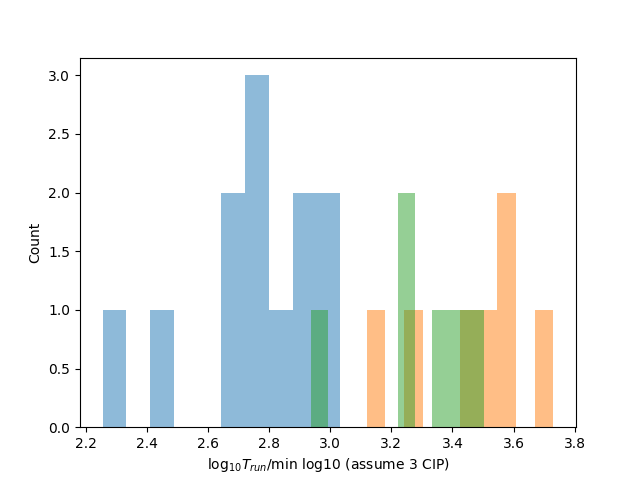}
\caption{\label{fig:ros_profile_exceptional} \textbf{Projected runtime for exceptional events}
A histogram of the runtime $T_{\rm run}$ (in minutes) for RIFT using IMRPhenomPv2 for GW190425 (blue),
GW190814 (orange), and GW190412 (green), \response{run multiple times for each event, with successive runs} using contemporary and near-contemporary code configurations.
}
\end{figure}

The two events GW190814 and GW190412 are asymmetric compact binary black hole mergers, whose posteriors exhibit strong correlations
between $\mc,\eta$, and component compact object spins.  As a result, the revised RIFT configuration enables
significantly more efficient performance for these events.
For context, %
during O3 both events required months of wallclock time and extensive human oversight, at least two orders of magnitude more effort than other contemporary O3 RIFT analyses.  
Now, both events can be analyzed automatically
with substantially reduced computational and wallclock time, with minimal
human oversight.    While detailed timing depends strongly on the waveform model and optimization settings used, we can
consistently produce results for both within days (for slow  models with complex physics) to even tens of minutes  (for simple physics and fast
models).
As our replication study adds no new scientific insight about these three events, we \response{do not} illustrate them here. 
Instead, Figure \ref{fig:ros_profile_exceptional} shows the estimated run duration  assuming  no resource congestion: the
cumulative CIP evaluation time, divided by the number of CIP workers used simultaneously (here, 3).
While this histogram shows only runtimes for  IMRPhenomPv2, RIFT's computational cost should be comparable for more
costly waveforms; see Appendix \ref{ap:prof} for further discussion.

\subsection{GW200115}
\begin{figure*}
\includegraphics[width=\columnwidth]{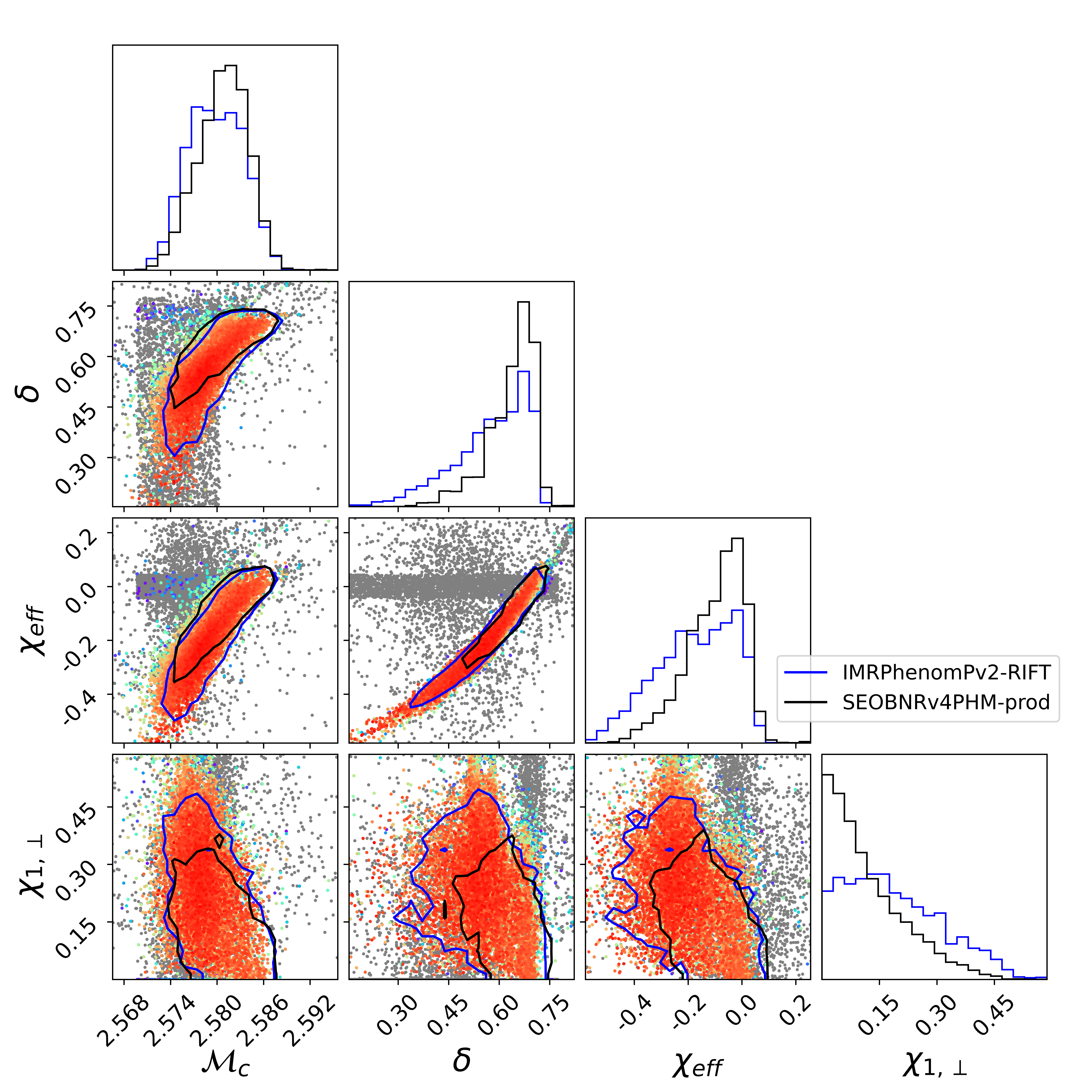}
\includegraphics[width=\columnwidth]{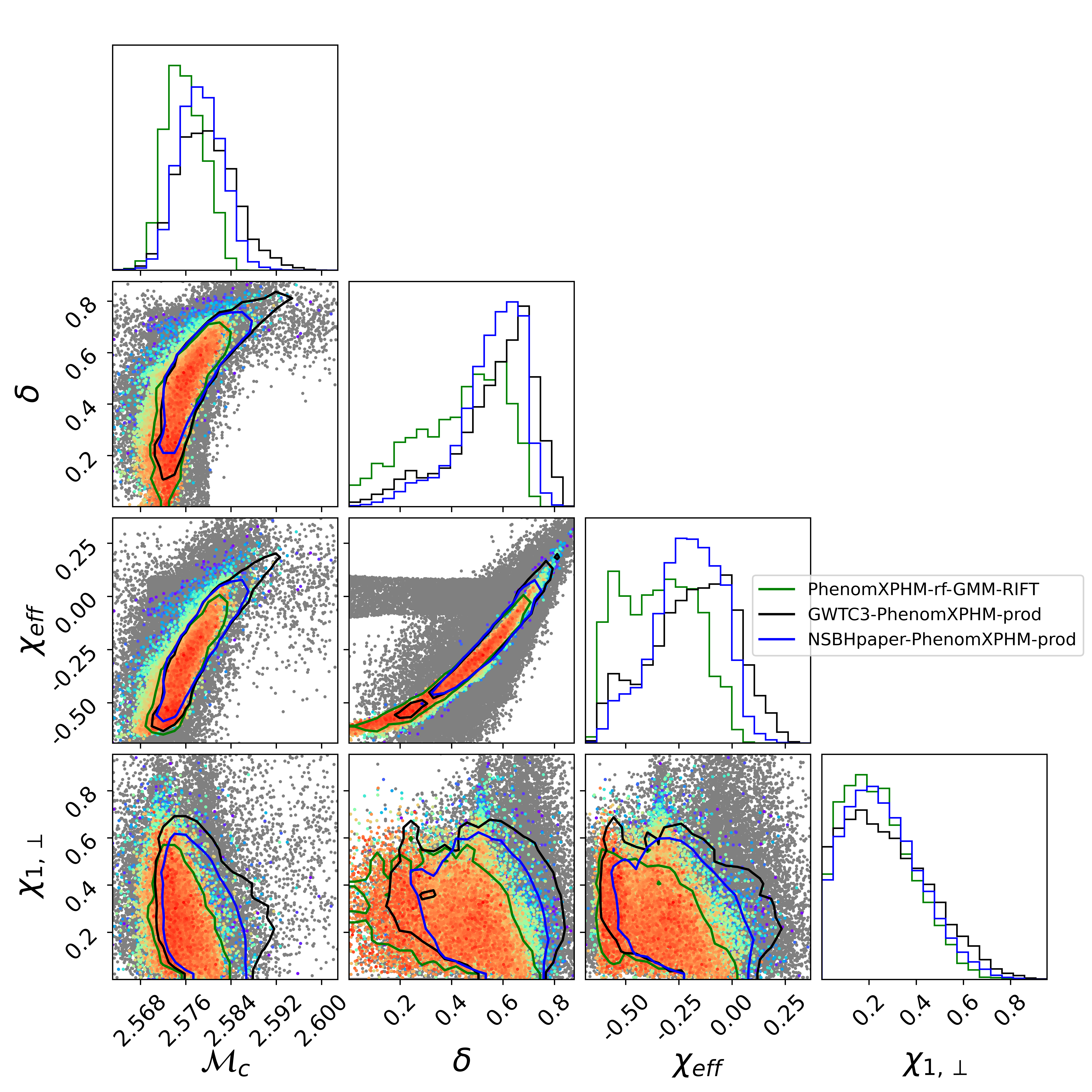}
\caption{\label{fig:gw200115}\textbf{Analysis of GW200115}:
Corner plot with likelihoods (points and color scale) for GW200115, analyzed with RIFT and IMRPhenomPv2 (left panel) and
IMRPhenomXPHM (right panel), \response{showing the parameters $\mc$, $\delta=(m_1-m_2)/M$, $\chi_{\rm eff}$ and the
  magnitude of the transverse component of the primary spin $\chi_{1,\perp} = \sqrt{\chi_{1,x}^2+\chi_{1,y}^2}$}.   Color scale
shows RIFT marginal likelihood evaluations versus intrinsic parameters; solid contours show 90\% credible intervals for
the two-dimensional marginal distributions; and the diagonal panels show one-dimensional marginal distributions of these
parameters.   In the left panel, for comparison the black solid curves and distributions show similar results derived with the published
SEOBNRv4PHM analysis.   In the right panel, for comparison the green and blue curves show previously-reported results
derived with IMRPhenomXPHM with different analysis settings and inference.
}
\end{figure*}

Based on the inferred mass of its secondary, the low mass asymmetric merger GW200115 is expected to be a neutron
star-black hole merger.   As with GW190814 and GW190412, the revised RIFT extensions and configurations presented in
this work enable dramatically more efficnet analysis, without human intervention.
The left panel in Figure \ref{fig:gw200115} shows an analysis with IMRPhenomPv2 of GW200115 
plotted against the SEOBNRv4PHM production run (black solid). The new run uses a faster set of interpolators (rf) and samplers (GMM) and 
a better and new coordinate system, which reduces the runtime to a matter of a couple of days compared to weeks. 

The right panel in Figure \ref{fig:gw200115} shows a similar reanalysis of GW200115 with IMRPhenomXPHM.  For comparison, the solid black and blue contours and distributions illustrate
previously-reported results, which incorporate calibration marginalization and  an alternative distance prior, and were
performed with a different analysis code.    As expected, the RIFT analysis presented here conforms as expected to the marginal
likelhoods shown in color scale.   This reanalysis
favors a higher secondary mass and a more negatively aligned spin. 
All differences between these calculations are modest, with largely overlapping support.

\subsection{GW151226}

The original published  analyses of GW151226  favored comparable binary masses, with a nominal posterior for
the two ordered variables $m_1>m_2$ as close to equal mass as would be expected given the strong degeneracy along lines
of constant $\mc$.  These results were corrobrated in GWTC-2 with reanalysis including direct comparison to numerical relativity simulations including
higher-order modes, albeit
at the time limited only to nonprecessing simulations \cite{gwtc2-nrsample-release}.  Several groups have published
reanalyses of these events (e.g., \cite{2021arXiv210505960M,gwastro-151226-RoryThraneFollowupHighQ,gwastro-151226-ChiaEtAl}), including a recent LVK reanalysis
\cite{LIGO-O3-O3a_final-catalog}. 

One   followup reinvestigation of this event using models with recent semianalytic waveform
models  have found modest support for higher mass ratio
\cite{gwastro-151226-ChiaEtAl}.  
They suggest  the high-mass-ratio configurations ($1/q >5$) could be
consistent with strong orbital precession.

\begin{figure*}
\includegraphics[width=\columnwidth]{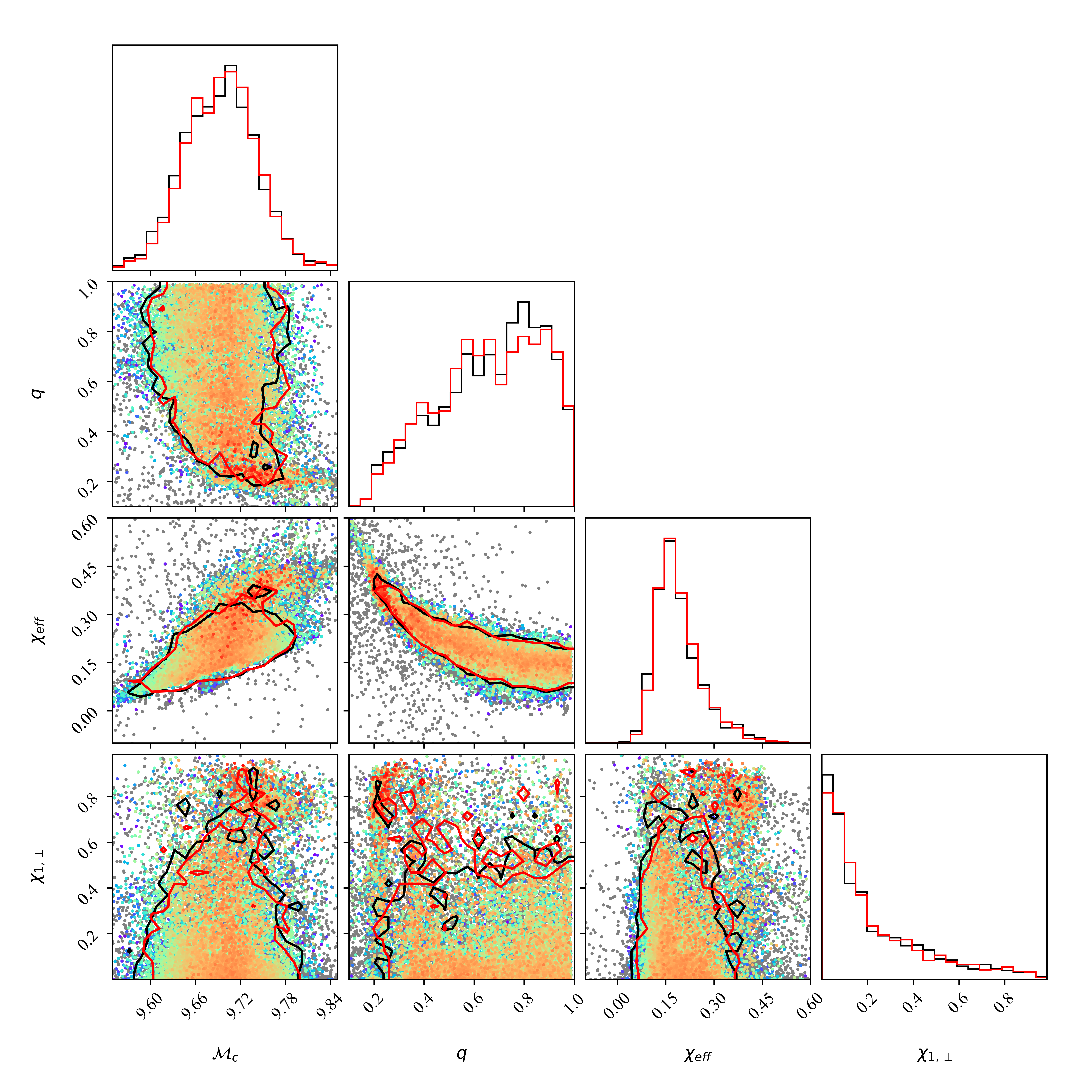}
\includegraphics[width=\columnwidth]{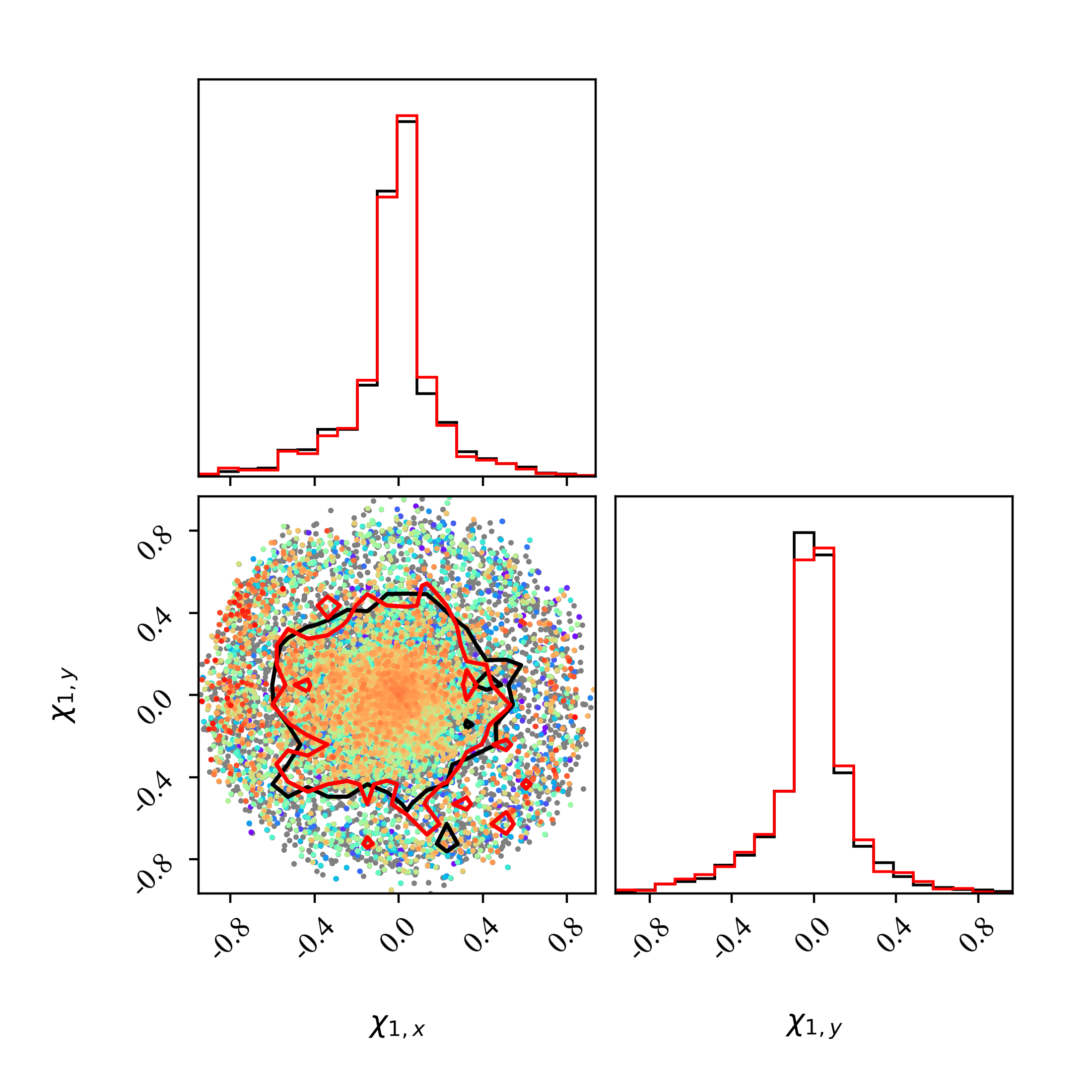}
\caption{\label{fig:gw151226}\textbf{Reanalysis of GW151226 with IMRPhenomXPHM}: 
Two corner plots
  with 90\% credible interval quantiles for the last two RIFT iterations, indicated by different line colors.  
\response{The left panel shows $\mc,q=m_2/m_1,\chi_{\rm eff}$ and 
the magnitude of the transverse dimensionless spin of the primary $\chi_{1,\perp} = \sqrt{\chi_{1,x}^2+\chi_{1,y}^2}$.
The right panel shows the two transverse components of the dimensionless spin of the primary $\chi_{1,x},\chi_{1,y}$.}
The color
  scale shows the likelihood range, over a dynamic range $\Delta \ln {\cal L}\le 15$; gray points indicate likelihood
  evaluations below this range.  
  While a small region of marginally higher likelihood exists at asymmetric mass ratio, large transverse spins, and
  positive $\chi_{\rm eff}$, overall our inferences still favor a conventional interpretation for this event.
}
\end{figure*}

The two panels of Figure \ref{fig:gw151226} shows an unsupervised RIFT reanalysis of this event with IMRPhenomXPHM, using our contemporary architecture: adaptive convergence, RIP and pseudo-cylindrical coordiantes,
et cetera as described in Section \ref{sec:recommend}.  
As previously, contours indicate 90\% credible intervals, while the colorscale indicates $\ln{\cal L}$; points colored
in light gray have $\ln {\cal L}$ farther than 15 away from the peak value.   
All analyses use settings comparable to our original analysis of this event: C02 data, with identical noise power
spectra estimates.    We have both reanalyzed these events with our customary current pipeline and also performed a
targeted analysis to densely evaluate the likelihood for $\eta \in [0.02, 0.08]$.
For our analysis with IMRPhenomXPHM, we adopt a reference frequency of $f_{\rm ref}=100\unit{Hz}$,
comparable to \cite{gwastro-151226-ChiaEtAl}.
We have corroborated our conclusions with a focused investigation of high-mass-ratio region.
In neither analysis do we find the posterior strongly supports a high $q$, strongly precessing interpretation of GW151226.

Looking more closely at the underlying marginal likelihoods, in both cases we find \emph{some} high-$q$, strongly-precessing configurations with significant (but nonexceptional) marginal
likelihood.  In both our unsupervised and targeted analyses, we find that, \emph{conditioned on the requirement of high mass ratio}, the posterior
distribution for transverse spin $\boldsymbol{\chi}_{1,\perp}$ suggests well-localized spin orientation, consistent with
the statements in \cite{gwastro-151226-ChiaEtAl}.  In other words, the high-$q$ region of the posterior modestly favors
transverse spins, and the specific orientation of these spins is
better constrained with
$f_{\rm ref}=100 \unit{Hz}$ than the fiducial 10 Hz.  At this reference frequency, the one-dimensional distribution of
$\phi_i$ and the two-dimensional distributions $\boldsymbol{\chi}_{1,\perp}$ show some modest asymmetries, in this mass
region.
While we concur with \cite{gwastro-151226-ChiaEtAl} that the choice of prior strongly suppresses the significance of the
transverse spins in the \emph{posterior}, the lack of exceptional precessing configurations with with high
\emph{marginal likelihoods} demonstrates that these are at best comparably likely to the well-explored comparable-mass
component of the posterior.

The differences in interpretation between this analysis and those of Chia et al can easily arise from relatively small
details,  For a relatively weak event like
GW151226, small perturbations to a nearly-flat, low-significance likelihood introduced by (for example) different
choices in data conditioning can easily produce large changes in the posterior. 
Closely examining their work, their figures suggest no support  for spin-orbit-induced modulations during the
\emph{insprial} phase:  their Fig. 6 suggests that their early-time maximum likelihood waveforms  are consistent with low transverse
spin.  In other words, their results are consistent with a viewing angle consistent with $\pm \mathbf{J}$, the total
angular momentum direction, along which minimal modulation is expected.   

\skipme{
\subsection{Computing evidence for zero spin or equal mass with RIFT}
\editremark{ONLY INCLUDE THIS IF AND ONLY IF WE CAN GET A FULL SET OF O3 likelihoods released, \response{do not} want to rerun all.
  transparent to do this; give selected examples, and data release, using above as exmaples; highlight no need
to explore multiple spin priors to do so, since RIFT result has all the likelihoods; contrast with \cite{2021PhRvD.104h3036O}}
}

\section{Conclusions}

We have described extensions of the RIFT parameter inference software ecosystem, including the specific choices adopted
during O3 and many new extensions proposed for post-O3 work.   In this work specifically, we introduce new coordinate
systems, fitting techniques,  integration techniques, and pipeline architectures which together significantly accelerate
the performance and scientific return of RIFT as a parameter inference tool. 
To highlight RIFT's capabilities, we briefly report on reanalyses of several pertinent GW observations.

The improvements highlighted here \response{do not} exhaust RIFT's capability, both in general or for ultra-low-latency analysis.
For example, previous studies proposed direct quadrature of both intrinsic and extrinsic variables using surrogate
models \cite{gwastro-PE-AlternativeArchitecturesROM}.  RIFT could likewise benefit from normalizing flows to accelerate
its Monte Carlo integration, in effect accelerating our importance sampling via approximate inference and benefitting
from several group's efforts to provide such preliminary estimates \cite{2021PhRvL.127x1103D,2020PhRvL.124d1102C,2020arXiv201012931D,2022NatPh..18..112G,2021PhLB..81536161K}.
Additionally and more broadly, RIFT can benefit from many additional improvements in its implementation details,
including more optimal coordinates; more use of GPU-accelerated computation; and streamlined pipeline architecture. 
Even with the existing codebase, RIFT offers novel low-latency capability, even with costly models.  For example, RIFT
could use a small number of very large iterations,  after loose targeting with preliminary estimates (e.g., from
nonprecessing inference).  Even the existing framework can complete inference within a handful of minutes for simple
models, with appropriate operating point choices.
We defer discussion of specific low-latency analysis frameworks to a dedicated publication.

\begin{acknowledgements}
The authors thank Ben Farr for many helpful comments on the manuscript.
JW is supported by NSF PHY-2110460. ROS, VD, and AY are supported by NSF PHY-2012057; ROS is also supported via NSF PHY-1912632
and AST-1909534.
DW is supported by NSF PHY-1912649, NSF PHY-2207728.
JR is supported by NSF RUI-2110441 and AST-2219109.
This material is based upon work supported by NSF's LIGO Laboratory which is a major facility fully funded by the
National Science Foundation.
The authors are grateful for computational resources provided by the LIGO Laboratory, supported by National Science
Foundation Grants PHY-0757058 and PHY-0823459, and by the International Gravitational Wave Observatory Network.
\end{acknowledgements}

\appendix
\section{Gaussian mixture model implementation}
Gaussian mixture models (GMMs) and expectation-maximization (EM) have been thoroughly described in the literature \cite{mm-stat-EM,mm-stat-EM-book,mm-stat-EM-theory}.
In this appendix, we 
summarize our implementation, emphasizing features needed for our work.

\noindent \emph{Expectation-maximization (EM) for GMMs}:
The EM algorithm fits $K$ Gaussians, each described by its mean $\bm{\mu}_k$,
its covariance $\bm{\Sigma}_k$, and its mixture weight $\pi_k$, such that the likelihood
of the model $\mathscr{L}$ is maximized. For each sample $\bm{x}_n$ and its corresponding
sample weight $w_n$,
\begin{equation}
    \mathscr{L} = \prod_n P(\bm{x}_n).
\end{equation}
$P(\bm{x})$ is the probability of a point $\bm{x}$ under the current model, or
\begin{equation}
    P(\bm{x}_n) = w_n \sum_{k} N(\bm{x}_n | \bm{\mu}_k, \bm{\Sigma}_k) \pi_k,
\end{equation}
where $N(\bm{x}_n | \bm{\mu}_k, \bm{\Sigma}_k)$ is the multivariate Gaussian density.
$P(\bm{x})$ can be split into the $K$ individual probabilities for each $\bm{x}$:
\begin{equation}
    p_{nk} = {{N(\bm{x}_n | \bm{\mu}_k, \bm{\Sigma}_k) \pi_k w_n} \over P(\bm{x}_n)}
\end{equation}
The above equations describe the expectation step (E-step) of the EM
algorithm. The means, covariances, and mixture weights are estimated from
$p_{nk}$ in the maximization step (M-step) as follows.
\begin{equation}
    \bm{\mu}_k = \sum_{n} p_{nk} \bm{x}_n \Big/ \sum_{n} p_{nk}
\end{equation}
\begin{equation}
    \bm{\Sigma}_k = \sum_{n} p_{nk} (\bm{x}_n - \bm{\mu}_k) (\bm{x}_n - \bm{\mu}_k) \Big/ \sum_{n} p_{nk}
\end{equation}
\begin{equation}
    \pi_k = {1 \over N} \sum_{n} p_{nk}
\end{equation}

\begin{equation}
    p_{nk} = {{N(\bm{x}_n | \bm{\mu}_k, \bm{\Sigma}_k) \pi_k w_n} \over P(\bm{x}_n)}
\end{equation}

The iterative EM algorithm is initilized by guessing initial values for the means,
covariances, and mixture weights. Alternating E-steps and M-steps are done until the
change in $\mathscr{L}$ between iterations is below a predetermined threshold.

The standard expectation-maximization algorithm has been modified to allow
iterative, online updates of the model with a new set of samples and weights. 
A new GMM is first trained using the new data.
Each component $i$ in the new model is then matched to a component $j$ in the old
model in such a way as to minimize the total Mahalanobis distance $M_D$ between
all of the means:
\begin{align}
    \nonumber \sum M_D = \sum_{i, j=0} (\sqrt{(\bm{\mu}_j - \bm{\mu}_i)^T \bm{\Sigma}_i^{-1}(\bm{\mu}_j - \bm{\mu}_i)} \\
    + \sqrt{(\bm{\mu}_i - \bm{\mu}_j)^T \bm{\Sigma}_j^{-1}(\bm{\mu}_i - \bm{\mu}_j)})
\end{align}
The combination of components which minimizes $\sum M_D$ is the one that is kept. \\

Once the components have been matched, they are combined. $N$ is the total
number of points the old model has been trained
on and $M$ is the number of samples in the new batch.
\begin{equation}
    \bm{\mu} = {{N \pi_j \bm{\mu}_j + M \pi_i \bm{\mu}_i} \over {N \pi_j + M \pi_i}}
\end{equation}
\begin{equation}
    \bm{\Sigma} = {{N \pi_j \bm{\Sigma}_j + M \pi_i \bm{\Sigma}_i} \over {N \pi_j + M \pi_i}}
    + {{N \pi_j \bm{\mu}_j \bm{\mu}_j^T + M \pi_i \bm{\mu}_i \bm{\mu}_i^T} \over {N \pi_j + M \pi_i}}
    - \bm{\mu \mu}^T
\end{equation}
\begin{equation}
    \pi = {{N \pi_j + M \pi_k} \over {N + M}}
\end{equation}

\noindent \emph{Accounting for finite domains}:
Monte Carlo sampling for parameter estimation requires samples from a finite, rectangular domain.
When sampling from a GMM, therefore, we must truncate the infinite-domain multivariate Gaussians to our finite domain.
There is no widely-used implementation of a truncated multivariate Gaussian, but we can take advantage of \texttt{Scipy}'s univariate \texttt{truncnorm} function.

In general, to generate a sample from a multivariate Gaussian from a distribution with covariance $\bm{\Sigma}$ and mean $\bm{\mu}$, we first generate a sample $\bm{x}$ from a Gaussian centered at the origin with covariance 1.
Our final sample is then
\begin{equation}
    \bm{x}' = \bm{\lambda}^{1/2} \bm{\phi} \bm{x} + \bm{\mu},
\end{equation}
where $\bm{\lambda}$ is a diagonal matrix of the eigenvalues of $\bm{\Sigma}$ and $\bm{\phi}$ is a matrix containing the corresponding eigenvectors of $\bm{\Sigma}$.

To generate truncated Gaussian samples, we then simply take our initial sample (with mean 0 and covariance 1) using \texttt{truncnorm}, and transform them to have the desired mean and covariance.
The problem with this approach, however, is that the bounds are transformed along with the samples, resulting in a parallelogram-shaped domain.
Our solution to this problem is to sample from the smallest rectangualar region that, when transformed, will contain the desired sampling domain; any samples that end up outside of this region are simply thrown out.

Each corner of our desired domain is transformed by $\bm{r}$, where
\begin{equation}
    \bm{r} = [\bm{\lambda}^{1/2} \bm{\phi}]^{-1}.
\end{equation}
For each dimension, the minimum and maximum transformed corner points are used as the bounds for our univariate \texttt{truncnorm} samples.
After transforming the samples, we throw out any that fall outside our original bounds.

\section{Monte Carlo integration and  independent samples}
\label{ap:mc}
RIFT  relies heavily on Monte Carlo integration.  This appendix provides a brief review, highlighting pertinent subtle issues
about results and convergence that arise in real applications.
In this section, we will consider an integral $S=\int L p(x) dx$ over some volume in  $x$ relative to a normalized
probability $p(x)$, and its alternative expression $S=\int L [p(x)/p_x(x)]p_x(x) dx$ relative to another probability
density $p_x(x)$ over $x$.   We will define the random variable $w= L p/p_x$, such that $S = \E{w}$ (averaging over the
distribution from $p_s$).   
The Monte Carlo approach to this integral involves drawing many samples $x_k$, evaluating $w_k$, and evaluating the
sample mean $\bar{w} = \sum_{k=1}^N w_k/N$.  According to the weak (and strong) laws of large numbers, the sample mean
will converge to $S$, so long as $\E{w}$ is finite (even if higher-order moments do not exist).   %
In the special case that $w$ has finite and known variance $V(w)\equiv \sigma_w^2$, the distribution of the sample mean will be asymptotically
normal, with a mean of $\E{w}$ and a variance of $\sigma_w^2/N$.   If the variance exists and can be well-approximated
by the sample variance, then the samples themselves provide an estimate for the integral and its error \cite{book-mm-NumericalRecipies}.

RIFT both uses the Monte Carlo integral (ILE) and, when appropriate (CIP),  the associated weighted samples themselves.
To introduce notation, the  weighted Monte Carlo integration methods output points $x_k$, likelihoods $L_k\ge 0$, and weights
$w_k=L(x_k) p(x_k)/p_s(x_k)$, where $x_k$ are fair draws from the sampling prior $p_s(x)$.  Using these outputs, we
evaluate overall Monte Carlo integrals; estimate marginal distributions; and resample to produce fair-draw outputs.
As each outcome involves a different expression of these samples, several measures of convergence and hence the ``number
of independent samples'' have been adopted, expressed in terms of normalized sample probabilities $p_k = w_k/\sum_q w_q$.

Historically, RIFT adopts a very conservative account of the number of independent samples \cite{gwastro-PE-AlternativeArchitectures}:
\begin{align}
\label{eq:neff}
n_{\rm eff} = \frac{\sum p_k}{\max \{p_k \} } %
\end{align}
The value $1/n_{\rm eff}$ is the largest discontinuous jump in the estimator $\hat{P}(<x) = \sum_k p_k \theta(x_k-x)$
for any one-dimensional cumulative probability distribution $P(<x)$ derived from the full samples.  
Alternatively, the sample size can be defined using the estimated moments of the weight distribution.  For example, one
estimate of the  effective sample size is \cite{kish} 
\begin{align}
\label{eq:neff_ESS}
n_{\rm eff, ESS} = \frac{ (\sum_k p_k)^2}{\sum_k p_k^2}
\end{align}
An alternative choice grounded in the Monte Carlo integral error standardizes the sample size to the sample variance.
As the natural count of independent samples scales as $N/V(w)$,
 an alternative estimate for the number of independent samples based on the Monte Carlo variance estimate \cite{2019RNAAS...3...66F}:
\begin{align}
\frac{\left\langle w\right\rangle^2}{n_{\rm var}} \equiv  \frac{1}{N}  \left [ - \left\langle w\right\rangle^2 +  \frac{1}{N-1} \sum_k w_k^2    \right]
\end{align}
In other words, $n_{\rm eff}$ is the ratio of the sample mean (squared) and the sample variance, times the number of
points drawn: $n_{\rm eff} = N \bar{w}^2/s^2$ where $s^2$ denotes the sample variance.
This accounting of the number of independent points can be dramatically larger than the conservative estimate
of Eq. (\ref{eq:neff}), depending on the integrand, or nearly zero for scenarios where the variance diverges, as discussed below.
Finally, the entropy of the probability weights $H(p) = \sum_k p_k \ln (1/p_k)$  is maximized at $\ln n$ when all the
probability weights are equal.  Motivated by the maximum value of entropy, we define 
\begin{align}
n_{\rm eff,H} \equiv \exp( H(p))
\end{align}
This information-theory-based estimate of the number of independent evaluations can be slightly less conservative than
$n_{\rm eff}$.
As a practical illustration of these sample size conventions, we introduce a one-dimensional toy model: $L=1$, $p(x)=1$ and
$p_x(x)=\alpha x^{\alpha-1}$ for $x\in[0,1]$, with $\alpha>0$.  For example, these pair of priors might represent an attempt to rescale a
single spin's volumetric sampling density (i.e., the case $\alpha=3$ and $x=|\chi_1|$) to reflect a physical uniform
spin magnitude prior.   In this scenario, $w=p/p_s=1/\alpha x^{\alpha-1}$, which is defined over  $[1/\alpha,\infty)$
  for $\alpha>1$ and over $[0,1/\alpha)$ for $\alpha<1$.    All comments below are easily
verified by simple numerical experiments.

\begin{itemize}
\item \emph{Monte Carlo integral}:  As required by the strong and weak law of large numbers, $\sum_k w_k/N$ is nearly
  unity almost always for large $N$.  At fixed sample size but changing $\alpha$,   the standard naive  Monte Carlo uncertainty estimate $s/\sqrt{N}$
  increases, reflecting the rarity of  sample points sufficiently close to the small
  region near $x\simeq 0$ which dominates the integral.

  Nominally we would expect the Monte Carlo uncertainty to scale as $V(w)/N$.  For $\alpha \in (0,2)$ the variance is
  integrable, but for $\alpha>2$ the lower limit diverges:
 \[
\E{w^2} = \int_0^1 dx/p_s(x) = \frac{1}{\alpha(2-\alpha)} x^{2-\alpha} \bigg\vert_{0}^2
\]
Therefore, the analytic expression $V(w)=(1-\alpha)^2/\alpha(2-\alpha)$ for the variance is only well-defined for
$\alpha<2$ -- in particular, excluding the highly-desirable scenario of reweighting from a volumetric to a uniform spin magnitude prior!  Nonetheless, above this threshold and in the regime of a formally divergent variance, the conventional
estimate for Monte Carlo error based on the \emph{sample} variance is a reasonable estimate of the true error scale for
many $\alpha<10$.
These divergences in the moments of $w$ do \emph{not} limit the efficacy of Monte Carlo integration, whose convergence
is asymptotically protected by the law of large numbers.

\item \emph{Moment-based size}:  We proposed two sample size estimates based on (sample) means and variances of the
  distribution.   As noted above, for $\alpha>2$, the true second moment and variances diverge.  Nonetheless, in
  empirical experiments with our toy problem with $\alpha>2$, both $n_{\rm eff}$ and $n_{\rm var}$ exhibit qualitative
  consistent behavior relative to the other two sample size estimates discussed below.  

\item \emph{Default (max-sample) size}: The single-most-significant sample provides a conservative estimate for the
  effective sample size which manifestly must remain finite and comprehensible.  For non-normalized draws $w_k$ such that $\E{w}=1$, our default estimate for $N_{\rm eff}$ is
  approximately $N/\text{max}_k w_k $.  For the scenario with $\alpha>1$, roughly speaking since the nearest sample has probability 
  $1/N$, this means $n_{\rm eff} \simeq N \alpha x_{\rm min}^{\alpha-1} \simeq \alpha N x^\alpha/x \simeq \alpha/x$, and
  thus  $n_{\rm eff} = N/w_{\rm max} = \alpha N^{1/\alpha}$  -- in other words, a
  few times the natural number $N^{1/\alpha}$ of points expected nearby.
  
More formally for $\alpha>1$, the cumualtive distribution of the largest value $w_{\rm max}$:
  out  of $N$ samples of $w$ is
  $P(<w_{\rm max})^N$.  Because of reordering between $x$ and $w$ for $\alpha>1$, the
  cumulative distribution function for $w$ is easily expressed in terms of the cumulative distribution of $u=x^\alpha$ : $P(<w) =
  P(>u(w)) = 1- u = 1-(1/\alpha w)^{(\alpha/(1-\alpha))}$.  In the limit of large $N$, the median value of $w_{\rm max}$ can be
  estimated by solving $1/2 = P(<w_{\rm max})^N$, leading to a simple approximate expression for the median value of
  $n_{\rm eff}$ for this one-dimensional rescaling: 
\begin{align}
\text{median}_x(n_{\rm eff}) \simeq  N \alpha \left(\frac{N}{\ln 2}\right)^{-(\alpha-1)/\alpha}  \simeq O(N^{1/\alpha})
\end{align}

\item \emph{Entropy-based size}: Finally, because our entropy size estimate involves a Monte Carlo estimate of the
  distribution entropy, we anticipate that the finiteness of the entropy ensures the entropy-based sample size is stable
  and well-posed for all $\alpha>0$. 
\end{itemize}

As illustrated by the discussion above, these different sample size estimates can have dramatically different behavior,
including different scaling with $N$!   For our toy problem, empirically $n_{\rm eff,H}$ is a nearly constant fraction
of $N$; $n_{\rm eff}$ scales as $N^{1/\alpha}$, with a very unfavorable prefactor; and $n_{\rm eff,var}\simeq n_{\rm
  eff, ESS}$ scale roughly in between (e.g., comparable to $N^{2/\alpha}$).
In particular, these expressions suggest that both our probability distributions (with error scale set by  $1/n_{\rm
  eff}$ in their cumulative) and evidence (with error scale set by $1/\sqrt{n_{\rm var}}$) have uncertainties scaling as
$N^{-1/\alpha}$ for this toy problem.  This expected but extremely unfavorable scaling has straightforward implications
for our reweighting strategies: in short, avoid spin reweighting whenever possible, unless drawing samples from a
distribution with similar singular behavior near the origin.

\section{How resources determine possible operating points}
\label{ap:prof}

As synthetic and real sources accumulate, RIFT users typically need to perform extremely large numbers of source
inferences.  The rate, latency, and accuracy of these inferences depend on the available resources, waveform model cost,
and population of signals being analyzed.  Different science objectives and available resources can produce dramatically
different choices for how to operate the RIFT pipeline.  In this section., we briefly outline how these choices impact
RIFT analysis throughput and overall cost, highlighting a few expected use cases.

Generally, RIFT involves two sets of calculations, potentially provided by distinct pools of resources: ILE evaluations,
provided by $N_I$ resources (e.g., low-cost GPUs); and CIP posterior generation, provided by $N_C$ resources (e.g., usually
modestly memory-rich CPUs).  Both ILE and CIP are characterized by a typical runtime. 
Each ILE marginal likelihood evaluation for a specific model (i.e., approximation, mode
list, starting frequency) and at a fiducial accuracy $\epsilon$ (i.e., the relative error in their marginal likelihood)
requires a time $\tau_I$, ranging from a second to a few minutes. 
Each CIP posterior generation worker requires a time $\tau_C$ that depends strongly on the approach used, dimensionality
of the space, and model complexity.  In this appendix we consider CIP configurations with runtimes from seconds to
hours.  
Uusally, both ILE and CIP involve Monte Carlo integration, so their runtime nominally increases as $1/\epsilon^2$.  
RIFT  employs many instances of ILE and CIP simultaneously.

The typical wait time and  total pipeline resource usage follows by accounting for the total cost needed for all stages of the analysis.
We assume a full analysis requires  $n_I\simeq O(10^3)$ likelihood
evaluations, organized roughly into $n_{it}$ chunks of size $n_I/n_{it}$.     After each chunk, a number $n_{c}$ CIP workers will each
independently generate a fraction of the overall posterior; the total time needed to complete posterior generation can
be appreciably reduced by employing many CIP workers simultaneously.
When many RIFT analyses are performed simultaneously, the overall resource usage
per event can be estimated ignoring the pipeline's serialization of ILE and CIP stages.
In this circumstance,  the resource usage $T_R$  and average analysis time $T_W$
\begin{subequations}
\label{eq:resources}
\begin{align}
T_R &=  \tau_I n_I + \tau_Cn_C n_{it} \\
T_R/10^4 \unit{s} & \simeq  (\tau_i/\unit{s})(n_I/10^4)  + 10 (\tau_c/\unit{h})(n_C/3)(n_{\rm it}/10) \\
T_W &=  \frac{\tau_I n_I}{N_I} + \frac{\tau_Cn_C n_{it}}{N_C}
\end{align}
where the first expression provides the total time needed to perform an analysis, while the second estimates the
effective analysis duration given the resources available. The total resource usage needed to perform ${\cal N}$ analyses is just ${\cal N} T_R$.   
The two types of resources contribute equally to the average analysis wait time when
\begin{align}
N_{C,\rm match} & \equiv N_I \frac{\tau_C n_C n_{it}}{n_i \tau_I} \\
 &\simeq 10  N_I  \frac{ (\tau_c/\unit{h})(n_C/3)(n_{\rm it}/10)}{(\tau_i/\unit{h})(n_I/10^4)} \nonumber
\end{align}
More concretely, a RIFT analysis pool could need roughly $N_{c,\rm match}/N_I\simeq $ ten times as many CPUs as GPUs, to maintain
a steady state, given these fiducial timescales $\tau_I,\tau_C$.
Finally, ignoring resource contention limits and recognizing that each ILE job  in fact evaluates
$w\simeq O(10)$ likelihood evaluations in series, the user time needed to complete a single targeted analysis with a larger
number $n_C'$ of CIP instances could be as short as
\begin{align}
\label{eq:TU}
T_U &=n_{it} \left[\tau_I w + \frac{\tau_C }{ n_{C}'/n_C} \right]
\end{align}
\end{subequations}
where we assume each worker performs a fraction $\tau_C n_C/n_C'$ of the overall work of generating the posterior.  
The runtime $\tau_I$ depends strongly on the maximum mode order $\ell_{\rm max}$, if waveform generation costs are
subdominant to the costs of evaluating the likelihood many times.  Because the RIFT likelihood depends on matrix
multiplications over arrays of modes, in this regime the ILE runtime will scale roughly as
the square of the number of waveform spherical harmonic modes $h_{lm}$ used in the analysis:
\begin{align}
\tau_I 
&\simeq \tau_{I,\rm ref} \frac{1}{2}\sum_{\ell=2}^{\ell_{\rm max}} (2\ell +1)  
=  \tau_{I,\rm rm ref}\frac{2}{3}(\ell_{\rm max}+1)^3 - \frac{L+31}{6} \\
& \simeq \tau_{I,\rm ref} 80 (\ell_{\rm max}/4)^3
\end{align}
For contemporary hardware and GPU-accelerated  integration in ILE, we observe $\tau_I$ between 30-90 seconds for $\ell_{\rm max}=4$ and
$\tau_{I,\rm ref}$ less than one second for a simple nonprecessing model.

In practice, users will not achieve even these modest benchmarks on $T_W,T_U$ due to resource contention, queuing time,
and cluster mismaps.  For example, with typically $5000$ likelihood evaluations used in the first iteration and $w\simeq
20$ likelihood evaluations per worker, only a small fraction of the $5000/w \simeq 250$ ILE jobs needed can be queued simultaneously, as
usually $N_I \ll 250$.  As these first short likelihood evaluations finish, the time needed to queue new jobs to replace
them often substantially exceeds their duration unless $\tau_I$ is exceptionally long or $w$ large, both factors
contributing to overall run latency.

\subsection{Default operating choices}
Our default operating point choices reflect the fiducial scalings in Eq. (\ref{eq:resources}), appropriate to precessing
black hole binaries analyzed with a fast waveform approximation.   Conflating the impact
of our hardware and queue priority environments,  we effectively have access to relatively many low-cost GPUs  (e.g.,
tens of GPUs per user), but have less frequent access to the high-memory nodes we usually use for CIP (e.g., tens of
non-GPU cores per user).  Otherwise, our typical analyses' inputs are compraable to the fiducial scalings above:   $N_I\simeq 2\times 10^4$ marginal likelihood evaluations
to achieve our target accuracy, with $n_{\rm it}$ between $5$ and $10$.  
As a result, our analyses' wait times are invariably CIP-constrained, as $N_C \ll N_{C,\rm match}$; total resource usage
is likewise CIP dominated, with
 $T_R\simeq N_C \tau_c n_{\rm it}$ between a few tens to $O(100)$ hours per run; and effective wait times $T_W\simeq \tau_C n_C n_{it}/N_C$
of order a few to several hours, or even tens of hours for larger $\tau_C$.   The user wait time $T_U$ for any specific analysis will be smaller in direct proportion to
the number of workers employed.  
An individual with access to these resources can maintain roughly $N_C/n_{it}n_C$ analyses simultaneously in a steady state; for our
fiducial single user, this number is of order unity.
While we scaled the discussion above to individual users, a large organization with more resources (e.g., $N_I\simeq
200,N_C >  2000$) and control over queue priority can achieve correspondingly higher throughput simply by allocating
more resources and priority to RIFT operations.   Such high resources should be sufficient \emph{in principle} to complete even
costly analyses with larger values of $\tau_I,\tau_C$ in roughly tens of minutes on average \cite{gwastro-PENR-RIFT-GPU}.  

In these circumstances, operating point choices which maximize $N_C, \varepsilon$ and minimize $\tau_C n_{it} n_C$ have immediate
return on overall cost and latency.   For example,  the number of matching CIP-capable resources  $N_C$ can be enhanced with lower memory requirements or alternative
computing pools (e.g., the open science grid).   The number of iterations and $\tau_C$ can be reduced by well-adapted
coordinates and prior settings.   Three extreme examples of low $\tau_C$ involve nonprecessing binaries (for which CIP can
often complete within minutes); Gaussian-based posterior generation (for which CIP can complete within about one
minute); and AMR-based grid placement (for which $\tau_C$ completes in seconds).   The user efficiency $\varepsilon$ can be increased with careful planning and extensive
automation.   In a resource-saturated environment,  increasing the number $n_C$
of workers per job does not change throughput, just the latency $T_U$ of each analysis.

Finally, we emphasize that user mishaps, poor planning,  and cluster mischance usually dominate unused time. A  typical single user will
usually  complete only a small fraction $\varepsilon$ of intended analyses in their final form, with the overwhelming
majority associated with exploratory work, preliminary analysis, and validation.

\subsection{High-resource, low-latency configuration}
If a highly-resourced organizations targets large-scale automated low-latency analysis with the conventional RIFT
pipeline, the achievable latency would nominally  eventually be limited by the first term in Eq. (\ref{eq:TU}):  $n_{it} \tau_I w$,
associated with the runtime needed to serially perform $n_{it}$ instances of ILE in series, each evaluating the
likelihood $w$ times.  
In practice, however, several sources of pipeline overhead will contribute to added lag, such as the startup time for CIP and ILE.

\subsection{Extremely low-latency configurations}
The lowest possible latencies $T_U$  can be achieved using a simple waveform model (i.e., low $\tau_I$) with limited
waveform physics (i.e., small $\tau_C$), small numbers of evaluations $w$ per ILE worker, and few iterations $n_{\rm
  it}$ needed to
achieve the target accuracy goal.    As an example, the AMR-based strategy \cite{2022arXiv220105263R}  referenced above
is designed to have $w\simeq 1$ and likelihood evaluation times  $\tau_I$ of order  tens of seconds (i.e., larger than
the steady-state limit due to startup and file access overhead).  Using high-priority queuing and with AMR grid
placement requiring  $\tau_C$ of order seconds, conceivably an AMR approach should perform followup within a minute or less.

\subsection{GPU-limited configuration}
A configuration with large $\tau_I$ or relatively small $N_I$ can produce an unusual ILE-limited configuration.  These
circumstances can arise for analyses with many higher order modes, as $\tau_I\propto \ell_{\rm max}^3$, or with few
available high-speed GPU resources needed to achieve accelerated integration.    These circumstances also require that
$N_I/ \tau_I$ is larger for a GPU configuration (small $N_I$ but also small $\tau_I$) is still large compared to the
corresponding product for a CPU configuration (larger $N_I$ but much larger $\tau_I$).  In these circumstances, the
typical analysis time will be dominated by likelihood evaluations ( $T_W\simeq \tau_I n_I/N_I$). 
As a concrete example, a user performing analyses of higher-order-mode models with a small GPU pool ($N_I\simeq 10$)
could have $\tau_I\simeq 1 \unit{min}$, implying a typical analysis wait time of $T_W\simeq 50\unit{h}$, ignoring the
smaller contribution from CIP to the overall analysis time.

\subsection{Extremely high-cost waveforms}
RIFT has in the past operated successfully with waveforms requiring hours to generate. 
Even for relatively modern waveform generators, the generation of waveforms for very low-mass binaries could be costly and
produce large data products, owing to the signal's duration and the potential need to adopt a high sampling rate to resolve
high-frequency higher-order modes.  
When the waveform generation cost dominates all other considerations, RIFT should employ the largest possible pool of
resources for $N_I$: both GPU and non-GPU resources should be included.    Similarly, each worker should evidently only
analyze one observation ($w\simeq 1$) to reduce latency $T_U \simeq n_{it} \tau_I$.   
In this configuration, the cost per analysis can substantially increase: $T_R\simeq n_{\rm it} \tau_I \simeq 2\times
10^3\unit{h}\; (\tau/\unit{4 \rm min})(n_{it}/10)$.
We emphasize that a high waveform cost does not preclude low-latency analysis, if $\tau_I$ is sufficiently small compared to the target
latency.  

\section{Numerical and adaptation approaches needed for strong signals}
\label{ap:strong_sources}
The main text describes our customary recommendations for RIFT, appropriate to the vast majority of sources with
signal-to-noise $\rho $ below $\simeq 30$.   In this section, we address additional numerical, operating point, and
algorithmic choices more appropriate to signals with high or very high amplitudes.

\subsection{Estimates of signal strength}
The intrinsic source signal-to-noise ratio has a well-understood  impact on the complexity and scale of the likelihood
${\cal L}_{\rm full}$ and posterior.    In this subsection, we will use $\rho$ to denote the true signal amplitude,
defined such that \emph{in the absence of noise} $\rho^2/2  = \text{max}_{\lambda,\theta} {\cal L}_{\rm full}$;
$\rho_{\rm hint}$ will be an estimated signal amplitude, provided by the search pipelines which discovered the event
candidate; and $\rho_{\rm guess}(\lambda)$ is a guess described below designed to estimate
$\text{max}_\theta \sqrt{2{\cal L}_{\rm full}(\lambda,\theta)}$ for a specific set of source parameters $\lambda$.  

\begin{widetext}
Our estimate $\rho_{\rm guess}(\lambda)$ is expressed in terms of the factors entering into the full likelihood used
within ILE
\cite{gwastro-PE-AlternativeArchitectures}:  
\begin{align}
 \label{eq:loglikelihood}
\ln & {\cal L}_{\rm full}(\lambda, \theta) 
 =  -\frac{1}{2}\sum_k \qmstateproduct{h_k(\lambda,\theta)-d_k}{h_k(\lambda,\theta)-d_k}_k  %
 - \qmstateproduct{d_k}{d_k}_k \\
&=\sum_k \sum_{lm}(F_k \Y{-2}_{lm})^* Q_{k,lm}(\lambda,t_k)  \label{eq:def:lnL:Decomposed}\\
&   -\frac{(D_{\rm ref}/D)^2}{4}\sum_k
\left[
{
|F_k|^2 [\Y{-2}_{lm}]^*\Y{-2}_{l'm'} U_{k,lm,lm'}(\lambda)
}
 {
+  \text{Re} \left( F_k^2 \Y{-2}_{lm} \Y{-2}_{l'm'} V_{k,lm,l'm'} \right)
}
\right]
\end{align}
\end{widetext}
where the pertinent factors are expressed in terms of inner products of the signal modes $h_{lm}$ with each other or
with the data:
\begin{subequations}
\label{eq:QUV}
\begin{align}
Q_{k,\ell m}(\lambda,t_k) &\equiv \qmstateproduct{h_{\ell m}(\lambda,t_k)}{d}_k \nonumber\\
&= 2 \int_{|f|>f_{\rm low}} \frac{df}{S_{n,k}(|f|)} e^{2\pi i f t_k} \tilde{h}_{\ell m}^*(\lambda;f) \tilde{d}(f)\ , \\
{ U_{k,\ell m,\ell' m'}(\lambda)}& = \qmstateproduct{h_{\ell m}}{h_{\ell'm'}}_k\ , \\
V_{k,\ell m,\ell' m'}(\lambda)& = \qmstateproduct{h_{\ell m}^*}{h_{\ell'm'}}_k  \ .
\end{align}
\end{subequations}
Our order-of-magnitude estimate $\rho_{\rm guess}$ follows by approximating this likelihood expression, omtting $V$;
eliding the impact of extrinsic angular factors $F_+,Y_{lm}$; and ignoring timing-related triangulation effects:
\begin{align}
\ln {\cal L}_{\rm full} \simeq  \frac{D_{\rm ref}}{D} Q(\lambda)   - \frac{(D_{\rm ref}/D)^2}{4} U(\lambda)
\end{align} 
Maximizing this expression over the single remaining extrinsic parameter $D$ produces an order-of-magnitude estimate for
the maximum value:
\begin{align}
\text{max}_\theta \ln {\cal L}(\lambda,\theta) \simeq O(1) \frac{U}{Q^2}
\end{align}
To account for all pertinent interferometers and modes symmetrically, we therefore define $\rho_{\rm guess}$ as follows:
\begin{align}
(\rho_{\rm guess}*2.3)^2 \simeq  \sum_k \sum_{lm} \frac{|U_{k,lm,lm}(\lambda)|}{\text{max}_| |Q_{k,lm}(t)|^2}
\end{align}
In this expression, the factor of $2.3$ has been chosen empirically, to produce estimates which correspond closely to
cases with known $\rho$.

\subsection{Choices for finite-precision floating point arithmetic and overflow}
RIFT performs Monte Carlo integrals such as Eq. (\ref{eq:lnL:MonteCarlo}) over functions of order $e^{\rho^2/2}$.  For
loud signals, these large integrands can easily produce numerical overflow.  For example, since a conventional 64-bit double-precision
floating point number can express numbers between $\simeq e^{\pm 308}$, while a conventional 128-bit quad-precision floating
number can express numbers over roughly twice that dynamic range, a source with amplitude louder than $\rho > \sqrt{2
  \ln (10)\times 308} \simeq  37.5 $ (for single
precision) or 53 (for quad precision) would produce a peak likelihood ${\cal L}_{\rm full}(\lambda,\theta)$ which
overflows the precision of available arithmetic.

RIFT's Monte Carlo integration suite offers a range of solutions, balancing stability against speed.  The two new
integrators (GMM and AC) both can operate in a conventional overflow-protected mode, where all likelihoods are expressed
as logs and all sums appearing in integrals like Eq. (\ref{eq:lnL:MonteCarlo}) are performed via  the ``logsumexp'' function $g(\mathbf{x}) = \ln
\sum_k e^{x_k}$.   This stability comes at increased cost, primarily for the AC integrator which may need to transfer data
between the CPU and GPU to perform this calculation.  
For most soures with modest amplitudes, however, RIFT can safely operate all its integrators with raw floating point
numbers.  To mitigate the impact of overflow, the user can choose to offset the floating point precision window,
multiplying the likelihoods by a factor $e^{-{\cal O}}$.  Customarily, we choose ${\cal O} \simeq \rho_{\rm hint}^2/2 -
O(\text{few})$ or ${\cal O} \simeq \rho_{\rm guess}^2/2 - O(\text{few})$, to ensure evaluations in the support of the
posterior avoid overflow.  This workaround allows us to mildly stretch the window available for analysis with raw floating point operations.
The pertinent limits for CPU-enabled operation of all our Monte Carlo integrators in raw floating-point mode, both in
CIP and ILE, are usually set by quad-precision floating point arithmetic.   However, when using GPU acceleration, the AC
integrator is currently limited by double-precision arithmetic, a constraint which limits GPU-accelerated AC integration
in ILE with raw floating-point numbers to signals of $ \rho \lesssim 35$. 

\subsection{More flexible sampling models}
As described in the text, our default extrinsic integration strategy \emph{does not adapt} in several dimensions,
limiting adaptation usually to sky location.  This brute-force approach ensures ILE and RIFT will correctly cover the
complex, correlated, often multimodal extrinsic posteriors arising ubiquitously for weak sources.  For strong sources,
however, our brute-force approach becomes untenable.  Rather, to have any chance to find the small fraction $\simeq
\rho^{-d_{\rm eff}}$ of the prior extrinsic volume where the posterior has support, where $d_{\rm eff}$ counts the
number of extrinsic dimensions, we must adapt our sampling distribution in all dimensions simultaneously, using
well-chosen coordinates.

For these reasons, four the loudest signals, we recommend GMM sampling, using sky- and phase-rotated coordinates, with
distance marginalization.

\section{Targets for future improvement}
While we've substantially extneded RIFT relative to the O2 and O3 editions, 
RIFT could be easily improved in several ways.

\subsection{Conventional convergence criteria}
While most other inference codes have standardized on a target $n_{\rm eff, var}$, RIFT's hodepodge of convergence tests
and diagnostics can produce uneven-quality posteriors over parameter space.  We should report and use evidence-based
convergence diagnostics for the iterate-to-convergence step, and consistently report $n_{\rm eff,var}$ from CIP at all
stages.
Both of these updates require architectural changes: our pipeline currently only passes samples to our convergence tests, not
evidences (or evidence histories).

RIFT should also adopt a much longer, user-selected iterate-to-convergence cap.  Our experience suggests that 10
iterations will be more than enough; if more iterations are required, the user should reconsider their choices, as
they've probably made an error or adopted options that are poorly suited to their problem.  However, most end-users want a
black-box framework which will iterate to convergence no matter how long it takes.

\subsection{Miscellaneous technical improvements}

\noindent \emph{Better coordinates}: Our interpolations and thus RIFT can be prone to under-exploring regions near the
hard $q\simeq 1$ boundary.  Initial grids and mass ratio coordinates that further emphasize this region should be explored.

\noindent \emph{Better integration (general)}: Two of our adaptive Monte Carlo integrators (AC and default) adopt largely
ad hoc choices for the number of sampling bins (i.e., $100$ bins for each adaptive coordinate).  This arbitrary
dimension-independent  choice places severe limits on our ability to adaptively sample in many dimensions $d_a$.
The GMM integrators adopt ad-hoc choices for the number of components, and those components are initiallzed randomly
without information deduced from previous analysis or function data.
For example, the adaptive CIP integrators are re-initialized and independently adapt for
each iteration and for each worker, not efficiently exploiting the many previous iterations to initialize an adaptive
sampler.  Particularly for the GMM sampler and during the convergence phase, such initialization could help improve convergence.

\noindent \emph{Better integration (ILE)}: Despite heavy use of GPU optimization, our Monte Carlo integration of the
extrinsic likelihood could be substantially improved.  For example,  recent work \cite{2022arXiv220703508R}  strongly suggests
that the posterior (and hence our Monte Carlo integration) can be substantially simplified by suitable coordinates.
They demonstrate that careful use of
reference frequency, polarization coordinate, and emission polar angle can dramatically simplify the phase posterior.
By contrast, we're presently sampling uniformly over these two angles, introducing substantial inefficnecy at high
amplitude.
ILE integration generally only  adapts in a small subset of the available dimensions.

\noindent \emph{Integration target (CIP,convergence)}: 
The integration sample size target $n_{eff}$ for each individual
CIP worker and the overall output isn't self-consistently chosen with the target accuracy threshold used to assess
convergence.   The fiducial threshold of $10^{-2}$ applied to Eq. (\ref{eq:test_metric:KL}), or equivalent thresholds
applied to other metrics like the JS divergence,  should be user-adjustable, using
some clearly understood empirical relationship between this threshold and a target accuracy goal for the final
posterior.   As several other groups have adopted JS divergence to assess convergence, we should adjust our convergence
criteria to use this diagnostic.  
The target accuracy threshold should be adaptively tightened, and the number of raw Monte Carlo samples $N$ increased,
over the course of an analysis,  rather than fix the threshold and maximum number of evaluations for iterations.

\noindent \emph{Overall infrastructure}: ILE and CIP should use a task-based parallelism architecture, offloading
startup costs and management to the scheduler and better-enabling ongoing use of resources.   Too often nodes are
under-used for ILE integration, while too few CIP instances are instantiated given integration needs.

We should generalize our approach to allow for conditional priors, such as a mass-dependent prior on the tidal
deformability $\lambda$ or mass ratio.

\subsection{Caveats and stability considerations}
RIFT's code settings and use cases are carefully tailored to match the capabilities of the fitting and Monte Carlo
integration algorithms used.    Previously in Section \ref{sec:sub:o3_problems}, we described several inefficiencies and limitations of the code elements
used in RIFT's O3-era operation.   In this section, we briefly highlight ways in which the \emph{new} components of RIFT could be misused or misbehave, as
an aid to diagnosing potential analysis problems.

\noindent \emph{GMM integrator stability with correlated sampling}: The GMM integrator was designed to adapt efficiently
to \emph{correlated} dimensions, including multiple correlated components.   However, this flexibility if employed
unchecked can easily wildly overfit, with the EM algorithm producing singular covariance matricies.   For this reason, at present we hardcode the number
of components for both ILE and CIP, depending on the use case.  Additionally, for ILE we only allow pairwise
correlation, using physics-based motivation. 

Since our production configuration uses correlated GMM sampling, we emphasize the ways in which this configuration can
misbehave.  For low-mass and high-mass ratio binaries, the strongly correlated posteriors can produce singular
covariance matricies, requiring the sampler to reset.  If this sampler reset occurs at an inopportune time, just prior
to the end of a run, an individual worker's output is more likely to be ``spoiled''.

\bibliography{paperexport,LIGO-publications,gw-astronomy-mergers,gw-astronomy-mergers-approximations}

\begin{thebibliography}{105}
\expandafter\ifx\csname natexlab\endcsname\relax\def\natexlab#1{#1}\fi
\expandafter\ifx\csname bibnamefont\endcsname\relax
  \def\bibnamefont#1{#1}\fi
\expandafter\ifx\csname bibfnamefont\endcsname\relax
  \def\bibfnamefont#1{#1}\fi
\expandafter\ifx\csname citenamefont\endcsname\relax
  \def\citenamefont#1{#1}\fi
\expandafter\ifx\csname url\endcsname\relax
  \def\url#1{\texttt{#1}}\fi
\expandafter\ifx\csname urlprefix\endcsname\relax\def\urlprefix{URL }\fi
\providecommand{\bibinfo}[2]{#2}
\providecommand{\eprint}[2][]{\url{#2}}

\bibitem[{\citenamefont{{LIGO Scientific Collaboration}
  et~al.}(2015)\citenamefont{{LIGO Scientific Collaboration}, {Aasi}, {Abbott},
  {Abbott}, {Abbott}, {Abernathy}, {Ackley}, {Adams}, {Adams}, {Addesso}
  et~al.}}]{2015CQGra..32g4001L}
\bibinfo{author}{\bibnamefont{{LIGO Scientific Collaboration}}},
  \bibinfo{author}{\bibfnamefont{J.}~\bibnamefont{{Aasi}}},
  \bibinfo{author}{\bibfnamefont{B.~P.} \bibnamefont{{Abbott}}},
  \bibinfo{author}{\bibfnamefont{R.}~\bibnamefont{{Abbott}}},
  \bibinfo{author}{\bibfnamefont{T.}~\bibnamefont{{Abbott}}},
  \bibinfo{author}{\bibfnamefont{M.~R.} \bibnamefont{{Abernathy}}},
  \bibinfo{author}{\bibfnamefont{K.}~\bibnamefont{{Ackley}}},
  \bibinfo{author}{\bibfnamefont{C.}~\bibnamefont{{Adams}}},
  \bibinfo{author}{\bibfnamefont{T.}~\bibnamefont{{Adams}}},
  \bibinfo{author}{\bibfnamefont{P.}~\bibnamefont{{Addesso}}},
  \bibnamefont{et~al.}, \bibinfo{journal}{Classical and Quantum Gravity}
  \textbf{\bibinfo{volume}{32}}, \bibinfo{eid}{074001} (\bibinfo{year}{2015}),
  \eprint{1411.4547}.

\bibitem[{\citenamefont{{Accadia} and {et
  al}}(2012)}]{gw-detectors-Virgo-original-preferred}
\bibinfo{author}{\bibfnamefont{T.}~\bibnamefont{{Accadia}}} \bibnamefont{and}
  \bibinfo{author}{\bibnamefont{{et al}}}, \bibinfo{journal}{Journal of
  Instrumentation} \textbf{\bibinfo{volume}{7}}, \bibinfo{pages}{P03012}
  (\bibinfo{year}{2012}),
  \urlprefix\url{http://iopscience.iop.org/1748-0221/7/03/P03012}.

\bibitem[{\citenamefont{{Acernese} et~al.}(2015)\citenamefont{{Acernese},
  {Agathos}, {Agatsuma}, {Aisa}, {Allemandou}, {Allocca}, {Amarni}, {Astone},
  {Balestri}, {Ballardin} et~al.}}]{2015CQGra..32b4001A}
\bibinfo{author}{\bibfnamefont{F.}~\bibnamefont{{Acernese}}},
  \bibinfo{author}{\bibfnamefont{M.}~\bibnamefont{{Agathos}}},
  \bibinfo{author}{\bibfnamefont{K.}~\bibnamefont{{Agatsuma}}},
  \bibinfo{author}{\bibfnamefont{D.}~\bibnamefont{{Aisa}}},
  \bibinfo{author}{\bibfnamefont{N.}~\bibnamefont{{Allemandou}}},
  \bibinfo{author}{\bibfnamefont{A.}~\bibnamefont{{Allocca}}},
  \bibinfo{author}{\bibfnamefont{J.}~\bibnamefont{{Amarni}}},
  \bibinfo{author}{\bibfnamefont{P.}~\bibnamefont{{Astone}}},
  \bibinfo{author}{\bibfnamefont{G.}~\bibnamefont{{Balestri}}},
  \bibinfo{author}{\bibfnamefont{G.}~\bibnamefont{{Ballardin}}},
  \bibnamefont{et~al.}, \bibinfo{journal}{Classical and Quantum Gravity}
  \textbf{\bibinfo{volume}{32}}, \bibinfo{eid}{024001} (\bibinfo{year}{2015}),
  \eprint{1408.3978}.

\bibitem[{\citenamefont{{Akutsu} et~al.}(2021)\citenamefont{{Akutsu}, {Ando},
  {Arai}, {Arai}, {Araki}, {Araya}, {Aritomi}, {Aso}, {Bae}, {Bae}
  et~al.}}]{2021PTEP.2021eA101A}
\bibinfo{author}{\bibfnamefont{T.}~\bibnamefont{{Akutsu}}},
  \bibinfo{author}{\bibfnamefont{M.}~\bibnamefont{{Ando}}},
  \bibinfo{author}{\bibfnamefont{K.}~\bibnamefont{{Arai}}},
  \bibinfo{author}{\bibfnamefont{Y.}~\bibnamefont{{Arai}}},
  \bibinfo{author}{\bibfnamefont{S.}~\bibnamefont{{Araki}}},
  \bibinfo{author}{\bibfnamefont{A.}~\bibnamefont{{Araya}}},
  \bibinfo{author}{\bibfnamefont{N.}~\bibnamefont{{Aritomi}}},
  \bibinfo{author}{\bibfnamefont{Y.}~\bibnamefont{{Aso}}},
  \bibinfo{author}{\bibfnamefont{S.}~\bibnamefont{{Bae}}},
  \bibinfo{author}{\bibfnamefont{Y.}~\bibnamefont{{Bae}}},
  \bibnamefont{et~al.}, \bibinfo{journal}{Progress of Theoretical and
  Experimental Physics} \textbf{\bibinfo{volume}{2021}}, \bibinfo{eid}{05A101}
  (\bibinfo{year}{2021}), \eprint{2005.05574}.

\bibitem[{\citenamefont{{The LIGO Scientific Collaboration and the Virgo
  Collaboration}}(2016)}]{DiscoveryPaper}
\bibinfo{author}{\bibnamefont{{The LIGO Scientific Collaboration and the Virgo
  Collaboration}}}, \bibinfo{journal}{\prl} \textbf{\bibinfo{volume}{16}},
  \bibinfo{pages}{061102} (\bibinfo{year}{2016}).

\bibitem[{\citenamefont{{Abbott et al.\ (The LIGO Scientific Collaboration and
  the Virgo Collaboration)}}(2016)}]{LIGO-O1-BBH}
\bibinfo{author}{\bibfnamefont{B.}~\bibnamefont{{Abbott et al.\ (The LIGO
  Scientific Collaboration and the Virgo Collaboration)}}},
  \bibinfo{journal}{\prx} \textbf{\bibinfo{volume}{6}}, \bibinfo{pages}{041015}
  (\bibinfo{year}{2016}), \eprint{1606.04856},
  \urlprefix\url{https://journals.aps.org/prx/abstract/10.1103/PhysRevX.6.041015}.

\bibitem[{\citenamefont{{The LIGO Scientific Collaboration}
  et~al.}(2017{\natexlab{a}})\citenamefont{{The LIGO Scientific Collaboration},
  {the Virgo Collaboration}, {Abbott}, {Abbott}, {Abbott}, {Acernese},
  {Ackley}, {Adams}, {Adams}, {Addesso} et~al.}}]{LIGO-GW170817-bns}
\bibinfo{author}{\bibnamefont{{The LIGO Scientific Collaboration}}},
  \bibinfo{author}{\bibnamefont{{the Virgo Collaboration}}},
  \bibinfo{author}{\bibfnamefont{B.~P.} \bibnamefont{{Abbott}}},
  \bibinfo{author}{\bibfnamefont{R.}~\bibnamefont{{Abbott}}},
  \bibinfo{author}{\bibfnamefont{T.~D.} \bibnamefont{{Abbott}}},
  \bibinfo{author}{\bibfnamefont{F.}~\bibnamefont{{Acernese}}},
  \bibinfo{author}{\bibfnamefont{K.}~\bibnamefont{{Ackley}}},
  \bibinfo{author}{\bibfnamefont{C.}~\bibnamefont{{Adams}}},
  \bibinfo{author}{\bibfnamefont{T.}~\bibnamefont{{Adams}}},
  \bibinfo{author}{\bibfnamefont{P.}~\bibnamefont{{Addesso}}},
  \bibnamefont{et~al.}, \bibinfo{journal}{\prl} \textbf{\bibinfo{volume}{119}},
  \bibinfo{pages}{161101} (\bibinfo{year}{2017}{\natexlab{a}}).

\bibitem[{\citenamefont{{Abbott}
  et~al.}(2021{\natexlab{a}})\citenamefont{{Abbott}, {Abbott}, {Abraham},
  {Acernese}, {Ackley}, {Adams}, {Adams}, {Adhikari}, {Adya}, {Affeldt}
  et~al.}}]{LIGO-O3-NSBH}
\bibinfo{author}{\bibfnamefont{R.}~\bibnamefont{{Abbott}}},
  \bibinfo{author}{\bibfnamefont{T.~D.} \bibnamefont{{Abbott}}},
  \bibinfo{author}{\bibfnamefont{S.}~\bibnamefont{{Abraham}}},
  \bibinfo{author}{\bibfnamefont{F.}~\bibnamefont{{Acernese}}},
  \bibinfo{author}{\bibfnamefont{K.}~\bibnamefont{{Ackley}}},
  \bibinfo{author}{\bibfnamefont{A.}~\bibnamefont{{Adams}}},
  \bibinfo{author}{\bibfnamefont{C.}~\bibnamefont{{Adams}}},
  \bibinfo{author}{\bibfnamefont{R.~X.} \bibnamefont{{Adhikari}}},
  \bibinfo{author}{\bibfnamefont{V.~B.} \bibnamefont{{Adya}}},
  \bibinfo{author}{\bibfnamefont{C.}~\bibnamefont{{Affeldt}}},
  \bibnamefont{et~al.}, \bibinfo{journal}{\apjl}
  \textbf{\bibinfo{volume}{915}}, \bibinfo{eid}{L5}
  (\bibinfo{year}{2021}{\natexlab{a}}), \eprint{2106.15163}.

\bibitem[{\citenamefont{{The LIGO Scientific Collaboration}
  et~al.}()\citenamefont{{The LIGO Scientific Collaboration}, {the Virgo
  Collaboration}, {Abbott}, {Abbott}, {Abbott}, {Abraham}, {Acernese},
  {Ackley}, {Adams}, {Adya} et~al.}}]{LIGO-O3-O3b-catalog}
\bibinfo{author}{\bibnamefont{{The LIGO Scientific Collaboration}}},
  \bibinfo{author}{\bibnamefont{{the Virgo Collaboration}}},
  \bibinfo{author}{\bibfnamefont{B.~P.} \bibnamefont{{Abbott}}},
  \bibinfo{author}{\bibfnamefont{R.}~\bibnamefont{{Abbott}}},
  \bibinfo{author}{\bibfnamefont{T.~D.} \bibnamefont{{Abbott}}},
  \bibinfo{author}{\bibfnamefont{S.}~\bibnamefont{{Abraham}}},
  \bibinfo{author}{\bibfnamefont{F.}~\bibnamefont{{Acernese}}},
  \bibinfo{author}{\bibfnamefont{K.}~\bibnamefont{{Ackley}}},
  \bibinfo{author}{\bibfnamefont{C.}~\bibnamefont{{Adams}}},
  \bibinfo{author}{\bibfnamefont{V.~B.} \bibnamefont{{Adya}}},
  \bibnamefont{et~al.}, \bibinfo{journal}{Available as LIGO-P2000318}  (????),
  \urlprefix\url{https://dcc.ligo.org/LIGO-P2000318}.

\bibitem[{\citenamefont{{The LIGO Scientific Collaboration}
  et~al.}(2021{\natexlab{a}})\citenamefont{{The LIGO Scientific Collaboration},
  {the Virgo Collaboration}, {Abbott}, {Abbott}, {Abbott}, {Abraham},
  {Acernese}, {Ackley}, {Adams}, {Adya} et~al.}}]{LIGO-O3-O3a_final-catalog}
\bibinfo{author}{\bibnamefont{{The LIGO Scientific Collaboration}}},
  \bibinfo{author}{\bibnamefont{{the Virgo Collaboration}}},
  \bibinfo{author}{\bibfnamefont{B.~P.} \bibnamefont{{Abbott}}},
  \bibinfo{author}{\bibfnamefont{R.}~\bibnamefont{{Abbott}}},
  \bibinfo{author}{\bibfnamefont{T.~D.} \bibnamefont{{Abbott}}},
  \bibinfo{author}{\bibfnamefont{S.}~\bibnamefont{{Abraham}}},
  \bibinfo{author}{\bibfnamefont{F.}~\bibnamefont{{Acernese}}},
  \bibinfo{author}{\bibfnamefont{K.}~\bibnamefont{{Ackley}}},
  \bibinfo{author}{\bibfnamefont{C.}~\bibnamefont{{Adams}}},
  \bibinfo{author}{\bibfnamefont{V.~B.} \bibnamefont{{Adya}}},
  \bibnamefont{et~al.}, \bibinfo{journal}{Available as LIGO-P2100063}
  (\bibinfo{year}{2021}{\natexlab{a}}),
  \urlprefix\url{https://dcc.ligo.org/LIGO-P2100063/public}.

\bibitem[{\citenamefont{{Abbott} et~al.}(2016)\citenamefont{{Abbott}, {Abbott},
  {Abbott}, {Abernathy}, {Acernese}, {Ackley}, {Adams}, {Adams}, {Addesso},
  {Adhikari} et~al.}}]{2016LRR....19....1A}
\bibinfo{author}{\bibfnamefont{B.~P.} \bibnamefont{{Abbott}}},
  \bibinfo{author}{\bibfnamefont{R.}~\bibnamefont{{Abbott}}},
  \bibinfo{author}{\bibfnamefont{T.~D.} \bibnamefont{{Abbott}}},
  \bibinfo{author}{\bibfnamefont{M.~R.} \bibnamefont{{Abernathy}}},
  \bibinfo{author}{\bibfnamefont{F.}~\bibnamefont{{Acernese}}},
  \bibinfo{author}{\bibfnamefont{K.}~\bibnamefont{{Ackley}}},
  \bibinfo{author}{\bibfnamefont{C.}~\bibnamefont{{Adams}}},
  \bibinfo{author}{\bibfnamefont{T.}~\bibnamefont{{Adams}}},
  \bibinfo{author}{\bibfnamefont{P.}~\bibnamefont{{Addesso}}},
  \bibinfo{author}{\bibfnamefont{R.~X.} \bibnamefont{{Adhikari}}},
  \bibnamefont{et~al.}, \bibinfo{journal}{Living Reviews in Relativity}
  \textbf{\bibinfo{volume}{19}}, \bibinfo{eid}{1} (\bibinfo{year}{2016}).

\bibitem[{\citenamefont{{Abbott} et~al.}(2017)\citenamefont{{Abbott}, {Abbott},
  {Abbott}, {Acernese}, {Ackley}, {Adams}, {Adams}, {Addesso}, {Adhikari},
  {Adya} et~al.}}]{2017PhRvL.118v1101A}
\bibinfo{author}{\bibfnamefont{B.~P.} \bibnamefont{{Abbott}}},
  \bibinfo{author}{\bibfnamefont{R.}~\bibnamefont{{Abbott}}},
  \bibinfo{author}{\bibfnamefont{T.~D.} \bibnamefont{{Abbott}}},
  \bibinfo{author}{\bibfnamefont{F.}~\bibnamefont{{Acernese}}},
  \bibinfo{author}{\bibfnamefont{K.}~\bibnamefont{{Ackley}}},
  \bibinfo{author}{\bibfnamefont{C.}~\bibnamefont{{Adams}}},
  \bibinfo{author}{\bibfnamefont{T.}~\bibnamefont{{Adams}}},
  \bibinfo{author}{\bibfnamefont{P.}~\bibnamefont{{Addesso}}},
  \bibinfo{author}{\bibfnamefont{R.~X.} \bibnamefont{{Adhikari}}},
  \bibinfo{author}{\bibfnamefont{V.~B.} \bibnamefont{{Adya}}},
  \bibnamefont{et~al.}, \bibinfo{journal}{Physical Review Letters}
  \textbf{\bibinfo{volume}{118}}, \bibinfo{eid}{221101} (\bibinfo{year}{2017}),
  \eprint{1706.01812}.

\bibitem[{\citenamefont{{The LIGO Scientific Collaboration}
  et~al.}(2017{\natexlab{b}})\citenamefont{{The LIGO Scientific Collaboration},
  {the Virgo Collaboration}, {Abbott}, {Abbott}, {Abbott}, {Acernese},
  {Ackley}, {Adams}, {Adams}, {Addesso} et~al.}}]{LIGO-GW170814}
\bibinfo{author}{\bibnamefont{{The LIGO Scientific Collaboration}}},
  \bibinfo{author}{\bibnamefont{{the Virgo Collaboration}}},
  \bibinfo{author}{\bibfnamefont{B.~P.} \bibnamefont{{Abbott}}},
  \bibinfo{author}{\bibfnamefont{R.}~\bibnamefont{{Abbott}}},
  \bibinfo{author}{\bibfnamefont{T.~D.} \bibnamefont{{Abbott}}},
  \bibinfo{author}{\bibfnamefont{F.}~\bibnamefont{{Acernese}}},
  \bibinfo{author}{\bibfnamefont{K.}~\bibnamefont{{Ackley}}},
  \bibinfo{author}{\bibfnamefont{C.}~\bibnamefont{{Adams}}},
  \bibinfo{author}{\bibfnamefont{T.}~\bibnamefont{{Adams}}},
  \bibinfo{author}{\bibfnamefont{P.}~\bibnamefont{{Addesso}}},
  \bibnamefont{et~al.}, \bibinfo{journal}{\prl} \textbf{\bibinfo{volume}{119}},
  \bibinfo{eid}{141101} (\bibinfo{year}{2017}{\natexlab{b}}),
  \eprint{1709.09660}.

\bibitem[{\citenamefont{{The LIGO Scientific Collaboration}
  et~al.}(2017{\natexlab{c}})\citenamefont{{The LIGO Scientific Collaboration},
  {the Virgo Collaboration}, {Abbott}, {Abbott}, {Abbott}, {Acernese},
  {Ackley}, {Adams}, {Adams}, {Addesso} et~al.}}]{LIGO-GW170608}
\bibinfo{author}{\bibnamefont{{The LIGO Scientific Collaboration}}},
  \bibinfo{author}{\bibnamefont{{the Virgo Collaboration}}},
  \bibinfo{author}{\bibfnamefont{B.~P.} \bibnamefont{{Abbott}}},
  \bibinfo{author}{\bibfnamefont{R.}~\bibnamefont{{Abbott}}},
  \bibinfo{author}{\bibfnamefont{T.~D.} \bibnamefont{{Abbott}}},
  \bibinfo{author}{\bibfnamefont{F.}~\bibnamefont{{Acernese}}},
  \bibinfo{author}{\bibfnamefont{K.}~\bibnamefont{{Ackley}}},
  \bibinfo{author}{\bibfnamefont{C.}~\bibnamefont{{Adams}}},
  \bibinfo{author}{\bibfnamefont{T.}~\bibnamefont{{Adams}}},
  \bibinfo{author}{\bibfnamefont{P.}~\bibnamefont{{Addesso}}},
  \bibnamefont{et~al.}, \bibinfo{journal}{\apjl}
  \textbf{\bibinfo{volume}{851}}, \bibinfo{eid}{L35}
  (\bibinfo{year}{2017}{\natexlab{c}}).

\bibitem[{\citenamefont{{The LIGO Scientific Collaboration}
  et~al.}(2018{\natexlab{a}})\citenamefont{{The LIGO Scientific Collaboration},
  {The Virgo Collaboration}, {Abbott}, {Abbott}, {Abbott}, {Acernese},
  {Ackley}, {Adams}, {Adams}, {Addesso} et~al.}}]{LIGO-O2-Catalog}
\bibinfo{author}{\bibnamefont{{The LIGO Scientific Collaboration}}},
  \bibinfo{author}{\bibnamefont{{The Virgo Collaboration}}},
  \bibinfo{author}{\bibfnamefont{B.~P.} \bibnamefont{{Abbott}}},
  \bibinfo{author}{\bibfnamefont{R.}~\bibnamefont{{Abbott}}},
  \bibinfo{author}{\bibfnamefont{T.~D.} \bibnamefont{{Abbott}}},
  \bibinfo{author}{\bibfnamefont{F.}~\bibnamefont{{Acernese}}},
  \bibinfo{author}{\bibfnamefont{K.}~\bibnamefont{{Ackley}}},
  \bibinfo{author}{\bibfnamefont{C.}~\bibnamefont{{Adams}}},
  \bibinfo{author}{\bibfnamefont{T.}~\bibnamefont{{Adams}}},
  \bibinfo{author}{\bibfnamefont{P.}~\bibnamefont{{Addesso}}},
  \bibnamefont{et~al.}, \bibinfo{journal}{\prx} \textbf{\bibinfo{volume}{9}},
  \bibinfo{eid}{031040} (\bibinfo{year}{2018}{\natexlab{a}}).

\bibitem[{\citenamefont{{Pankow} et~al.}(2015)\citenamefont{{Pankow}, {Brady},
  {Ochsner}, and {O'Shaughnessy}}}]{gwastro-PE-AlternativeArchitectures}
\bibinfo{author}{\bibfnamefont{C.}~\bibnamefont{{Pankow}}},
  \bibinfo{author}{\bibfnamefont{P.}~\bibnamefont{{Brady}}},
  \bibinfo{author}{\bibfnamefont{E.}~\bibnamefont{{Ochsner}}},
  \bibnamefont{and}
  \bibinfo{author}{\bibfnamefont{R.}~\bibnamefont{{O'Shaughnessy}}},
  \bibinfo{journal}{\prd} \textbf{\bibinfo{volume}{92}}, \bibinfo{eid}{023002}
  (\bibinfo{year}{2015}),
  \urlprefix\url{http://adsabs.harvard.edu/abs/2015PhRvD..92b3002P}.

\bibitem[{\citenamefont{{Lange} et~al.}(2018)\citenamefont{{Lange},
  {O'Shaughnessy}, and {Rizzo}}}]{gwastro-PENR-RIFT}
\bibinfo{author}{\bibfnamefont{J.}~\bibnamefont{{Lange}}},
  \bibinfo{author}{\bibfnamefont{R.}~\bibnamefont{{O'Shaughnessy}}},
  \bibnamefont{and} \bibinfo{author}{\bibfnamefont{M.}~\bibnamefont{{Rizzo}}},
  \bibinfo{journal}{Submitted to PRD; available at arxiv:1805.10457}
  (\bibinfo{year}{2018}).

\bibitem[{\citenamefont{{Veitch} et~al.}(2015)\citenamefont{{Veitch},
  {Raymond}, {Farr}, {Farr}, {Graff}, {Vitale}, {Aylott}, {Blackburn},
  {Christensen}, {Coughlin} et~al.}}]{gw-astro-PE-lalinference-v1}
\bibinfo{author}{\bibfnamefont{J.}~\bibnamefont{{Veitch}}},
  \bibinfo{author}{\bibfnamefont{V.}~\bibnamefont{{Raymond}}},
  \bibinfo{author}{\bibfnamefont{B.}~\bibnamefont{{Farr}}},
  \bibinfo{author}{\bibfnamefont{W.~M.} \bibnamefont{{Farr}}},
  \bibinfo{author}{\bibfnamefont{P.}~\bibnamefont{{Graff}}},
  \bibinfo{author}{\bibfnamefont{S.}~\bibnamefont{{Vitale}}},
  \bibinfo{author}{\bibfnamefont{B.}~\bibnamefont{{Aylott}}},
  \bibinfo{author}{\bibfnamefont{K.}~\bibnamefont{{Blackburn}}},
  \bibinfo{author}{\bibfnamefont{N.}~\bibnamefont{{Christensen}}},
  \bibinfo{author}{\bibfnamefont{M.}~\bibnamefont{{Coughlin}}},
  \bibnamefont{et~al.}, \bibinfo{journal}{\prd} \textbf{\bibinfo{volume}{91}},
  \bibinfo{pages}{042003} (\bibinfo{year}{2015}),
  \urlprefix\url{http://link.aps.org/doi/10.1103/PhysRevD.91.042003}.

\bibitem[{\citenamefont{{Hannam} et~al.}(2014)\citenamefont{{Hannam},
  {Schmidt}, {Boh{\'e}}, {Haegel}, {Husa}, {Ohme}, {Pratten}, and
  {P{\"u}rrer}}}]{gwastro-mergers-IMRPhenomP}
\bibinfo{author}{\bibfnamefont{M.}~\bibnamefont{{Hannam}}},
  \bibinfo{author}{\bibfnamefont{P.}~\bibnamefont{{Schmidt}}},
  \bibinfo{author}{\bibfnamefont{A.}~\bibnamefont{{Boh{\'e}}}},
  \bibinfo{author}{\bibfnamefont{L.}~\bibnamefont{{Haegel}}},
  \bibinfo{author}{\bibfnamefont{S.}~\bibnamefont{{Husa}}},
  \bibinfo{author}{\bibfnamefont{F.}~\bibnamefont{{Ohme}}},
  \bibinfo{author}{\bibfnamefont{G.}~\bibnamefont{{Pratten}}},
  \bibnamefont{and}
  \bibinfo{author}{\bibfnamefont{M.}~\bibnamefont{{P{\"u}rrer}}},
  \bibinfo{journal}{\prl} \textbf{\bibinfo{volume}{113}}, \bibinfo{eid}{151101}
  (\bibinfo{year}{2014}).

\bibitem[{\citenamefont{{Khan} et~al.}(2019)\citenamefont{{Khan},
  {Chatziioannou}, {Hannam}, and {Ohme}}}]{gwastro-mergers-IMRPhenomPv3}
\bibinfo{author}{\bibfnamefont{S.}~\bibnamefont{{Khan}}},
  \bibinfo{author}{\bibfnamefont{K.}~\bibnamefont{{Chatziioannou}}},
  \bibinfo{author}{\bibfnamefont{M.}~\bibnamefont{{Hannam}}}, \bibnamefont{and}
  \bibinfo{author}{\bibfnamefont{F.}~\bibnamefont{{Ohme}}},
  \bibinfo{journal}{\prd} \textbf{\bibinfo{volume}{100}}, \bibinfo{eid}{024059}
  (\bibinfo{year}{2019}), \eprint{1809.10113}.

\bibitem[{\citenamefont{{Boh{\'e}} et~al.}(2017)\citenamefont{{Boh{\'e}},
  {Shao}, {Taracchini}, {Buonanno}, {Babak}, {Harry}, {Hinder}, {Ossokine},
  {P{\"u}rrer}, {Raymond} et~al.}}]{gwastro-SEOBNRv4}
\bibinfo{author}{\bibfnamefont{A.}~\bibnamefont{{Boh{\'e}}}},
  \bibinfo{author}{\bibfnamefont{L.}~\bibnamefont{{Shao}}},
  \bibinfo{author}{\bibfnamefont{A.}~\bibnamefont{{Taracchini}}},
  \bibinfo{author}{\bibfnamefont{A.}~\bibnamefont{{Buonanno}}},
  \bibinfo{author}{\bibfnamefont{S.}~\bibnamefont{{Babak}}},
  \bibinfo{author}{\bibfnamefont{I.~W.} \bibnamefont{{Harry}}},
  \bibinfo{author}{\bibfnamefont{I.}~\bibnamefont{{Hinder}}},
  \bibinfo{author}{\bibfnamefont{S.}~\bibnamefont{{Ossokine}}},
  \bibinfo{author}{\bibfnamefont{M.}~\bibnamefont{{P{\"u}rrer}}},
  \bibinfo{author}{\bibfnamefont{V.}~\bibnamefont{{Raymond}}},
  \bibnamefont{et~al.}, \bibinfo{journal}{\prd} \textbf{\bibinfo{volume}{95}},
  \bibinfo{eid}{044028} (\bibinfo{year}{2017}), \eprint{1611.03703}.

\bibitem[{\citenamefont{{Varma} et~al.}(2019)\citenamefont{{Varma}, {Field},
  {Scheel}, {Blackman}, {Gerosa}, {Stein}, {Kidder}, and
  {Pfeiffer}}}]{2019PhRvR...1c3015V}
\bibinfo{author}{\bibfnamefont{V.}~\bibnamefont{{Varma}}},
  \bibinfo{author}{\bibfnamefont{S.~E.} \bibnamefont{{Field}}},
  \bibinfo{author}{\bibfnamefont{M.~A.} \bibnamefont{{Scheel}}},
  \bibinfo{author}{\bibfnamefont{J.}~\bibnamefont{{Blackman}}},
  \bibinfo{author}{\bibfnamefont{D.}~\bibnamefont{{Gerosa}}},
  \bibinfo{author}{\bibfnamefont{L.~C.} \bibnamefont{{Stein}}},
  \bibinfo{author}{\bibfnamefont{L.~E.} \bibnamefont{{Kidder}}},
  \bibnamefont{and} \bibinfo{author}{\bibfnamefont{H.~P.}
  \bibnamefont{{Pfeiffer}}}, \bibinfo{journal}{Physical Review Research}
  \textbf{\bibinfo{volume}{1}}, \bibinfo{eid}{033015} (\bibinfo{year}{2019}),
  \eprint{1905.09300}.

\bibitem[{\citenamefont{{Pratten} et~al.}(2021)\citenamefont{{Pratten},
  {Garc{\'\i}a-Quir{\'o}s}, {Colleoni}, {Ramos-Buades}, {Estell{\'e}s},
  {Mateu-Lucena}, {Jaume}, {Haney}, {Keitel}, {Thompson}
  et~al.}}]{gwastro-mergers-IMRPhenomXP}
\bibinfo{author}{\bibfnamefont{G.}~\bibnamefont{{Pratten}}},
  \bibinfo{author}{\bibfnamefont{C.}~\bibnamefont{{Garc{\'\i}a-Quir{\'o}s}}},
  \bibinfo{author}{\bibfnamefont{M.}~\bibnamefont{{Colleoni}}},
  \bibinfo{author}{\bibfnamefont{A.}~\bibnamefont{{Ramos-Buades}}},
  \bibinfo{author}{\bibfnamefont{H.}~\bibnamefont{{Estell{\'e}s}}},
  \bibinfo{author}{\bibfnamefont{M.}~\bibnamefont{{Mateu-Lucena}}},
  \bibinfo{author}{\bibfnamefont{R.}~\bibnamefont{{Jaume}}},
  \bibinfo{author}{\bibfnamefont{M.}~\bibnamefont{{Haney}}},
  \bibinfo{author}{\bibfnamefont{D.}~\bibnamefont{{Keitel}}},
  \bibinfo{author}{\bibfnamefont{J.~E.} \bibnamefont{{Thompson}}},
  \bibnamefont{et~al.}, \bibinfo{journal}{\prd} \textbf{\bibinfo{volume}{103}},
  \bibinfo{eid}{104056} (\bibinfo{year}{2021}), \eprint{2004.06503}.

\bibitem[{\citenamefont{{Ossokine} et~al.}(2020)\citenamefont{{Ossokine},
  {Buonanno}, {Marsat}, {Cotesta}, {Babak}, {Dietrich}, {Haas}, {Hinder},
  {Pfeiffer}, {P{\"u}rrer} et~al.}}]{2020PhRvD.102d4055O}
\bibinfo{author}{\bibfnamefont{S.}~\bibnamefont{{Ossokine}}},
  \bibinfo{author}{\bibfnamefont{A.}~\bibnamefont{{Buonanno}}},
  \bibinfo{author}{\bibfnamefont{S.}~\bibnamefont{{Marsat}}},
  \bibinfo{author}{\bibfnamefont{R.}~\bibnamefont{{Cotesta}}},
  \bibinfo{author}{\bibfnamefont{S.}~\bibnamefont{{Babak}}},
  \bibinfo{author}{\bibfnamefont{T.}~\bibnamefont{{Dietrich}}},
  \bibinfo{author}{\bibfnamefont{R.}~\bibnamefont{{Haas}}},
  \bibinfo{author}{\bibfnamefont{I.}~\bibnamefont{{Hinder}}},
  \bibinfo{author}{\bibfnamefont{H.~P.} \bibnamefont{{Pfeiffer}}},
  \bibinfo{author}{\bibfnamefont{M.}~\bibnamefont{{P{\"u}rrer}}},
  \bibnamefont{et~al.}, \bibinfo{journal}{\prd} \textbf{\bibinfo{volume}{102}},
  \bibinfo{eid}{044055} (\bibinfo{year}{2020}), \eprint{2004.09442}.

\bibitem[{\citenamefont{{Ashton} et~al.}(2019)\citenamefont{{Ashton},
  {H{\"u}bner}, {Lasky}, {Talbot}, {Ackley}, {Biscoveanu}, {Chu}, {Divakarla},
  {Easter}, {Goncharov} et~al.}}]{gwastro-pe-bilby-2018}
\bibinfo{author}{\bibfnamefont{G.}~\bibnamefont{{Ashton}}},
  \bibinfo{author}{\bibfnamefont{M.}~\bibnamefont{{H{\"u}bner}}},
  \bibinfo{author}{\bibfnamefont{P.~D.} \bibnamefont{{Lasky}}},
  \bibinfo{author}{\bibfnamefont{C.}~\bibnamefont{{Talbot}}},
  \bibinfo{author}{\bibfnamefont{K.}~\bibnamefont{{Ackley}}},
  \bibinfo{author}{\bibfnamefont{S.}~\bibnamefont{{Biscoveanu}}},
  \bibinfo{author}{\bibfnamefont{Q.}~\bibnamefont{{Chu}}},
  \bibinfo{author}{\bibfnamefont{A.}~\bibnamefont{{Divakarla}}},
  \bibinfo{author}{\bibfnamefont{P.~J.} \bibnamefont{{Easter}}},
  \bibinfo{author}{\bibfnamefont{B.}~\bibnamefont{{Goncharov}}},
  \bibnamefont{et~al.}, \bibinfo{journal}{\apjs}
  \textbf{\bibinfo{volume}{241}}, \bibinfo{eid}{27} (\bibinfo{year}{2019}),
  \eprint{1811.02042}.

\bibitem[{\citenamefont{{The LIGO Scientific Collaboration}
  et~al.}(2020{\natexlab{a}})\citenamefont{{The LIGO Scientific Collaboration},
  {the Virgo Collaboration}, {Abbott}, {Abbott}, {Abbott}, {Abraham},
  {Acernese}, {Ackley}, {Adams}, {Adya} et~al.}}]{LIGO-O3-GW190521-discovery}
\bibinfo{author}{\bibnamefont{{The LIGO Scientific Collaboration}}},
  \bibinfo{author}{\bibnamefont{{the Virgo Collaboration}}},
  \bibinfo{author}{\bibfnamefont{B.~P.} \bibnamefont{{Abbott}}},
  \bibinfo{author}{\bibfnamefont{R.}~\bibnamefont{{Abbott}}},
  \bibinfo{author}{\bibfnamefont{T.~D.} \bibnamefont{{Abbott}}},
  \bibinfo{author}{\bibfnamefont{S.}~\bibnamefont{{Abraham}}},
  \bibinfo{author}{\bibfnamefont{F.}~\bibnamefont{{Acernese}}},
  \bibinfo{author}{\bibfnamefont{K.}~\bibnamefont{{Ackley}}},
  \bibinfo{author}{\bibfnamefont{C.}~\bibnamefont{{Adams}}},
  \bibinfo{author}{\bibfnamefont{V.~B.} \bibnamefont{{Adya}}},
  \bibnamefont{et~al.}, \bibinfo{journal}{\prl} \textbf{\bibinfo{volume}{125}},
  \bibinfo{eid}{101102} (\bibinfo{year}{2020}{\natexlab{a}}).

\bibitem[{\citenamefont{{The LIGO Scientific Collaboration}
  et~al.}(2020{\natexlab{b}})\citenamefont{{The LIGO Scientific Collaboration},
  {the Virgo Collaboration}, {Abbott}, {Abbott}, {Abbott}, {Abraham},
  {Acernese}, {Ackley}, {Adams}, {Adya}
  et~al.}}]{LIGO-O3-GW190521-implications}
\bibinfo{author}{\bibnamefont{{The LIGO Scientific Collaboration}}},
  \bibinfo{author}{\bibnamefont{{the Virgo Collaboration}}},
  \bibinfo{author}{\bibfnamefont{B.~P.} \bibnamefont{{Abbott}}},
  \bibinfo{author}{\bibfnamefont{R.}~\bibnamefont{{Abbott}}},
  \bibinfo{author}{\bibfnamefont{T.~D.} \bibnamefont{{Abbott}}},
  \bibinfo{author}{\bibfnamefont{S.}~\bibnamefont{{Abraham}}},
  \bibinfo{author}{\bibfnamefont{F.}~\bibnamefont{{Acernese}}},
  \bibinfo{author}{\bibfnamefont{K.}~\bibnamefont{{Ackley}}},
  \bibinfo{author}{\bibfnamefont{C.}~\bibnamefont{{Adams}}},
  \bibinfo{author}{\bibfnamefont{V.~B.} \bibnamefont{{Adya}}},
  \bibnamefont{et~al.}, \bibinfo{journal}{\apjl}
  \textbf{\bibinfo{volume}{900}}, \bibinfo{eid}{L13}
  (\bibinfo{year}{2020}{\natexlab{b}}), \eprint{2009.01190}.

\bibitem[{\citenamefont{{The LIGO Scientific Collaboration}
  et~al.}(2021{\natexlab{b}})\citenamefont{{The LIGO Scientific Collaboration},
  {the Virgo Collaboration}, {Abbott}, {Abbott}, {Abraham}, {Acernese},
  {Ackley}, {Adams}, {Adams}, {Adhikari} et~al.}}]{LIGO-O3-O3a-catalog}
\bibinfo{author}{\bibnamefont{{The LIGO Scientific Collaboration}}},
  \bibinfo{author}{\bibnamefont{{the Virgo Collaboration}}},
  \bibinfo{author}{\bibfnamefont{R.}~\bibnamefont{{Abbott}}},
  \bibinfo{author}{\bibfnamefont{T.~D.} \bibnamefont{{Abbott}}},
  \bibinfo{author}{\bibfnamefont{S.}~\bibnamefont{{Abraham}}},
  \bibinfo{author}{\bibfnamefont{F.}~\bibnamefont{{Acernese}}},
  \bibinfo{author}{\bibfnamefont{K.}~\bibnamefont{{Ackley}}},
  \bibinfo{author}{\bibfnamefont{A.}~\bibnamefont{{Adams}}},
  \bibinfo{author}{\bibfnamefont{C.}~\bibnamefont{{Adams}}},
  \bibinfo{author}{\bibfnamefont{R.~X.} \bibnamefont{{Adhikari}}},
  \bibnamefont{et~al.}, \bibinfo{journal}{Physical Review X}
  \textbf{\bibinfo{volume}{11}}, \bibinfo{eid}{021053}
  (\bibinfo{year}{2021}{\natexlab{b}}), \eprint{2010.14527}.

\bibitem[{\citenamefont{Christensen and Meyer}(2022)}]{RevModPhys.94.025001}
\bibinfo{author}{\bibfnamefont{N.}~\bibnamefont{Christensen}} \bibnamefont{and}
  \bibinfo{author}{\bibfnamefont{R.}~\bibnamefont{Meyer}},
  \bibinfo{journal}{Rev. Mod. Phys.} \textbf{\bibinfo{volume}{94}},
  \bibinfo{pages}{025001} (\bibinfo{year}{2022}),
  \urlprefix\url{https://link.aps.org/doi/10.1103/RevModPhys.94.025001}.

\bibitem[{\citenamefont{{Wysocki} et~al.}(2019)\citenamefont{{Wysocki},
  {O'Shaughnessy}, {Lange}, and {Fang}}}]{gwastro-PENR-RIFT-GPU}
\bibinfo{author}{\bibfnamefont{D.}~\bibnamefont{{Wysocki}}},
  \bibinfo{author}{\bibfnamefont{R.}~\bibnamefont{{O'Shaughnessy}}},
  \bibinfo{author}{\bibfnamefont{J.}~\bibnamefont{{Lange}}}, \bibnamefont{and}
  \bibinfo{author}{\bibfnamefont{Y.-L.~L.} \bibnamefont{{Fang}}},
  \bibinfo{journal}{\prd} \textbf{\bibinfo{volume}{99}}, \bibinfo{eid}{084026}
  (\bibinfo{year}{2019}), \eprint{1902.04934}.

\bibitem[{\citenamefont{{Jan} et~al.}(2020{\natexlab{a}})\citenamefont{{Jan},
  {Yelikar}, {Lange}, and {O'Shaughnessy}}}]{2020PhRvD.102l4069J}
\bibinfo{author}{\bibfnamefont{A.~Z.} \bibnamefont{{Jan}}},
  \bibinfo{author}{\bibfnamefont{A.~B.} \bibnamefont{{Yelikar}}},
  \bibinfo{author}{\bibfnamefont{J.}~\bibnamefont{{Lange}}}, \bibnamefont{and}
  \bibinfo{author}{\bibfnamefont{R.}~\bibnamefont{{O'Shaughnessy}}},
  \bibinfo{journal}{\prd} \textbf{\bibinfo{volume}{102}}, \bibinfo{eid}{124069}
  (\bibinfo{year}{2020}{\natexlab{a}}), \eprint{2011.03571}.

\bibitem[{\citenamefont{{Wysocki}
  et~al.}(2020{\natexlab{a}})\citenamefont{{Wysocki}, {O'Shaughnessy}, {Wade},
  and {Lange}}}]{gwastro-PopulationReconstruct-EOSStack-Wysocki2019}
\bibinfo{author}{\bibfnamefont{D.}~\bibnamefont{{Wysocki}}},
  \bibinfo{author}{\bibfnamefont{R.}~\bibnamefont{{O'Shaughnessy}}},
  \bibinfo{author}{\bibfnamefont{L.}~\bibnamefont{{Wade}}}, \bibnamefont{and}
  \bibinfo{author}{\bibfnamefont{J.}~\bibnamefont{{Lange}}},
  \bibinfo{journal}{Submitted to PRD; available as arxiv:2001.01747}
  (\bibinfo{year}{2020}{\natexlab{a}}),
  \urlprefix\url{https://arxiv.org/abs/2001.01747}.

\bibitem[{\citenamefont{{Al-Mamun} et~al.}(2021)\citenamefont{{Al-Mamun},
  {Steiner}, {N{\"a}ttil{\"a}}, {Lange}, {O'Shaughnessy}, {Tews}, {Gandolfi},
  {Heinke}, and {Han}}}]{gwastro-nsnuc-SteinerJoint-2020}
\bibinfo{author}{\bibfnamefont{M.}~\bibnamefont{{Al-Mamun}}},
  \bibinfo{author}{\bibfnamefont{A.~W.} \bibnamefont{{Steiner}}},
  \bibinfo{author}{\bibfnamefont{J.}~\bibnamefont{{N{\"a}ttil{\"a}}}},
  \bibinfo{author}{\bibfnamefont{J.}~\bibnamefont{{Lange}}},
  \bibinfo{author}{\bibfnamefont{R.}~\bibnamefont{{O'Shaughnessy}}},
  \bibinfo{author}{\bibfnamefont{I.}~\bibnamefont{{Tews}}},
  \bibinfo{author}{\bibfnamefont{S.}~\bibnamefont{{Gandolfi}}},
  \bibinfo{author}{\bibfnamefont{C.}~\bibnamefont{{Heinke}}}, \bibnamefont{and}
  \bibinfo{author}{\bibfnamefont{S.}~\bibnamefont{{Han}}},
  \bibinfo{journal}{\prl} \textbf{\bibinfo{volume}{126}}, \bibinfo{eid}{061101}
  (\bibinfo{year}{2021}), \eprint{2008.12817}.

\bibitem[{\citenamefont{{The LIGO Scientific Collaboration}
  et~al.}(2020{\natexlab{c}})\citenamefont{{The LIGO Scientific Collaboration},
  {the Virgo Collaboration}, {Abbott}, {Abbott}, {Abbott}, {Abraham},
  {Acernese}, {Ackley}, {Adams}, {Adya} et~al.}}]{LIGO-O3-GW190412}
\bibinfo{author}{\bibnamefont{{The LIGO Scientific Collaboration}}},
  \bibinfo{author}{\bibnamefont{{the Virgo Collaboration}}},
  \bibinfo{author}{\bibfnamefont{B.~P.} \bibnamefont{{Abbott}}},
  \bibinfo{author}{\bibfnamefont{R.}~\bibnamefont{{Abbott}}},
  \bibinfo{author}{\bibfnamefont{T.~D.} \bibnamefont{{Abbott}}},
  \bibinfo{author}{\bibfnamefont{S.}~\bibnamefont{{Abraham}}},
  \bibinfo{author}{\bibfnamefont{F.}~\bibnamefont{{Acernese}}},
  \bibinfo{author}{\bibfnamefont{K.}~\bibnamefont{{Ackley}}},
  \bibinfo{author}{\bibfnamefont{C.}~\bibnamefont{{Adams}}},
  \bibinfo{author}{\bibfnamefont{V.~B.} \bibnamefont{{Adya}}},
  \bibnamefont{et~al.}, \bibinfo{journal}{\prd} \textbf{\bibinfo{volume}{102}},
  \bibinfo{eid}{043015} (\bibinfo{year}{2020}{\natexlab{c}}).

\bibitem[{\citenamefont{{The LIGO Scientific Collaboration}
  et~al.}(2020{\natexlab{d}})\citenamefont{{The LIGO Scientific Collaboration},
  {the Virgo Collaboration}, {Abbott}, {Abbott}, {Abbott}, and {et
  al}}}]{LIGO-GW170817-EOSrank}
\bibinfo{author}{\bibnamefont{{The LIGO Scientific Collaboration}}},
  \bibinfo{author}{\bibnamefont{{the Virgo Collaboration}}},
  \bibinfo{author}{\bibfnamefont{B.~P.} \bibnamefont{{Abbott}}},
  \bibinfo{author}{\bibfnamefont{R.}~\bibnamefont{{Abbott}}},
  \bibinfo{author}{\bibfnamefont{T.~D.} \bibnamefont{{Abbott}}},
  \bibnamefont{and} \bibinfo{author}{\bibnamefont{{et al}}},
  \bibinfo{journal}{Classical and Quantum Gravity}
  \textbf{\bibinfo{volume}{37}}, \bibinfo{eid}{045006}
  (\bibinfo{year}{2020}{\natexlab{d}}), \eprint{1908.01012}.

\bibitem[{\citenamefont{{Udall} et~al.}(2021)\citenamefont{{Udall}, {Brandt},
  {Manchanda}, {Arulanandan}, {Clark}, {Lange}, {O'Shaughnessy}, and
  {Cadonati}}}]{gwastro-RIFT-runmon}
\bibinfo{author}{\bibfnamefont{R.}~\bibnamefont{{Udall}}},
  \bibinfo{author}{\bibfnamefont{J.}~\bibnamefont{{Brandt}}},
  \bibinfo{author}{\bibfnamefont{G.}~\bibnamefont{{Manchanda}}},
  \bibinfo{author}{\bibfnamefont{A.}~\bibnamefont{{Arulanandan}}},
  \bibinfo{author}{\bibfnamefont{J.}~\bibnamefont{{Clark}}},
  \bibinfo{author}{\bibfnamefont{J.}~\bibnamefont{{Lange}}},
  \bibinfo{author}{\bibfnamefont{R.}~\bibnamefont{{O'Shaughnessy}}},
  \bibnamefont{and}
  \bibinfo{author}{\bibfnamefont{L.}~\bibnamefont{{Cadonati}}},
  \bibinfo{journal}{arXiv e-prints} \bibinfo{eid}{arXiv:2110.10243}
  (\bibinfo{year}{2021}), \eprint{2110.10243}.

\bibitem[{\citenamefont{Cornish}(2021{\natexlab{a}})}]{PhysRevD.104.104054}
\bibinfo{author}{\bibfnamefont{N.~J.} \bibnamefont{Cornish}},
  \bibinfo{journal}{Phys. Rev. D} \textbf{\bibinfo{volume}{104}},
  \bibinfo{pages}{104054} (\bibinfo{year}{2021}{\natexlab{a}}),
  \urlprefix\url{https://link.aps.org/doi/10.1103/PhysRevD.104.104054}.

\bibitem[{\citenamefont{Cornish}(2021{\natexlab{b}})}]{PhysRevD.103.104057}
\bibinfo{author}{\bibfnamefont{N.~J.} \bibnamefont{Cornish}},
  \bibinfo{journal}{Phys. Rev. D} \textbf{\bibinfo{volume}{103}},
  \bibinfo{pages}{104057} (\bibinfo{year}{2021}{\natexlab{b}}),
  \urlprefix\url{https://link.aps.org/doi/10.1103/PhysRevD.103.104057}.

\bibitem[{\citenamefont{Morisaki and Raymond}(2020)}]{PhysRevD.102.104020}
\bibinfo{author}{\bibfnamefont{S.}~\bibnamefont{Morisaki}} \bibnamefont{and}
  \bibinfo{author}{\bibfnamefont{V.}~\bibnamefont{Raymond}},
  \bibinfo{journal}{Phys. Rev. D} \textbf{\bibinfo{volume}{102}},
  \bibinfo{pages}{104020} (\bibinfo{year}{2020}),
  \urlprefix\url{https://link.aps.org/doi/10.1103/PhysRevD.102.104020}.

\bibitem[{\citenamefont{{Yelikar} et~al.}(2023)\citenamefont{{Yelikar},
  {Delfavero}, and {O'Shaughnessy}}}]{2023arXiv230101337Y}
\bibinfo{author}{\bibfnamefont{A.~B.} \bibnamefont{{Yelikar}}},
  \bibinfo{author}{\bibfnamefont{V.}~\bibnamefont{{Delfavero}}},
  \bibnamefont{and}
  \bibinfo{author}{\bibfnamefont{R.}~\bibnamefont{{O'Shaughnessy}}},
  \bibinfo{journal}{arXiv e-prints} \bibinfo{eid}{arXiv:2301.01337}
  (\bibinfo{year}{2023}), \eprint{2301.01337}.

\bibitem[{\citenamefont{{O'Shaughnessy} and {others}}()}]{code-rift-newrepo}
\bibinfo{author}{\bibfnamefont{L.~J.} \bibnamefont{{O'Shaughnessy}},
  \bibfnamefont{R.}} \bibnamefont{and} \bibinfo{author}{\bibnamefont{{others}}}
  (????), \urlprefix\url{http://git.ligo.org/rapidpe-rift/rift}.

\bibitem[{\citenamefont{{Abbott}
  et~al.}(2021{\natexlab{b}})\citenamefont{{Abbott}, {Abbott}, {Abraham},
  {Acernese}, {Ackley}, {Adams}, {Adhikari}, {Adya}, {Affeldt}, {Agathos}
  et~al.}}]{2021SoftX..1300658A}
\bibinfo{author}{\bibfnamefont{R.}~\bibnamefont{{Abbott}}},
  \bibinfo{author}{\bibfnamefont{T.~D.} \bibnamefont{{Abbott}}},
  \bibinfo{author}{\bibfnamefont{S.}~\bibnamefont{{Abraham}}},
  \bibinfo{author}{\bibfnamefont{F.}~\bibnamefont{{Acernese}}},
  \bibinfo{author}{\bibfnamefont{K.}~\bibnamefont{{Ackley}}},
  \bibinfo{author}{\bibfnamefont{C.}~\bibnamefont{{Adams}}},
  \bibinfo{author}{\bibfnamefont{R.~X.} \bibnamefont{{Adhikari}}},
  \bibinfo{author}{\bibfnamefont{V.~B.} \bibnamefont{{Adya}}},
  \bibinfo{author}{\bibfnamefont{C.}~\bibnamefont{{Affeldt}}},
  \bibinfo{author}{\bibfnamefont{M.}~\bibnamefont{{Agathos}}},
  \bibnamefont{et~al.}, \bibinfo{journal}{SoftwareX}
  \textbf{\bibinfo{volume}{13}}, \bibinfo{eid}{100658}
  (\bibinfo{year}{2021}{\natexlab{b}}), \eprint{1912.11716}.

\bibitem[{\citenamefont{Lepage}(1980)}]{Lepage:1980dq}
\bibinfo{author}{\bibfnamefont{G.~P.} \bibnamefont{Lepage}},
  \bibinfo{journal}{Newman Laboratory of Nuclear studies report CLNS-80/447}
  (\bibinfo{year}{1980}),
  \urlprefix\url{https://lib-extopc.kek.jp/preprints/PDF/1980/8006/8006210.pdf}.

\bibitem[{\citenamefont{{Lepage}}(2021)}]{2021JCoPh.43910386L}
\bibinfo{author}{\bibfnamefont{G.~P.} \bibnamefont{{Lepage}}},
  \bibinfo{journal}{Journal of Computational Physics}
  \textbf{\bibinfo{volume}{439}}, \bibinfo{eid}{110386} (\bibinfo{year}{2021}),
  \eprint{2009.05112}.

\bibitem[{\citenamefont{Press et~al.}()\citenamefont{Press, Teukolsky,
  Flannery, and Vetterling}}]{book-mm-NumericalRecipies}
\bibinfo{author}{\bibnamefont{Press}},
  \bibinfo{author}{\bibnamefont{Teukolsky}},
  \bibinfo{author}{\bibnamefont{Flannery}}, \bibnamefont{and}
  \bibinfo{author}{\bibnamefont{Vetterling}}, \emph{\bibinfo{title}{Numerical
  recipies}} (????), \urlprefix\url{http://www.nr.com}.

\bibitem[{\citenamefont{Pedregosa et~al.}(2011)\citenamefont{Pedregosa,
  Varoquaux, Gramfort, Michel, Thirion, Grisel, Blondel, Prettenhofer, Weiss,
  Dubourg et~al.}}]{scikit-learn}
\bibinfo{author}{\bibfnamefont{F.}~\bibnamefont{Pedregosa}},
  \bibinfo{author}{\bibfnamefont{G.}~\bibnamefont{Varoquaux}},
  \bibinfo{author}{\bibfnamefont{A.}~\bibnamefont{Gramfort}},
  \bibinfo{author}{\bibfnamefont{V.}~\bibnamefont{Michel}},
  \bibinfo{author}{\bibfnamefont{B.}~\bibnamefont{Thirion}},
  \bibinfo{author}{\bibfnamefont{O.}~\bibnamefont{Grisel}},
  \bibinfo{author}{\bibfnamefont{M.}~\bibnamefont{Blondel}},
  \bibinfo{author}{\bibfnamefont{P.}~\bibnamefont{Prettenhofer}},
  \bibinfo{author}{\bibfnamefont{R.}~\bibnamefont{Weiss}},
  \bibinfo{author}{\bibfnamefont{V.}~\bibnamefont{Dubourg}},
  \bibnamefont{et~al.}, \bibinfo{journal}{Journal of Machine Learning Research}
  \textbf{\bibinfo{volume}{12}}, \bibinfo{pages}{2825} (\bibinfo{year}{2011}),
  \urlprefix\url{https://arxiv.org/abs/1201.0490}.

\bibitem[{\citenamefont{{Delfavero}}(2019)}]{gwastro-mergers-RIFT-DelfaveroThesis}
\bibinfo{author}{\bibfnamefont{M.}~\bibnamefont{{Delfavero}}},
  \emph{\bibinfo{title}{{Assessing the Convergence of Iterative Parameter
  Estimation}}} (\bibinfo{year}{2019}), \bibinfo{note}{mS thesis for RIT,
  available as https://scholarworks.rit.edu/theses/10153/}.

\bibitem[{\citenamefont{{O'Shaughnessy}
  et~al.}(2017)\citenamefont{{O'Shaughnessy}, {Blackman}, and
  {Field}}}]{gwastro-PE-AlternativeArchitecturesROM}
\bibinfo{author}{\bibfnamefont{R.}~\bibnamefont{{O'Shaughnessy}}},
  \bibinfo{author}{\bibfnamefont{J.}~\bibnamefont{{Blackman}}},
  \bibnamefont{and} \bibinfo{author}{\bibfnamefont{S.}~\bibnamefont{{Field}}},
  \bibinfo{journal}{\cqg}  (\bibinfo{year}{2017}),
  \urlprefix\url{http://iopscience.iop.org/article/10.1088/1361-6382/aa7649}.

\bibitem[{\citenamefont{{Husa} et~al.}(2016)\citenamefont{{Husa}, {Khan},
  {Hannam}, {P{\"u}rrer}, {Ohme}, {Forteza}, and
  {Boh{\'e}}}}]{2016PhRvD..93d4006H}
\bibinfo{author}{\bibfnamefont{S.}~\bibnamefont{{Husa}}},
  \bibinfo{author}{\bibfnamefont{S.}~\bibnamefont{{Khan}}},
  \bibinfo{author}{\bibfnamefont{M.}~\bibnamefont{{Hannam}}},
  \bibinfo{author}{\bibfnamefont{M.}~\bibnamefont{{P{\"u}rrer}}},
  \bibinfo{author}{\bibfnamefont{F.}~\bibnamefont{{Ohme}}},
  \bibinfo{author}{\bibfnamefont{X.~J.} \bibnamefont{{Forteza}}},
  \bibnamefont{and}
  \bibinfo{author}{\bibfnamefont{A.}~\bibnamefont{{Boh{\'e}}}},
  \bibinfo{journal}{\prd} \textbf{\bibinfo{volume}{93}}, \bibinfo{eid}{044006}
  (\bibinfo{year}{2016}), \eprint{1508.07250}.

\bibitem[{\citenamefont{{Khan} et~al.}(2016)\citenamefont{{Khan}, {Husa},
  {Hannam}, {Ohme}, {P{\"u}rrer}, {Forteza}, and
  {Boh{\'e}}}}]{2016PhRvD..93d4007K}
\bibinfo{author}{\bibfnamefont{S.}~\bibnamefont{{Khan}}},
  \bibinfo{author}{\bibfnamefont{S.}~\bibnamefont{{Husa}}},
  \bibinfo{author}{\bibfnamefont{M.}~\bibnamefont{{Hannam}}},
  \bibinfo{author}{\bibfnamefont{F.}~\bibnamefont{{Ohme}}},
  \bibinfo{author}{\bibfnamefont{M.}~\bibnamefont{{P{\"u}rrer}}},
  \bibinfo{author}{\bibfnamefont{X.~J.} \bibnamefont{{Forteza}}},
  \bibnamefont{and}
  \bibinfo{author}{\bibfnamefont{A.}~\bibnamefont{{Boh{\'e}}}},
  \bibinfo{journal}{\prd} \textbf{\bibinfo{volume}{93}}, \bibinfo{eid}{044007}
  (\bibinfo{year}{2016}), \eprint{1508.07253}.

\bibitem[{\citenamefont{{Cotesta} et~al.}(2018)\citenamefont{{Cotesta},
  {Buonanno}, {Boh{\'e}}, {Taracchini}, {Hinder}, and
  {Ossokine}}}]{2018PhRvD..98h4028C}
\bibinfo{author}{\bibfnamefont{R.}~\bibnamefont{{Cotesta}}},
  \bibinfo{author}{\bibfnamefont{A.}~\bibnamefont{{Buonanno}}},
  \bibinfo{author}{\bibfnamefont{A.}~\bibnamefont{{Boh{\'e}}}},
  \bibinfo{author}{\bibfnamefont{A.}~\bibnamefont{{Taracchini}}},
  \bibinfo{author}{\bibfnamefont{I.}~\bibnamefont{{Hinder}}}, \bibnamefont{and}
  \bibinfo{author}{\bibfnamefont{S.}~\bibnamefont{{Ossokine}}},
  \bibinfo{journal}{\prd} \textbf{\bibinfo{volume}{98}}, \bibinfo{eid}{084028}
  (\bibinfo{year}{2018}), \eprint{1803.10701}.

\bibitem[{\citenamefont{{Cook} et~al.}(2006)\citenamefont{{Cook}, {Gelman}, and
  {Rubin}}}]{mm-stats-PP}
\bibinfo{author}{\bibfnamefont{S.}~\bibnamefont{{Cook}}},
  \bibinfo{author}{\bibfnamefont{A.}~\bibnamefont{{Gelman}}}, \bibnamefont{and}
  \bibinfo{author}{\bibfnamefont{D.}~\bibnamefont{{Rubin}}},
  \bibinfo{journal}{Journal of Computational and Graphical Statistics}
  \textbf{\bibinfo{volume}{15}}, \bibinfo{pages}{675} (\bibinfo{year}{2006}),
  \urlprefix\url{https://www.tandfonline.com/doi/abs/10.1198/106186006X136976}.

\bibitem[{\citenamefont{{Sidery} et~al.}(2014)\citenamefont{{Sidery}, {Aylott},
  {Christensen}, {Farr}, {Farr}, {Feroz}, {Gair}, {Grover}, {Graff}, {Hanna}
  et~al.}}]{gwastro-skyloc-Sidery2013}
\bibinfo{author}{\bibfnamefont{T.}~\bibnamefont{{Sidery}}},
  \bibinfo{author}{\bibfnamefont{B.}~\bibnamefont{{Aylott}}},
  \bibinfo{author}{\bibfnamefont{N.}~\bibnamefont{{Christensen}}},
  \bibinfo{author}{\bibfnamefont{B.}~\bibnamefont{{Farr}}},
  \bibinfo{author}{\bibfnamefont{W.}~\bibnamefont{{Farr}}},
  \bibinfo{author}{\bibfnamefont{F.}~\bibnamefont{{Feroz}}},
  \bibinfo{author}{\bibfnamefont{J.}~\bibnamefont{{Gair}}},
  \bibinfo{author}{\bibfnamefont{K.}~\bibnamefont{{Grover}}},
  \bibinfo{author}{\bibfnamefont{P.}~\bibnamefont{{Graff}}},
  \bibinfo{author}{\bibfnamefont{C.}~\bibnamefont{{Hanna}}},
  \bibnamefont{et~al.}, \bibinfo{journal}{\prd} \textbf{\bibinfo{volume}{89}},
  \bibinfo{eid}{084060} (\bibinfo{year}{2014}), \eprint{1312.6013}.

\bibitem[{\citenamefont{{Jan} et~al.}(2020{\natexlab{b}})\citenamefont{{Jan},
  {Yelikar}, {Lange}, and
  {O'Shaughnessy}}}]{gwastro-RIFT-systematics-AnjaliAasim-2020}
\bibinfo{author}{\bibfnamefont{A.~Z.} \bibnamefont{{Jan}}},
  \bibinfo{author}{\bibfnamefont{A.~B.} \bibnamefont{{Yelikar}}},
  \bibinfo{author}{\bibfnamefont{J.}~\bibnamefont{{Lange}}}, \bibnamefont{and}
  \bibinfo{author}{\bibfnamefont{R.}~\bibnamefont{{O'Shaughnessy}}},
  \bibinfo{journal}{\prd} \textbf{\bibinfo{volume}{102}}, \bibinfo{eid}{124069}
  (\bibinfo{year}{2020}{\natexlab{b}}), \eprint{2011.03571}.

\bibitem[{\citenamefont{Thain et~al.}(2005)\citenamefont{Thain, Tannenbaum, and
  Livny}}]{condor-practice}
\bibinfo{author}{\bibfnamefont{D.}~\bibnamefont{Thain}},
  \bibinfo{author}{\bibfnamefont{T.}~\bibnamefont{Tannenbaum}},
  \bibnamefont{and} \bibinfo{author}{\bibfnamefont{M.}~\bibnamefont{Livny}},
  \bibinfo{journal}{Concurrency - Practice and Experience}
  \textbf{\bibinfo{volume}{17}}, \bibinfo{pages}{323} (\bibinfo{year}{2005}).

\bibitem[{\citenamefont{Bockelman et~al.}(2020)\citenamefont{Bockelman, Livny,
  Lin, and Prelz}}]{BOCKELMAN2020101213}
\bibinfo{author}{\bibfnamefont{B.}~\bibnamefont{Bockelman}},
  \bibinfo{author}{\bibfnamefont{M.}~\bibnamefont{Livny}},
  \bibinfo{author}{\bibfnamefont{B.}~\bibnamefont{Lin}}, \bibnamefont{and}
  \bibinfo{author}{\bibfnamefont{F.}~\bibnamefont{Prelz}},
  \bibinfo{journal}{Journal of Computational Science}  (\bibinfo{year}{2020}),
  ISSN \bibinfo{issn}{1877-7503},
  \urlprefix\url{http://www.sciencedirect.com/science/article/pii/S1877750320305147}.

\bibitem[{\citenamefont{Bockelman et~al.}(2015)\citenamefont{Bockelman,
  Cartwright, Frey, Fajardo, Lin, Selmeci, Tannenbaum, and
  Zvada}}]{1742-6596-664-6-062003}
\bibinfo{author}{\bibfnamefont{B.}~\bibnamefont{Bockelman}},
  \bibinfo{author}{\bibfnamefont{T.}~\bibnamefont{Cartwright}},
  \bibinfo{author}{\bibfnamefont{J.}~\bibnamefont{Frey}},
  \bibinfo{author}{\bibfnamefont{E.~M.} \bibnamefont{Fajardo}},
  \bibinfo{author}{\bibfnamefont{B.}~\bibnamefont{Lin}},
  \bibinfo{author}{\bibfnamefont{M.}~\bibnamefont{Selmeci}},
  \bibinfo{author}{\bibfnamefont{T.}~\bibnamefont{Tannenbaum}},
  \bibnamefont{and} \bibinfo{author}{\bibfnamefont{M.}~\bibnamefont{Zvada}},
  \bibinfo{journal}{Journal of Physics: Conference Series}
  \textbf{\bibinfo{volume}{664}}, \bibinfo{pages}{062003}
  (\bibinfo{year}{2015}),
  \urlprefix\url{http://stacks.iop.org/1742-6596/664/i=6/a=062003}.

\bibitem[{\citenamefont{{Williams}}()}]{code-asimov}
\bibinfo{author}{\bibfnamefont{D.}~\bibnamefont{{Williams}}},
  \bibinfo{journal}{Available at https://git.ligo.org/asimov/asimov}  (????),
  \urlprefix\url{https://git.ligo.org/asimov/asimov}.

\bibitem[{\citenamefont{{Cho} et~al.}(2013)\citenamefont{{Cho}, {Ochsner},
  {O'Shaughnessy}, {Kim}, and
  {Lee}}}]{gwastro-mergers-HeeSuk-FisherMatrixWithAmplitudeCorrections}
\bibinfo{author}{\bibfnamefont{H.}~\bibnamefont{{Cho}}},
  \bibinfo{author}{\bibfnamefont{E.}~\bibnamefont{{Ochsner}}},
  \bibinfo{author}{\bibfnamefont{R.}~\bibnamefont{{O'Shaughnessy}}},
  \bibinfo{author}{\bibfnamefont{C.}~\bibnamefont{{Kim}}}, \bibnamefont{and}
  \bibinfo{author}{\bibfnamefont{C.}~\bibnamefont{{Lee}}},
  \bibinfo{journal}{\prd} \textbf{\bibinfo{volume}{87}}, \bibinfo{pages}{02400}
  (\bibinfo{year}{2013}),
  \urlprefix\url{http://xxx.lanl.gov/abs/arXiv:1209.4494}.

\bibitem[{\citenamefont{{O'Shaughnessy}
  et~al.}(2014)\citenamefont{{O'Shaughnessy}, {Farr}, {Ochsner}, {Cho},
  {Raymond}, {Kim}, and {Lee}}}]{gwastro-mergers-HeeSuk-CompareToPE-Precessing}
\bibinfo{author}{\bibfnamefont{R.}~\bibnamefont{{O'Shaughnessy}}},
  \bibinfo{author}{\bibfnamefont{B.}~\bibnamefont{{Farr}}},
  \bibinfo{author}{\bibfnamefont{E.}~\bibnamefont{{Ochsner}}},
  \bibinfo{author}{\bibfnamefont{H.-S.} \bibnamefont{{Cho}}},
  \bibinfo{author}{\bibfnamefont{V.}~\bibnamefont{{Raymond}}},
  \bibinfo{author}{\bibfnamefont{C.}~\bibnamefont{{Kim}}}, \bibnamefont{and}
  \bibinfo{author}{\bibfnamefont{C.-H.} \bibnamefont{{Lee}}},
  \bibinfo{journal}{\prd} \textbf{\bibinfo{volume}{89}},
  \bibinfo{pages}{102005} (\bibinfo{year}{2014}),
  \urlprefix\url{http://link.aps.org/doi/10.1103/PhysRevD.89.102005}.

\bibitem[{\citenamefont{Lee et~al.}(2022)\citenamefont{Lee, Morisaki, and
  Tagoshi}}]{gwastro-mergers-SoichiroCoordinates-2022}
\bibinfo{author}{\bibfnamefont{E.}~\bibnamefont{Lee}},
  \bibinfo{author}{\bibfnamefont{S.}~\bibnamefont{Morisaki}}, \bibnamefont{and}
  \bibinfo{author}{\bibfnamefont{H.}~\bibnamefont{Tagoshi}},
  \bibinfo{journal}{Phys. Rev. D} \textbf{\bibinfo{volume}{105}},
  \bibinfo{pages}{124057} (\bibinfo{year}{2022}),
  \urlprefix\url{https://link.aps.org/doi/10.1103/PhysRevD.105.124057}.

\bibitem[{\citenamefont{{Blanchet}}(2014)}]{lrr-Blanchet-PN}
\bibinfo{author}{\bibfnamefont{L.}~\bibnamefont{{Blanchet}}},
  \bibinfo{journal}{Living Reviews in Relativity}
  \textbf{\bibinfo{volume}{17}}, \bibinfo{pages}{2} (\bibinfo{year}{2014}),
  \eprint{1310.1528}, \urlprefix\url{http://xxx.lanl.gov/abs/arXiv:1310.1528}.

\bibitem[{\citenamefont{{Arun} et~al.}(2009)\citenamefont{{Arun}, {Buonanno},
  {Faye}, and
  {Ochsner}}}]{gw-astro-mergers-approximations-SpinningPNHigherHarmonics}
\bibinfo{author}{\bibfnamefont{K.~G.} \bibnamefont{{Arun}}},
  \bibinfo{author}{\bibfnamefont{A.}~\bibnamefont{{Buonanno}}},
  \bibinfo{author}{\bibfnamefont{G.}~\bibnamefont{{Faye}}}, \bibnamefont{and}
  \bibinfo{author}{\bibfnamefont{E.}~\bibnamefont{{Ochsner}}},
  \bibinfo{journal}{\prd} \textbf{\bibinfo{volume}{79}},
  \bibinfo{pages}{104023} (\bibinfo{year}{2009}), \eprint{0810.5336}.

\bibitem[{\citenamefont{{Poisson} and {Will}}(1995)}]{1995PhRvD..52..848P}
\bibinfo{author}{\bibfnamefont{E.}~\bibnamefont{{Poisson}}} \bibnamefont{and}
  \bibinfo{author}{\bibfnamefont{C.~M.} \bibnamefont{{Will}}},
  \bibinfo{journal}{\prd} \textbf{\bibinfo{volume}{52}}, \bibinfo{pages}{848}
  (\bibinfo{year}{1995}).

\bibitem[{\citenamefont{{Romero-Shaw} et~al.}(2020)\citenamefont{{Romero-Shaw},
  {Talbot}, {Biscoveanu}, {D'Emilio}, {Ashton}, {Berry}, {Coughlin},
  {Galaudage}, {Hoy}, {H{\"u}bner} et~al.}}]{2020MNRAS.499.3295R}
\bibinfo{author}{\bibfnamefont{I.~M.} \bibnamefont{{Romero-Shaw}}},
  \bibinfo{author}{\bibfnamefont{C.}~\bibnamefont{{Talbot}}},
  \bibinfo{author}{\bibfnamefont{S.}~\bibnamefont{{Biscoveanu}}},
  \bibinfo{author}{\bibfnamefont{V.}~\bibnamefont{{D'Emilio}}},
  \bibinfo{author}{\bibfnamefont{G.}~\bibnamefont{{Ashton}}},
  \bibinfo{author}{\bibfnamefont{C.~P.~L.} \bibnamefont{{Berry}}},
  \bibinfo{author}{\bibfnamefont{S.}~\bibnamefont{{Coughlin}}},
  \bibinfo{author}{\bibfnamefont{S.}~\bibnamefont{{Galaudage}}},
  \bibinfo{author}{\bibfnamefont{C.}~\bibnamefont{{Hoy}}},
  \bibinfo{author}{\bibfnamefont{M.}~\bibnamefont{{H{\"u}bner}}},
  \bibnamefont{et~al.}, \bibinfo{journal}{\mnras}
  \textbf{\bibinfo{volume}{499}}, \bibinfo{pages}{3295} (\bibinfo{year}{2020}),
  \eprint{2006.00714}.

\bibitem[{\citenamefont{Gerosa et~al.}(2021)\citenamefont{Gerosa, Mould,
  Gangardt, Schmidt, Pratten, and Thomas}}]{Gerosa_chipavg_2021}
\bibinfo{author}{\bibfnamefont{D.}~\bibnamefont{Gerosa}},
  \bibinfo{author}{\bibfnamefont{M.}~\bibnamefont{Mould}},
  \bibinfo{author}{\bibfnamefont{D.}~\bibnamefont{Gangardt}},
  \bibinfo{author}{\bibfnamefont{P.}~\bibnamefont{Schmidt}},
  \bibinfo{author}{\bibfnamefont{G.}~\bibnamefont{Pratten}}, \bibnamefont{and}
  \bibinfo{author}{\bibfnamefont{L.~M.} \bibnamefont{Thomas}},
  \bibinfo{journal}{Physical Review D} \textbf{\bibinfo{volume}{103}}
  (\bibinfo{year}{2021}), ISSN \bibinfo{issn}{2470-0029},
  \urlprefix\url{http://dx.doi.org/10.1103/PhysRevD.103.064067}.

\bibitem[{\citenamefont{{Henshaw} et~al.}(2022)\citenamefont{{Henshaw},
  {O'Shaughnessy}, and {Cadonati}}}]{precession_RIFT}
\bibinfo{author}{\bibfnamefont{C.}~\bibnamefont{{Henshaw}}},
  \bibinfo{author}{\bibfnamefont{R.}~\bibnamefont{{O'Shaughnessy}}},
  \bibnamefont{and}
  \bibinfo{author}{\bibfnamefont{L.}~\bibnamefont{{Cadonati}}},
  \bibinfo{journal}{Classical and Quantum Gravity}
  \textbf{\bibinfo{volume}{39}}, \bibinfo{eid}{125003} (\bibinfo{year}{2022}),
  \eprint{2201.05220}.

\bibitem[{\citenamefont{{De Renzis} et~al.}(2022)\citenamefont{{De Renzis},
  {Gerosa}, {Pratten}, {Schmidt}, and {Mould}}}]{2022arXiv220700030D}
\bibinfo{author}{\bibfnamefont{V.}~\bibnamefont{{De Renzis}}},
  \bibinfo{author}{\bibfnamefont{D.}~\bibnamefont{{Gerosa}}},
  \bibinfo{author}{\bibfnamefont{G.}~\bibnamefont{{Pratten}}},
  \bibinfo{author}{\bibfnamefont{P.}~\bibnamefont{{Schmidt}}},
  \bibnamefont{and} \bibinfo{author}{\bibfnamefont{M.}~\bibnamefont{{Mould}}},
  \bibinfo{journal}{arXiv e-prints} \bibinfo{eid}{arXiv:2207.00030}
  (\bibinfo{year}{2022}), \eprint{2207.00030}.

\bibitem[{\citenamefont{{Wysocki}
  et~al.}(2020{\natexlab{b}})\citenamefont{{Wysocki}, {O'Shaughnessy}, {Wade},
  and {Lange}}}]{2020arXiv200101747W}
\bibinfo{author}{\bibfnamefont{D.}~\bibnamefont{{Wysocki}}},
  \bibinfo{author}{\bibfnamefont{R.}~\bibnamefont{{O'Shaughnessy}}},
  \bibinfo{author}{\bibfnamefont{L.}~\bibnamefont{{Wade}}}, \bibnamefont{and}
  \bibinfo{author}{\bibfnamefont{J.}~\bibnamefont{{Lange}}},
  \bibinfo{journal}{arXiv e-prints} \bibinfo{eid}{arXiv:2001.01747}
  (\bibinfo{year}{2020}{\natexlab{b}}), \eprint{2001.01747}.

\bibitem[{\citenamefont{{Breiman}}(2001)}]{2001MachL..45....5B}
\bibinfo{author}{\bibfnamefont{L.}~\bibnamefont{{Breiman}}},
  \bibinfo{journal}{Machine Learning} \textbf{\bibinfo{volume}{45}},
  \bibinfo{pages}{5} (\bibinfo{year}{2001}).

\bibitem[{\citenamefont{Murphy}(2012)}]{book-Murphy-MachineLearning}
\bibinfo{author}{\bibfnamefont{K.~P.} \bibnamefont{Murphy}},
  \emph{\bibinfo{title}{Machine Learning: A Probabilistic Perspective}}
  (\bibinfo{publisher}{The MIT Press}, \bibinfo{year}{2012}), ISBN
  \bibinfo{isbn}{0262018020, 9780262018029}.

\bibitem[{\citenamefont{{Geurts} et~al.}(2006)\citenamefont{{Geurts}, {Ernst},
  and {Wehenkel}}}]{mm-stat-ExtraTrees}
\bibinfo{author}{\bibfnamefont{P.}~\bibnamefont{{Geurts}}},
  \bibinfo{author}{\bibfnamefont{D.}~\bibnamefont{{Ernst}}}, \bibnamefont{and}
  \bibinfo{author}{\bibfnamefont{L.}~\bibnamefont{{Wehenkel}}},
  \bibinfo{journal}{Machine learning} \textbf{\bibinfo{volume}{63}},
  \bibinfo{pages}{3} (\bibinfo{year}{2006}).

\bibitem[{\citenamefont{{Rasmussen} and {Williams}}(2006)}]{book-Rasmussen-GP}
\bibinfo{author}{\bibfnamefont{C.}~\bibnamefont{{Rasmussen}}} \bibnamefont{and}
  \bibinfo{author}{\bibfnamefont{C.}~\bibnamefont{{Williams}}},
  \emph{\bibinfo{title}{{Gaussian Processes for Machine Learning}}}
  (\bibinfo{publisher}{The MIT Press}, \bibinfo{year}{2006}).

\bibitem[{\citenamefont{{Bauer} et~al.}(2016)\citenamefont{{Bauer}, {van der
  Wilk}, and {Rasmussen}}}]{2016arXiv160604820B}
\bibinfo{author}{\bibfnamefont{M.}~\bibnamefont{{Bauer}}},
  \bibinfo{author}{\bibfnamefont{M.}~\bibnamefont{{van der Wilk}}},
  \bibnamefont{and} \bibinfo{author}{\bibfnamefont{C.~E.}
  \bibnamefont{{Rasmussen}}}, \bibinfo{journal}{arXiv e-prints}
  (\bibinfo{year}{2016}), \eprint{1606.04820}.

\bibitem[{\citenamefont{{{\'A}lvarez} et~al.}(2009)\citenamefont{{{\'A}lvarez},
  {Luengo}, {Titsias}, and {Lawrence}}}]{2009arXiv0912.3268A}
\bibinfo{author}{\bibfnamefont{M.~A.} \bibnamefont{{{\'A}lvarez}}},
  \bibinfo{author}{\bibfnamefont{D.}~\bibnamefont{{Luengo}}},
  \bibinfo{author}{\bibfnamefont{M.~K.} \bibnamefont{{Titsias}}},
  \bibnamefont{and} \bibinfo{author}{\bibfnamefont{N.~D.}
  \bibnamefont{{Lawrence}}}, \bibinfo{journal}{arXiv e-prints}
  (\bibinfo{year}{2009}), \eprint{0912.3268}.

\bibitem[{\citenamefont{{Hensman} et~al.}(2013)\citenamefont{{Hensman}, {Fusi},
  and {Lawrence}}}]{2013arXiv1309.6835H}
\bibinfo{author}{\bibfnamefont{J.}~\bibnamefont{{Hensman}}},
  \bibinfo{author}{\bibfnamefont{N.}~\bibnamefont{{Fusi}}}, \bibnamefont{and}
  \bibinfo{author}{\bibfnamefont{N.~D.} \bibnamefont{{Lawrence}}},
  \bibinfo{journal}{arXiv e-prints}  (\bibinfo{year}{2013}),
  \eprint{1309.6835}.

\bibitem[{\citenamefont{{Jankowiak} et~al.}(2019)\citenamefont{{Jankowiak},
  {Pleiss}, and {Gardner}}}]{2019arXiv191007123J}
\bibinfo{author}{\bibfnamefont{M.}~\bibnamefont{{Jankowiak}}},
  \bibinfo{author}{\bibfnamefont{G.}~\bibnamefont{{Pleiss}}}, \bibnamefont{and}
  \bibinfo{author}{\bibfnamefont{J.~R.} \bibnamefont{{Gardner}}},
  \bibinfo{journal}{arXiv e-prints}  (\bibinfo{year}{2019}),
  \eprint{1910.07123}.

\bibitem[{\citenamefont{{Rosenbrock}}(1960)}]{rosenbrock}
\bibinfo{author}{\bibfnamefont{H.~H.} \bibnamefont{{Rosenbrock}}},
  \bibinfo{journal}{The Computer Journal} \textbf{\bibinfo{volume}{3}},
  \bibinfo{pages}{175} (\bibinfo{year}{1960}).

\bibitem[{\citenamefont{{Fowlie} et~al.}(2020)\citenamefont{{Fowlie},
  {Handley}, and {Su}}}]{2020MNRAS.497.5256F}
\bibinfo{author}{\bibfnamefont{A.}~\bibnamefont{{Fowlie}}},
  \bibinfo{author}{\bibfnamefont{W.}~\bibnamefont{{Handley}}},
  \bibnamefont{and} \bibinfo{author}{\bibfnamefont{L.}~\bibnamefont{{Su}}},
  \bibinfo{journal}{\mnras} \textbf{\bibinfo{volume}{497}},
  \bibinfo{pages}{5256} (\bibinfo{year}{2020}), \eprint{2006.03371}.

\bibitem[{\citenamefont{{Dempster} et~al.}(1977)\citenamefont{{Dempster},
  {Laird}, and {Rubin}}}]{mm-stat-EM}
\bibinfo{author}{\bibfnamefont{A.}~\bibnamefont{{Dempster}}},
  \bibinfo{author}{\bibfnamefont{N.}~\bibnamefont{{Laird}}}, \bibnamefont{and}
  \bibinfo{author}{\bibfnamefont{D.}~\bibnamefont{{Rubin}}},
  \bibinfo{journal}{Journal of the Royal Statistical Society, Series B}
  \textbf{\bibinfo{volume}{39}}, \bibinfo{pages}{1} (\bibinfo{year}{1977}).

\bibitem[{\citenamefont{M.R. and {Chen}}(2010)}]{mm-stat-EM-theory}
\bibinfo{author}{\bibfnamefont{G.}~\bibnamefont{M.R.}} \bibnamefont{and}
  \bibinfo{author}{\bibfnamefont{Y.}~\bibnamefont{{Chen}}},
  \bibinfo{journal}{Foundations and Trends in Signal Processing}
  \textbf{\bibinfo{volume}{4}}, \bibinfo{pages}{223} (\bibinfo{year}{2010}).

\bibitem[{\citenamefont{{McLachlan} and Krishnan}(2008)}]{mm-stat-EM-book}
\bibinfo{author}{\bibfnamefont{G.}~\bibnamefont{{McLachlan}}} \bibnamefont{and}
  \bibinfo{author}{\bibfnamefont{T.}~\bibnamefont{Krishnan}},
  \emph{\bibinfo{title}{{The EM Algorithm and Extensions}}}
  (\bibinfo{publisher}{John Wiley and Sons}, \bibinfo{year}{2008}).

\bibitem[{\citenamefont{{Ashton} and {Talbot}}(2021)}]{2021MNRAS.507.2037A}
\bibinfo{author}{\bibfnamefont{G.}~\bibnamefont{{Ashton}}} \bibnamefont{and}
  \bibinfo{author}{\bibfnamefont{C.}~\bibnamefont{{Talbot}}},
  \bibinfo{journal}{\mnras} \textbf{\bibinfo{volume}{507}},
  \bibinfo{pages}{2037} (\bibinfo{year}{2021}), \eprint{2106.08730}.

\bibitem[{\citenamefont{{Thrane} and {Talbot}}(2020)}]{2020PASA...37...36T}
\bibinfo{author}{\bibfnamefont{E.}~\bibnamefont{{Thrane}}} \bibnamefont{and}
  \bibinfo{author}{\bibfnamefont{C.}~\bibnamefont{{Talbot}}},
  \bibinfo{journal}{\pasa} \textbf{\bibinfo{volume}{37}}, \bibinfo{eid}{e036}
  (\bibinfo{year}{2020}).

\bibitem[{\citenamefont{{Morisaki}}(2021)}]{gwastro-mergers-Soichiro-dmarg}
\bibinfo{author}{\bibfnamefont{S.}~\bibnamefont{{Morisaki}}},
  \bibinfo{journal}{LIGO DCC T2100485}  (\bibinfo{year}{2021}),
  \urlprefix\url{https://dcc.ligo.org/LIGO-T2100485}.

\bibitem[{\citenamefont{{Delfavero} et~al.}(2021)\citenamefont{{Delfavero},
  {O'Shaughnessy}, {Wysocki}, and
  {Yelikar}}}]{gwastro-mergers-GaussianLikelihoods-Delfavero2021}
\bibinfo{author}{\bibfnamefont{V.}~\bibnamefont{{Delfavero}}},
  \bibinfo{author}{\bibfnamefont{R.}~\bibnamefont{{O'Shaughnessy}}},
  \bibinfo{author}{\bibfnamefont{D.}~\bibnamefont{{Wysocki}}},
  \bibnamefont{and}
  \bibinfo{author}{\bibfnamefont{A.}~\bibnamefont{{Yelikar}}},
  \bibinfo{journal}{arXiv e-prints} \bibinfo{eid}{arXiv:2107.13082}
  (\bibinfo{year}{2021}), \eprint{2107.13082}.

\bibitem[{\citenamefont{{Rose} et~al.}(2022)\citenamefont{{Rose}, {Valsan},
  {Brady}, {Walsh}, and {Pankow}}}]{2022arXiv220105263R}
\bibinfo{author}{\bibfnamefont{C.~A.} \bibnamefont{{Rose}}},
  \bibinfo{author}{\bibfnamefont{V.}~\bibnamefont{{Valsan}}},
  \bibinfo{author}{\bibfnamefont{P.~R.} \bibnamefont{{Brady}}},
  \bibinfo{author}{\bibfnamefont{S.}~\bibnamefont{{Walsh}}}, \bibnamefont{and}
  \bibinfo{author}{\bibfnamefont{C.}~\bibnamefont{{Pankow}}},
  \bibinfo{journal}{arXiv e-prints} \bibinfo{eid}{arXiv:2201.05263}
  (\bibinfo{year}{2022}), \eprint{2201.05263}.

\bibitem[{\citenamefont{{The LIGO Scientific Collaboration}
  et~al.}(2018{\natexlab{b}})\citenamefont{{The LIGO Scientific Collaboration},
  {the Virgo Collaboration}, {Abbott}, {Abbott}, {Abbott}, {Acernese},
  {Ackley}, {Adams}, {Adams}, {Addesso} et~al.}}]{LIGO-GW170817-EOS}
\bibinfo{author}{\bibnamefont{{The LIGO Scientific Collaboration}}},
  \bibinfo{author}{\bibnamefont{{the Virgo Collaboration}}},
  \bibinfo{author}{\bibfnamefont{B.~P.} \bibnamefont{{Abbott}}},
  \bibinfo{author}{\bibfnamefont{R.}~\bibnamefont{{Abbott}}},
  \bibinfo{author}{\bibfnamefont{T.~D.} \bibnamefont{{Abbott}}},
  \bibinfo{author}{\bibfnamefont{F.}~\bibnamefont{{Acernese}}},
  \bibinfo{author}{\bibfnamefont{K.}~\bibnamefont{{Ackley}}},
  \bibinfo{author}{\bibfnamefont{C.}~\bibnamefont{{Adams}}},
  \bibinfo{author}{\bibfnamefont{T.}~\bibnamefont{{Adams}}},
  \bibinfo{author}{\bibfnamefont{P.}~\bibnamefont{{Addesso}}},
  \bibnamefont{et~al.}, \bibinfo{journal}{\prl} \textbf{\bibinfo{volume}{121}},
  \bibinfo{pages}{161101} (\bibinfo{year}{2018}{\natexlab{b}}).

\bibitem[{\citenamefont{{Capano} et~al.}(2020)\citenamefont{{Capano}, {Tews},
  {Brown}, {Margalit}, {De}, {Kumar}, {Brown}, {Krishnan}, and
  {Reddy}}}]{2020NatAs...4..625C}
\bibinfo{author}{\bibfnamefont{C.~D.} \bibnamefont{{Capano}}},
  \bibinfo{author}{\bibfnamefont{I.}~\bibnamefont{{Tews}}},
  \bibinfo{author}{\bibfnamefont{S.~M.} \bibnamefont{{Brown}}},
  \bibinfo{author}{\bibfnamefont{B.}~\bibnamefont{{Margalit}}},
  \bibinfo{author}{\bibfnamefont{S.}~\bibnamefont{{De}}},
  \bibinfo{author}{\bibfnamefont{S.}~\bibnamefont{{Kumar}}},
  \bibinfo{author}{\bibfnamefont{D.~A.} \bibnamefont{{Brown}}},
  \bibinfo{author}{\bibfnamefont{B.}~\bibnamefont{{Krishnan}}},
  \bibnamefont{and} \bibinfo{author}{\bibfnamefont{S.}~\bibnamefont{{Reddy}}},
  \bibinfo{journal}{Nature Astronomy} \textbf{\bibinfo{volume}{4}},
  \bibinfo{pages}{625} (\bibinfo{year}{2020}), \eprint{1908.10352}.

\bibitem[{\citenamefont{{Landry} and {Essick}}(2019)}]{2019PhRvD..99h4049L}
\bibinfo{author}{\bibfnamefont{P.}~\bibnamefont{{Landry}}} \bibnamefont{and}
  \bibinfo{author}{\bibfnamefont{R.}~\bibnamefont{{Essick}}},
  \bibinfo{journal}{\prd} \textbf{\bibinfo{volume}{99}}, \bibinfo{eid}{084049}
  (\bibinfo{year}{2019}), \eprint{1811.12529}.

\bibitem[{\citenamefont{{Legred} et~al.}(2021)\citenamefont{{Legred},
  {Chatziioannou}, {Essick}, {Han}, and {Landry}}}]{2021PhRvD.104f3003L}
\bibinfo{author}{\bibfnamefont{I.}~\bibnamefont{{Legred}}},
  \bibinfo{author}{\bibfnamefont{K.}~\bibnamefont{{Chatziioannou}}},
  \bibinfo{author}{\bibfnamefont{R.}~\bibnamefont{{Essick}}},
  \bibinfo{author}{\bibfnamefont{S.}~\bibnamefont{{Han}}}, \bibnamefont{and}
  \bibinfo{author}{\bibfnamefont{P.}~\bibnamefont{{Landry}}},
  \bibinfo{journal}{\prd} \textbf{\bibinfo{volume}{104}}, \bibinfo{eid}{063003}
  (\bibinfo{year}{2021}), \eprint{2106.05313}.

\bibitem[{\citenamefont{{Gorda} et~al.}(2022)\citenamefont{{Gorda},
  {Komoltsev}, and {Kurkela}}}]{2022arXiv220411877G}
\bibinfo{author}{\bibfnamefont{T.}~\bibnamefont{{Gorda}}},
  \bibinfo{author}{\bibfnamefont{O.}~\bibnamefont{{Komoltsev}}},
  \bibnamefont{and}
  \bibinfo{author}{\bibfnamefont{A.}~\bibnamefont{{Kurkela}}},
  \bibinfo{journal}{arXiv e-prints} \bibinfo{eid}{arXiv:2204.11877}
  (\bibinfo{year}{2022}), \eprint{2204.11877}.

\bibitem[{\citenamefont{{Abbott et al. (The LIGO Scientific Collaboration and
  the Virgo Collaboration)}}(2016)}]{NRPaper}
\bibinfo{author}{\bibfnamefont{B.}~\bibnamefont{{Abbott et al. (The LIGO
  Scientific Collaboration and the Virgo Collaboration)}}},
  \bibinfo{journal}{\prd} \textbf{\bibinfo{volume}{94}},
  \bibinfo{pages}{064035} (\bibinfo{year}{2016}),
  \urlprefix\url{http://link.aps.org/doi/10.1103/PhysRevD.94.064035}.

\bibitem[{\citenamefont{{LIGO Scientific
  Collaboration}}()}]{gwtc2-nrsample-release}
\bibinfo{author}{\bibnamefont{{LIGO Scientific Collaboration}}},
  \bibinfo{journal}{Available as LIGO-P1900124 from dcc.ligo.org}  (????),
  \urlprefix\url{https://dcc.ligo.org/LIGO-P1900124/public}.

\bibitem[{\citenamefont{{Mateu-Lucena}
  et~al.}(2021)\citenamefont{{Mateu-Lucena}, {Husa}, {Colleoni},
  {Estell{\'e}s}, {Garc{\'\i}a-Quir{\'o}s}, {Keitel}, {de Lluc Planas}, and
  {Ramos-Buades}}}]{2021arXiv210505960M}
\bibinfo{author}{\bibfnamefont{M.}~\bibnamefont{{Mateu-Lucena}}},
  \bibinfo{author}{\bibfnamefont{S.}~\bibnamefont{{Husa}}},
  \bibinfo{author}{\bibfnamefont{M.}~\bibnamefont{{Colleoni}}},
  \bibinfo{author}{\bibfnamefont{H.}~\bibnamefont{{Estell{\'e}s}}},
  \bibinfo{author}{\bibfnamefont{C.}~\bibnamefont{{Garc{\'\i}a-Quir{\'o}s}}},
  \bibinfo{author}{\bibfnamefont{D.}~\bibnamefont{{Keitel}}},
  \bibinfo{author}{\bibfnamefont{M.}~\bibnamefont{{de Lluc Planas}}},
  \bibnamefont{and}
  \bibinfo{author}{\bibfnamefont{A.}~\bibnamefont{{Ramos-Buades}}},
  \bibinfo{journal}{arXiv e-prints} \bibinfo{eid}{arXiv:2105.05960}
  (\bibinfo{year}{2021}), \eprint{2105.05960}.

\bibitem[{\citenamefont{{Vajpeyi} et~al.}(2022)\citenamefont{{Vajpeyi},
  {Smith}, and {Thrane}}}]{gwastro-151226-RoryThraneFollowupHighQ}
\bibinfo{author}{\bibfnamefont{A.}~\bibnamefont{{Vajpeyi}}},
  \bibinfo{author}{\bibfnamefont{R.}~\bibnamefont{{Smith}}}, \bibnamefont{and}
  \bibinfo{author}{\bibfnamefont{E.}~\bibnamefont{{Thrane}}},
  \bibinfo{journal}{arXiv e-prints} \bibinfo{eid}{arXiv:2203.13406}
  (\bibinfo{year}{2022}), \eprint{2203.13406}.

\bibitem[{\citenamefont{{Chia} et~al.}(2022)\citenamefont{{Chia}, {Olsen},
  {Roulet}, {Dai}, {Venumadhav}, {Zackay}, and
  {Zaldarriaga}}}]{gwastro-151226-ChiaEtAl}
\bibinfo{author}{\bibfnamefont{H.~S.} \bibnamefont{{Chia}}},
  \bibinfo{author}{\bibfnamefont{S.}~\bibnamefont{{Olsen}}},
  \bibinfo{author}{\bibfnamefont{J.}~\bibnamefont{{Roulet}}},
  \bibinfo{author}{\bibfnamefont{L.}~\bibnamefont{{Dai}}},
  \bibinfo{author}{\bibfnamefont{T.}~\bibnamefont{{Venumadhav}}},
  \bibinfo{author}{\bibfnamefont{B.}~\bibnamefont{{Zackay}}}, \bibnamefont{and}
  \bibinfo{author}{\bibfnamefont{M.}~\bibnamefont{{Zaldarriaga}}},
  \bibinfo{journal}{\prd} \textbf{\bibinfo{volume}{106}}, \bibinfo{eid}{024009}
  (\bibinfo{year}{2022}), \eprint{2105.06486}.

\bibitem[{\citenamefont{{Dax} et~al.}(2021)\citenamefont{{Dax}, {Green},
  {Gair}, {Macke}, {Buonanno}, and {Sch{\"o}lkopf}}}]{2021PhRvL.127x1103D}
\bibinfo{author}{\bibfnamefont{M.}~\bibnamefont{{Dax}}},
  \bibinfo{author}{\bibfnamefont{S.~R.} \bibnamefont{{Green}}},
  \bibinfo{author}{\bibfnamefont{J.}~\bibnamefont{{Gair}}},
  \bibinfo{author}{\bibfnamefont{J.~H.} \bibnamefont{{Macke}}},
  \bibinfo{author}{\bibfnamefont{A.}~\bibnamefont{{Buonanno}}},
  \bibnamefont{and}
  \bibinfo{author}{\bibfnamefont{B.}~\bibnamefont{{Sch{\"o}lkopf}}},
  \bibinfo{journal}{\prl} \textbf{\bibinfo{volume}{127}}, \bibinfo{eid}{241103}
  (\bibinfo{year}{2021}), \eprint{2106.12594}.

\bibitem[{\citenamefont{{Chua} and {Vallisneri}}(2020)}]{2020PhRvL.124d1102C}
\bibinfo{author}{\bibfnamefont{A.~J.~K.} \bibnamefont{{Chua}}}
  \bibnamefont{and}
  \bibinfo{author}{\bibfnamefont{M.}~\bibnamefont{{Vallisneri}}},
  \bibinfo{journal}{\prl} \textbf{\bibinfo{volume}{124}}, \bibinfo{eid}{041102}
  (\bibinfo{year}{2020}), \eprint{1909.05966}.

\bibitem[{\citenamefont{{Delaunoy} et~al.}(2020)\citenamefont{{Delaunoy},
  {Wehenkel}, {Hinderer}, {Nissanke}, {Weniger}, {Williamson}, and
  {Louppe}}}]{2020arXiv201012931D}
\bibinfo{author}{\bibfnamefont{A.}~\bibnamefont{{Delaunoy}}},
  \bibinfo{author}{\bibfnamefont{A.}~\bibnamefont{{Wehenkel}}},
  \bibinfo{author}{\bibfnamefont{T.}~\bibnamefont{{Hinderer}}},
  \bibinfo{author}{\bibfnamefont{S.}~\bibnamefont{{Nissanke}}},
  \bibinfo{author}{\bibfnamefont{C.}~\bibnamefont{{Weniger}}},
  \bibinfo{author}{\bibfnamefont{A.~R.} \bibnamefont{{Williamson}}},
  \bibnamefont{and} \bibinfo{author}{\bibfnamefont{G.}~\bibnamefont{{Louppe}}},
  \bibinfo{journal}{arXiv e-prints} \bibinfo{eid}{arXiv:2010.12931}
  (\bibinfo{year}{2020}), \eprint{2010.12931}.

\bibitem[{\citenamefont{{Gabbard} et~al.}(2022)\citenamefont{{Gabbard},
  {Messenger}, {Heng}, {Tonolini}, and {Murray-Smith}}}]{2022NatPh..18..112G}
\bibinfo{author}{\bibfnamefont{H.}~\bibnamefont{{Gabbard}}},
  \bibinfo{author}{\bibfnamefont{C.}~\bibnamefont{{Messenger}}},
  \bibinfo{author}{\bibfnamefont{I.~S.} \bibnamefont{{Heng}}},
  \bibinfo{author}{\bibfnamefont{F.}~\bibnamefont{{Tonolini}}},
  \bibnamefont{and}
  \bibinfo{author}{\bibfnamefont{R.}~\bibnamefont{{Murray-Smith}}},
  \bibinfo{journal}{Nature Physics} \textbf{\bibinfo{volume}{18}},
  \bibinfo{pages}{112} (\bibinfo{year}{2022}), \eprint{1909.06296}.

\bibitem[{\citenamefont{{Krastev} et~al.}(2021)\citenamefont{{Krastev}, {Gill},
  {Villar}, and {Berger}}}]{2021PhLB..81536161K}
\bibinfo{author}{\bibfnamefont{P.~G.} \bibnamefont{{Krastev}}},
  \bibinfo{author}{\bibfnamefont{K.}~\bibnamefont{{Gill}}},
  \bibinfo{author}{\bibfnamefont{V.~A.} \bibnamefont{{Villar}}},
  \bibnamefont{and} \bibinfo{author}{\bibfnamefont{E.}~\bibnamefont{{Berger}}},
  \bibinfo{journal}{Physics Letters B} \textbf{\bibinfo{volume}{815}},
  \bibinfo{eid}{136161} (\bibinfo{year}{2021}), \eprint{2012.13101}.

\bibitem[{\citenamefont{{Kish}}(1965)}]{kish}
\bibinfo{author}{\bibfnamefont{L.}~\bibnamefont{{Kish}}},
  \emph{\bibinfo{title}{{Survey sampling}}} (\bibinfo{publisher}{John Wiley \&
  Sons, Inc, London}, \bibinfo{year}{1965}), ISBN
  \bibinfo{isbn}{0-471-10949-5}.

\bibitem[{\citenamefont{{Farr}}(2019)}]{2019RNAAS...3...66F}
\bibinfo{author}{\bibfnamefont{W.~M.} \bibnamefont{{Farr}}},
  \bibinfo{journal}{Research Notes of the American Astronomical Society}
  \textbf{\bibinfo{volume}{3}}, \bibinfo{eid}{66} (\bibinfo{year}{2019}),
  \eprint{1904.10879}.

\bibitem[{\citenamefont{{Roulet} et~al.}(2022)\citenamefont{{Roulet}, {Olsen},
  {Mushkin}, {Islam}, {Venumadhav}, {Zackay}, and
  {Zaldarriaga}}}]{2022arXiv220703508R}
\bibinfo{author}{\bibfnamefont{J.}~\bibnamefont{{Roulet}}},
  \bibinfo{author}{\bibfnamefont{S.}~\bibnamefont{{Olsen}}},
  \bibinfo{author}{\bibfnamefont{J.}~\bibnamefont{{Mushkin}}},
  \bibinfo{author}{\bibfnamefont{T.}~\bibnamefont{{Islam}}},
  \bibinfo{author}{\bibfnamefont{T.}~\bibnamefont{{Venumadhav}}},
  \bibinfo{author}{\bibfnamefont{B.}~\bibnamefont{{Zackay}}}, \bibnamefont{and}
  \bibinfo{author}{\bibfnamefont{M.}~\bibnamefont{{Zaldarriaga}}},
  \bibinfo{journal}{arXiv e-prints} \bibinfo{eid}{arXiv:2207.03508}
  (\bibinfo{year}{2022}), \eprint{2207.03508}.

\end{thebibliography}
\end{document}